\documentstyle[emulateapj,psfig]{article}

\makeatletter

\newenvironment{inlinefigure}{%
\def\@captype{figure}%
\noindent\begin{minipage}{0.999\linewidth}\begin{center}}
{\end{center}\end{minipage}\smallskip}
\makeatother

\lefthead{Barger et al.}

\begin{document}
\title{A Highly Complete Spectroscopic Survey of the GOODS-N 
Field\altaffilmark{1,2,3}} 
\author{
A.~J.~Barger,$\!$\altaffilmark{4,5,6}
L.~L.~Cowie,$\!$\altaffilmark{6}
W.-H.~Wang$\!$\altaffilmark{7,8}
}

\altaffiltext{1}{Based in part on data obtained at the W. M. Keck
Observatory, which is operated as a scientific partnership among the
the California Institute of Technology, the University of
California, and NASA and was made possible by the generous financial
support of the W. M. Keck Foundation.}
\altaffiltext{2}{Based in part on data obtained at the 
Canada-France-Hawaii Telescope, which is operated by the National
Research Council of Canada, the Institut des Sciences de l'Univers
of the Centre National de la Recherche Scientifique, and the
University of Hawaii.}
\altaffiltext{3}{Based in part on data collected at the Subaru
Telescope, which is operated by the National Astronomical Observatory
of Japan.}
\altaffiltext{4}{Department of Astronomy, University of
Wisconsin-Madison, 475 North Charter Street, Madison, WI 53706.}
\altaffiltext{5}{Department of Physics and Astronomy,
University of Hawaii, 2505 Correa Road, Honolulu, HI 96822.}
\altaffiltext{6}{Institute for Astronomy, University of Hawaii,
2680 Woodlawn Drive, Honolulu, HI 96822.}
\altaffiltext{7}{Jansky Fellow.} 
\altaffiltext{8}{National Radio Astronomy Observatory (NRAO),
1003 Lopezville Road, Socorro, NM 87801. The NRAO is a facility 
of the National Science Foundation operated under cooperative
agreement by Associated Universities, Inc.}

\slugcomment{Published in The Astrophysical Journal (2008, 689, 687)}

\begin{abstract}
We present a table of redshifts for 2907 galaxies and stars
in the 145~arcmin$^2$ {\em HST\/} ACS GOODS-North,
making this the most spectroscopically complete redshift sample 
obtained to date in a field of this size. We also include the redshifts,
where available, in a table containing just under 7000 galaxies 
from the ACS area with $K_{s,{\rm AB}}<24.5$ measured from a deep 
$K_s$ image obtained with WIRCam on the CFHT, as well as in a 
table containing 1016 sources with NUV$_{\rm AB}<25$ and 478 sources 
with FUV$_{\rm AB}<25.5$ (there is considerable overlap)
measured from the deep GALEX images in the ACS area.
Finally, we include the redshifts, where available, in a table
containing the 1199 $24~\mu$m sources to 80~$\mu$Jy measured from 
the wider-area {\em Spitzer\/} GOODS-North.
The redshift identifications are greater than 90\%
complete to magnitudes of ${\rm F435W\/}_{\rm AB}=24.5$, 
${\rm F850LP\/}_{\rm AB}=23.3$, and $K_{s,{\rm AB}}=21.5$ 
and to 24~$\mu$m fluxes of 250~$\mu$Jy.
An extensive analysis of these data will appear in a parallel paper,
but here we determine how efficient color-selection techniques are 
at identifying high-redshift galaxies and Active Galactic Nuclei.
We also examine the feasibility of doing tomography 
of the intergalactic medium with a 30~m telescope.
\end{abstract}

\keywords{cosmology: observations --- galaxies: active --- galaxies: distances
          and redshifts --- galaxies: evolution --- galaxies: formation}

\section{Introduction}
\label{secintro}

Substantial progress has been made in tracing the formation 
and evolution of galaxies through extensive multiwavelength 
observations of extragalactic survey fields. However, the most 
challenging and time-consuming task is spectroscopically 
identifying the sources. This is especially true when one is 
trying to identify complete samples to faint magnitude limits. 
Although with existing multiwavelength data sets we can
construct spectral energy distributions (SEDs) and 
measure photometric redshifts for galaxies,  spectra
provide an enormous amount of additional information on the 
sources. We have therefore
carried out a very high quality, nearly spectroscopically 
complete, magnitude limited redshift survey of the Great 
Observatories Origins Deep Survey-North (GOODS-N; 
Giavalisco et al.\ 2004) field.
The GOODS-N is one of the most intensively
studied regions in the sky, and in many bandpasses it
has the deepest images ever obtained. It has also
been the target of extensive spectroscopic observations
over the years
(e.g., Cohen et al.\ 2000; Wirth et al.\ 2004; Cowie et al.\ 2004).
Thus, it is nearly ideal for the present study. 

In a parallel paper (Cowie \& Barger 2008) we use this
unique spectral database to place galaxies into 
the overall evolutionary scheme by combining spectral line and 
break diagnostics with colors and by analyzing metal evolution.
Here we present the data catalogs and use the redshift information 
to test the efficiency of using color-selection techniques to 
identify populations of high-redshift galaxies and 
Active Galactic Nuclei (AGNs). We also determine
which AGNs can be distinguished by their colors.
We then use the data set to examine the feasibility of doing 
intergalactic medium (IGM) tomography on a 30~m telescope. 

Techniques have been developed to pre-select high-redshift 
candidates for spectroscopic confirmation using simple 
photometric selection criteria. The Lyman Break Galaxy (LBG)
dropout technique (e.g., Cowie et al.\ 1988; Songaila et al. 2000;
Lilly et al. 2001; Steidel \& Hamilton 1993;
Steidel et al.\ 1995) is the most heavily 
used of these and has really 
opened up the field to the study of $z>3$ star-forming galaxies.
However, with the advent of wide-area near-infrared (NIR) arrays, 
other selections, including the Distant Red Galaxy (DRG) color 
selection (Franx et al.\ 2003; van Dokkum et al.\ 2003; 
$z\gtrsim 2$) and the $BzK$ color selection (Daddi et al.\ 2004; 
$1.4\lesssim z\lesssim 2.5$), have recently become popular. 
The above selections are, of course, limited to sources that 
can be observed in optical and NIR surveys, but they have produced 
a substantial sample of spectroscopically confirmed high-redshift 
galaxies and AGNs. 
H$^{-}$ methods using the mid-infrared (MIR) spectral shapes
have also been suggested (e.g., Simpson \& Eisenhardt 1999;
Sawicki 2002). Although these techniques have not been heavily 
used, they are of considerable interest for optically faint
sources.

With all of these avenues for finding high-redshift sources,
we would hope that there are not still significant populations 
that remain unidentified. 
Recently, however, Le F{\`e}vre et al.\ (2005) and
Paltani et al.\ (2007) have challenged this view using
the VIMOS Very Large Telescope Deep Survey (VVDS), a 
purely $I$ flux-selected sample (targets have magnitudes 
between $I_{AB}=17.5$ and $I_{AB}=24$) of $9295$ galaxies 
(``first epoch'') with measured redshifts in the range 
$0\le z \le 5$. These authors claim surface densities consistently 
larger than those of Steidel et al.\ (1996, 1999, 2004) 
by factors of $1.6-6.2$ at $z\approx3$, with the largest
factor applying to the brightest magnitudes.
In calculating the surface densities they corrected 
for the fraction of galaxies they did not observe ($\sim 76$\%) 
and for their estimate of the fraction of galaxies with incorrect 
redshifts. The need for the latter correction arises from their 
use of redshifts with all four of their reliability flags 
(flags 3 and 4 have very reliable measurements; flag 2 has a 
reasonably reliable measurement, but there is a non-negligible 
probability that the redshift is wrong; flag 1 has a tentative 
measurement, but there is a significant probability that the 
redshift is wrong). They concluded that the LBG selection 
techniques and photometric redshift studies are not able to 
identify the full population of high-redshift galaxies found
through spectroscopic observations of pure magnitude selected 
samples. This claim requires verification from a highly 
spectroscopically complete sample with very reliable redshift
identifications.

A 30~m telescope will revolutionize wide-field spectroscopy,
making sources that are currently 10 times too faint 
for moderate-to-high dispersion spectroscopy in the optical
accessible and mapping the distribution of galaxies 
over the redshift range $2\le z \le 4$. Of most interest,
however, is whether a 30~m telescope will allow dense sampling 
of the IGM through the use of background galaxies. This would
be an enormous improvement over the one-dimensional 
information now obtained through the study of rare high-redshift 
quasars, whose surface density on the sky at magnitudes 
accessible with Keck is very low.
To trace the three-dimensional distribution of diffuse gas
at high redshifts would require a high density of background
probes, but it needs to be determined whether the necessary
surface density of such probes exists.

The structure of the paper is as follows.
In \S\ref{secsurvey1} and \S\ref{secsurvey2}
we describe, respectively, our photometry and our spectroscopy.
In \S\ref{seccolor} we examine how well different color selection
techniques do, including LBG selection in \S\ref{secLBG},
$BzK$ selection in \S\ref{secK}, and H$^-$ selection in
\S\ref{sechminus}. In \S\ref{sectom} we consider whether IGM
tomography is feasible with a 30~m telescope, and in 
\S\ref{secsummary} we summarize our results.

We assume $\Omega_M=0.3$, $\Omega_\Lambda=0.7$, and
$H_0=70$~km~s$^{-1}$~Mpc$^{-1}$ throughout.
All magnitudes are given in the AB magnitude system,
where an AB magnitude is defined by
$m_{AB}=-2.5\log f_\nu - 48.60$.
Here $f_\nu$ is the flux of the source in units of
ergs~cm$^{-2}$~s$^{-1}$~Hz$^{-1}$.

\section{The Photometric Data}
\label{secsurvey1}

\subsection{Optical Imaging}

We take the optical photometric data from existing 
work. In particular, we take the $U$ magnitudes 
from Capak et al.\ (2004) and the F435W, F606W, F775W, and F850LP 
magnitudes from the {\em HST\/} Advanced Camera for Surveys (ACS) 
observations of the GOODS-N (Giavalisco et al.\ 2004).
There is an absolute offset in declination of 
$0\farcs4$ from the 20~cm VLA image of Richards (2000)
to the ACS and {\em Spitzer\/} images, so we applied a $-0\farcs4$
offset to the ACS and {\em Spitzer\/} catalogs to make them consistent 
with the ground-based optical and NIR data and the X-ray and radio 
data, all of which are already in the VLA astrometric frame.
All coordinates in the tables are given in the VLA astrometric frame.

\subsection{Near-Infrared Imaging}

We have obtained new ultradeep $K_s$ images from the Wide-field
Infrared Camera (WIRCam) on the CFHT 3.6~m telescope and from 
the Multi-Object Infrared Camera and Spectrograph (MOIRCS) on 
the Subaru 8.2~m telescope. WIRCam covers the entire ACS and 
{\em Spitzer\/} GOODS-N regions in one pointing. The WIRCam 
image is an $\sim40$~hr exposure and reaches a $5\sigma$ limit of 
$K_{s,{\rm AB}}=24.8$. MOIRCS has a smaller field of view, but we 
obtained an ultradeep mosaic which covers nearly the full ACS
area after including all available data in the Subaru archive.
The MOIRCS image has a very non-uniform sensitivity distribution,
but the image quality is extremely good ($0\farcs4-0\farcs5$
throughout). Although we cross-compared the WIRCam image with 
the MOIRCS image to eliminate artifacts and fragments of larger 
galaxies in constructing the catalogs, we measured the $K_s$-band
magnitudes from the WIRCam image alone. Below we describe the 
WIRCam and MOIRCS observations and data reduction.

\subsubsection{WIRCam Imaging}

WIRCam consists of four 2k~$\times~$2k HAWAII2-RG detectors
covering a field of view of $20\arcmin \times 20\arcmin$ with 
a $0\farcs3$ pixel scale. $K_s$-band imaging observations were 
carried out by our group in the semesters of 2006A and 2007A
and by a Canadian group led by Luc Simard in 2006A. The images 
were dithered to cover the detector gap and to obtain a uniform 
sensitivity distribution. Most of the observations were performed 
under photometric conditions with seeing between $0.6\arcsec$ 
and $1\arcsec$, and the total integration time was 39.7~hr.

The data were reduced in the Interactive Data Language (IDL)
using the SIMPLE Imaging and Mosaicking Pipeline 
(SIMPLE, W.-H. Wang 2008, in 
preparation\footnote{see also 
http://www.aoc.nrao.edu/\~{}whwang/idl/SIMPLE/index.htm}). 
Images within a dither set (typically 15~min in length) were
flattened using an iterative median sky flat in which a simple 
median sky was first derived to flatten the images and then a second 
median sky was derived by masking all of the detected objects 
using the flattened images. After the images were flattened, the
residual sky background was subtracted from each image with a 
smooth polynomial surface fitted to the masked image. Crosstalk 
effects between each of the 32 readout channels on a detector 
were removed by subtracting the median of the 32 64~$\times$~2k 
channels in the object-masked image. This removes most of the 
crosstalk and only weak residual features persist around
the few brightest stars in the field. The small areas 
contaminated by the residual crosstalk are mostly outside the 
ACS and {\em Spitzer\/} GOODS-N regions and are excluded in this 
work. The brightest cosmic ray hits were removed by a 
5~pixel$\times$5~pixel sigma filtering in each flattened image.

We used the SExtractor package (Bertin \& Arnouts 1996) to measure 
the object positions and fluxes in each flattened, sky subtracted, and 
crosstalk removed image in a dither set. The first-order derivative 
of the optical distortion function was derived by measuring the offsets 
of each object in the dither sequence as a function of location in the 
images. Absolute astrometry was obtained by matching the detected 
objects to a reference catalog constructed with brighter and compact
objects in the ACS catalog (Giavalisco et al.\ 2004; 
after correcting for the $0\farcs4$ offset between the
ACS and radio frames) and the SuprimeCam catalog (Capak et al.\ 2004).
The reduced exposures were then warped directly from the raw frames to a
common tangential sky plane with a sub-pixel accuracy. This
projection corrects for both optical distortion and absolute astrometry.
All projected images were weighted by their sky transparencies and
exposure times and were combined to form a large mosaic. When images 
from a dither set were combined, a sigma filter was applied to pixels 
that have the same sky position to further remove fainter cosmic rays 
and artifacts such as reflections inside the optics. In the image 
combination the SExtractor fluxes of objects in the images
were used to correct for variable extinction. Determining the absolute 
photometry of the WIRCam image is not straightforward
because there were no standard star observations.  Our initial
attempt was to calibrate the WIRCam image using 2MASS objects, but
this was not successful. We found that objects with sufficient 
signal-to-noise ratios in the 2MASS catalog are in the nonlinear 
regime of WIRCam. On the other hand, objects in the WIRCam linear 
regime are close to the detection limits of 2MASS where selection 
effects bias the 2MASS magnitudes. These make the calibration
not better than 5\%. Thus, we decided not to rely on 2MASS and
instead calibrated the WIRCam image using our MOIRCS image 
(see \S\ref{secmoircs}), 
which was calibrated with frequent standard star observations.

The final WIRCam $K_s$-band mosaic covers an area of 
$30\arcmin\times 30\arcmin$. Approximately 550~arcmin$^2$ has at 
least 10~hr of integration, and the deepest area has 39.7~hr 
of integration. The image FWHM is $\sim 0\farcs8$ and is quite uniform 
across the field. The rms astrometry error between the WIRCam catalog 
and the ACS/SuprimeCam reference catalog is $0\farcs13$. The typical 
$5\sigma$ limiting magnitude in the 550~arcmin$^2$ deep area is 
$K_{s,{\rm AB}}=24.8$.

\subsubsection{MOIRCS Imaging}
\label{secmoircs}

MOIRCS contains two 2k~$\times~$2k HAWAII2 detectors covering a
field of view of $\sim 4\arcmin \times 7\arcmin$ with a $0\farcs117$
pixel scale. Our group and Japanese groups led by various 
investigators have imaged the ACS and {\em Spitzer\/} GOODS-N regions 
with multiple pointings in the $K_s$ band.
Part of the Japanese data were published in Kajisawa et al.\ (2006).  
Here we include all our data, taken between Dec 2005 and Jan 2008, 
and all the data available in the Subaru archive, taken between 
Jan 2005 and May 2006. The majority of the observations were 
performed under photometric conditions with excellent seeing
of $0\farcs25-0\farcs6$. A very small fraction of the data has 
large extinctions of $>0.5$~mag or poor seeing of $>0\farcs7$, 
and these observations were excluded in this work.

The reduction of the MOIRCS images also used the SIMPLE package 
and is almost identical to the reduction of the WIRCam images 
described above. Here we only describe the differences. MOIRCS 
produces nearly circular fringes in roughly half of the images 
in $K_s$. The fringes were modeled in polar coordinates
where they are almost straight lines and were subtracted from 
the images in the original Cartesian coordinates. The MOIRCS 
images were calibrated by observing various UKIRT Faint Standards 
(Hawarden et al.\ 2001) at least every half night on
each detector. $K_s$ band magnitudes of the standard stars were
estimated by interpolating the $H$ and $K$ magnitudes in the 
UKIRT list assuming a power-law SED and using the  
$H$ and $K$ passbands for the UKIRT broadband filters
and the MOIRCS $K_s$ passband (see Tokunaga et al.\ 2002).
Data taken under non-photometric conditions and poorly calibrated
archive data were recalibrated with photometric data taken by 
our group.

The final MOIRCS $K_s$ mosaic contains more than 20 pointings with
different centers and position angles. The final mosaic 
has a total area of approximately 250~arcmin$^2$. Among all of the
detected objects in the MOIRCS mosaic, the median $5\sigma$ 
limit is $K_{s,{\rm AB}}=24.85$, comparable to that of the WIRCam 
image. However, because of the different strategies adopted by
the various groups, the MOIRCS coverage is extremely non-uniform.
The deepest MOIRCS area is around the Hubble Deep Field-North
(HDF-N) proper with a total
integration time of $\sim22$~hr and extremely deep $5\sigma$
limits of $K_{s,{\rm AB}}=26.2-26.7$. At some edges of the ACS and 
{\em Spitzer\/} GOODS-N regions the integration times are less than 
20~min and the limiting magnitudes become $<23.9$. The MOIRCS mosaic 
has a very high image quality of $0\farcs4-0\farcs5$ over the 
entire field of view. The rms astrometry error between the 
MOIRCS positions and the ACS/SuprimeCam reference catalog is 
$0\farcs08$.

\subsubsection{K$_s$ catalog}

We defined a sample of 6909 sources within the most uniformly 
covered ACS GOODS-N region (145~arcmin$^2$)
having $K_{s,{\rm AB}}<24.5$ in the WIRCam image,
where $K_{s,{\rm AB}}$ is a $3''$ diameter aperture magnitude 
corrected to a $6''$ diameter magnitude using a median offset
measured for the brighter objects.

We list these sources in Table~\ref{tab1} ordered by $K_s$ 
magnitude. We also give the flux and error from the WIRCAM
image measured with the AUTO magnitude of SExtractor,
which may provide a better approximation to the total
magnitude for brighter galaxies in the sample.
In constructing the catalog we eliminated a  
number of sources which lay within $3''$ of a brighter object 
and where reliable photometry and spectroscopy could not be 
obtained. We also excluded objects that lay within an $8''$ 
radius of the 9 brightest stars in the field for the same reason.
We eliminated artifacts and fragments of larger galaxies
by visual inspection of the images and by cross-comparing 
the independent MOIRCS and WIRCam data. In the table we give
the catalog number (col.~[1]), the right ascension and declination 
(col.~[2]$-$[3]), the flux and $1\sigma$ error in $\mu$Jy corresponding 
to the SExtractor AUTO magnitude (col.~[4]$-$[5]),
the $K_s$ magnitude (col.~[6]),
the SExtractor AUTO magnitudes measured in the four ACS bands
(col.~[7]$-$[10]), and the ground-based $U$-band magnitudes measured 
in a $3''$ diameter aperture and corrected to a $6''$ 
diameter magnitude using a median offset (col.~[11]). 
We describe the remaining five columns of Table~\ref{tab1} in
\S\ref{secsurvey2}.

\subsection{Mid-Infrared Imaging}

We took as our primary MIR sample the DR1+ 
Multi-Band Imaging Photometer for {\em Spitzer\/}
(MIPS; Rieke et al.\ 2004) $24~\mu$m source 
list of 1199 sources with fluxes above $80~\mu$Jy measured from 
the version 0.36 MIPS 24~$\mu$m map of the {\em Spitzer\/} Legacy Program.
This source list is a subset of a more extensive catalog which
will be presented by R.-R. Chary et al.\ (2008, in preparation).
The positions of corresponding 
{\em Spitzer\/} Infrared Array Camera (IRAC; Fazio et al.\ 2004) sources 
were used to measure the 24~$\mu$m fluxes, where possible.
We give this sample ordered by 24~$\mu$m flux in Table~\ref{tab2}. 
Assigning NIR and optical counterparts to the 
$24~\mu$m sources is complicated because of the uncertainties 
in the positions. We first applied the uniform offset of 
$-0\farcs4$ in declination to bring the absolute
astrometry into the VLA frame.
We next checked the relative astrometry by measuring
the positions of the $24~\mu$m sources on the $K_s$
images in cases where there was an unambiguous $K_s$ counterpart. 
There appears to be a slight distortion in the $24~\mu$m astrometry
relative to both the $K_s$ image and the 20~cm image, but
the effect is not large ($\lesssim 0\farcs5$ over
the entire image). We used a linear fit to remove this.
We then measured the position of the nearest $K_s$
source. In most cases the counterpart is unambiguous;
all but 65 of the sources have a $K_s$ counterpart within
$1''$. This reflects the use of the IRAC source positions
as priors in generating the $24~\mu$m catalog.
However, in some cases the $24~\mu$m source lies 
between two NIR galaxies, likely as a consequence
of blending. In these cases we assigned the
source to the nearest galaxy, or, in the most ambiguous
cases, to the brightest nearby galaxy using the IRAC
images, where available, or the $K_s$ image otherwise.
We give the initial position of each $24~\mu$m source
(after astrometric correction) in columns (1) and (2) of 
Table~\ref{tab2}, and we give the position of the 
adopted counterpart in columns (3) and (4). The 14 sources 
with separations greater than $1\farcs5$ from the counterpart
are marked with the letter ``w'' after column (11).
These counterparts should be treated with caution.
The associated $24~\mu$m fluxes may be overestimated
if the offset is a consequence of blending.
We also list in Table~\ref{tab2} the 24~$\mu$m flux (col.~[5]), 
the $K_s$ magnitude (col.~[6]), and the four ACS magnitudes 
(cols. [7]$-$[10]). Where the object lies off the 
{\em HST\/} image, we give the $z'$, $I$, and $B$
ground-based magnitudes instead (Capak et al.\ 2004). 
In these cases the F606W column is left blank. We give
the $U$-band magnitude in column~(11). We describe the 
remaining five columns of Table~\ref{tab2} in \S\ref{secsurvey2}.

\subsection{Near-Ultraviolet and Far-Ultraviolet Imaging}

The Galaxy Evolution Explorer (GALEX) mission 
(Martin et al.\ 2005) obtained a deep 150~ks exposure of
the ACS GOODS-N region in early 2004. We measured the 
near-ultraviolet (NUV; 2371~\AA\ central wavelength) and 
far-ultraviolet (FUV; 1528~\AA\ central wavelength) magnitudes 
from these images, which we obtained from the Multimission 
Archive at STScI (MAST). We used as a prior the positions
of the F435W$_{\rm AB}=26$ galaxy sample.
Given the large point spread function ($4\farcs5-6''$ FWHM)
of GALEX, we used an $8''$ diameter aperture to measure the
magnitudes using the GALEX zeropoints of 20.08 for the
NUV image and 18.82 for the FUV image. 
For the brighter objects (F435W$_{\rm AB}=20-23.5$~mag) we 
measured the median offset between the $8''$ magnitudes and the
magnitudes that we measured in a $24''$ diameter aperture. 
We then used this to correct all of the $8''$ magnitudes to 
approximate total magnitudes. 
The magnitudes agree on average to within 0.05~mag with the 
SExtractor based magnitudes given in the GALEX NUV$+$FUV
merged catalog 
for the region. We measured the noise level by randomly 
positioning apertures on blank areas of the sky and measuring 
the dispersion. We found $1\sigma$ limits of 26.8 for the NUV
image and 27.4 for the FUV image. 

GALEX has a large point-spread function, so contamination
by neighbors is a serious problem. Thus, in generating an 
NUV$+$FUV sample of isolated galaxies, we eliminated sources
which were closer than $8''$ to a brighter GALEX source at 
the same wavelength, or 
where, based on a visual inspection, the position was clearly
contaminated by the wings of a nearby bright source.
A substantial fraction of the galaxies in the optical
sample are eliminated by this isolation requirement.
The final sample contains 1016 sources with NUV$_{\rm AB}<25$
and 478 sources with FUV$_{\rm AB}<25.5$ (with considerable 
overlap between the two). We list these sources 
in Table~\ref{tab3} ordered by NUV magnitude. We give the right 
ascension and declination (cols.~[1] and [2]), 
the NUV magnitude (col.~[3]), and the FUV magnitude (col.~[4]).
Where we find a negative flux in the aperture, 
we show the magnitude corresponding to
the absolute flux with a minus sign in front.
We also give the 24~$\mu$m flux (col.~[5]), 
the $K_s$ magnitude (col.~[6]), the four ACS magnitudes 
(cols. [7]$-$[10]), and the $U$-band magnitude (col.~[11]). 
We describe the remaining five columns of Table~\ref{tab3}
in \S\ref{secsurvey2}.

\subsection{X-ray and Radio Imaging}

Richards (2000) presented a catalog of 20~cm sources detected
in the Very Large Array (VLA)\footnote{The VLA is a facility
of the NRAO.} map of the HDF-N, which
covers a $40'$ diameter region with an effective resolution
of $1.8''$. The absolute radio positions are known to 
$0.1''-0.2''$ rms. We cross-identified our catalogs with the
radio catalog to obtain 20~cm fluxes.

Alexander et al.\ (2003) presented the 2~Ms X-ray image
of the {\em Chandra\/} Deep Field-North (CDF-N), which they 
aligned with the Richards (2000)
VLA image. Near the aim point the X-ray data reach
limiting fluxes of $f_{2-8~{\rm keV}} \approx1.4\times
10^{-16}$~ergs~cm$^{-2}$~s$^{-1}$ and
$f_{0.5-2~{\rm keV}} \approx 1.5 \times
10^{-17}$~ergs~cm$^{-2}$~s$^{-1}$.
We use these data to identify AGNs, which we define on
energetic grounds as any source more luminous than 
$10^{42}$~ergs~s$^{-1}$ (Zezas et al.\ 1998;
Moran et al.\ 1999) calculated in either the rest-frame 
$0.5-2$~keV (soft) or rest-frame $2-8$~keV (hard) band. 
Trouille et al.\ (2008) present a table of redshifts
for the full X-ray sample, including sources which lie 
outside the ACS or {\em Spitzer\/} GOODS-N regions.

\section{The Spectroscopic Data}
\label{secsurvey2}

Our primary goal for this paper was to obtain the most complete 
and homogeneous 
spectral database possible for the GOODS-N field. Over the years
a number of groups have made observations of this region,
first primarily using the Low-Resolution Imaging 
Spectrograph (LRIS; Oke et al.\ 1995) on the Keck~I telescope 
(these data are summarized in Cohen et al.\ 2000), and later
using the large-format Deep Extragalactic Imaging Multi-Object 
Spectrograph (DEIMOS; Faber et al.\ 2003) on the Keck~II 10~m 
telescope. Wirth et al.\ (2004; Team Keck Treasury 
Redshift Survey or TKRS) and Cowie et al.\ (2004) presented
large samples of magnitude-selected redshifts obtained using DEIMOS.
In addition, Reddy et al.\ (2006) gave a substantial sample of 
color-selected redshifts, Chapman et al.\ (2004, 2005) and 
Swinbank et al.\ (2004) presented a small number of 
radio/submillimeter redshifts, and Treu et al.\ (2005)
measured redshifts for a sample of spheroids.
Barger et al.\ (2003), Trouille et al.\ (2008), and 
Barger et al.\ (2007) carried out extensive observations of 
the X-ray and 20~cm sources in the CDF-N.

Here we have added to these samples by observing all of the 
missing or unidentified galaxies in the F435W, F850LP, 
$K_s$, and 24~$\mu$m selected samples. In order to provide
a uniform spectral database, we also reobserved sources
where the original spectra were of poor quality or where 
previous redshifts were obtained with instruments other 
than DEIMOS, as well as where the existing redshift 
identifications were unconvincing (a small number).

We made our observations over a number of DEIMOS runs 
between 2004 and 2007. We used the 600 lines per mm grating, 
giving a resolution of $3.5$~\AA\ and a wavelength coverage of 
$5300$~\AA, which was also the configuration used in the TKRS
observations. The spectra were centered at an average wavelength 
of $7200$~\AA, though the exact wavelength range for each 
spectrum depends on the slit position. Each $\sim 1$~hr exposure 
was broken into three subsets, with the objects stepped along 
the slit by $1\farcs5$ in each direction. Unidentified sources
were continuously reobserved giving maximum exposure times of 
up to 7~hr. The spectra were reduced in the same
way as previous LRIS spectra (Cowie et al.\ 1996).
Only spectra that could be confidently identified based on
multiple emission and/or absorption lines were included in the
sample. A number of spectra were identified based on the 
doublet structure of the [OII]~3727~\AA\ line, which is 
resolved in the spectra.

In Table~\ref{tab4} we list the redshifts for 2907 sources in 
the ACS GOODS-N field. Only sources with $K_{s,{\rm AB}}<24.5$ or 
F850LP$_{\rm AB}<26$ are included, which omits a small 
number of emission line galaxies with known redshifts in the 
area. (A table of emission line galaxies will be given in 
E. M. Hu et al.\ 2008,
in preparation.) The table is ordered by right ascension. 
We give the right ascension and declination 
(cols.~[1] and [2]), the $K_s$, F850LP, F775W, F606W, F435W, 
and $U$ magnitudes (cols.~[3]$-$[8]), the spectroscopic redshift 
(col.~[9]), the source of the redshift (col.~[10]; see Table~\ref{tab6}), 
the 24~$\mu$m flux in $\mu$Jy (col.~[11]), and the rest-frame 
$2-8$~keV (col.~[12]) and $0.5-2$~keV (col.~[13]) X-ray 
luminosities in units of $10^{40}$~ergs~s$^{-1}$. Four stars 
were classified primarily on compactness and on their unambiguous
stellar colors. These are shown without a source
number in the tables and may be removed if desired.

We summarize the fraction of spectroscopically identified objects 
as a function of color and magnitude in Table~\ref{tab5}.
The redshift identifications are greater than 90\%
complete to magnitudes of ${\rm F435W\/}_{\rm AB}=24.5$, 
${\rm F850LP\/}_{\rm AB}=23.3$, and $K_{s,{\rm AB}}=21.5$. 

In Table~\ref{tab6} we list the published catalogs (the journal
references are given in col.~[5]) that we used in assembling our 
various redshift tables. We have assigned a unique source number 
(col.~[1]) to each catalog. It is this source number which 
appears in column~(10) of Table~\ref{tab4}. We also summarize 
in Table~\ref{tab6} the number of redshifts present in each 
catalog (col.~[2]) and the number of redshifts that
appear only in that catalog (``unique'', col.~[3]). We give a cumulative
redshift count of all the identified galaxies in column~(4). 
Galaxies from sources 1 and 2 have uniform DEIMOS spectra, while 
galaxies from source 3 have LRIS spectra. The galaxies from these 
three sources form our {\em spectral\/} database.

%
%
\begin{inlinefigure}
\vskip 0.4cm
\centerline{\psfig{figure=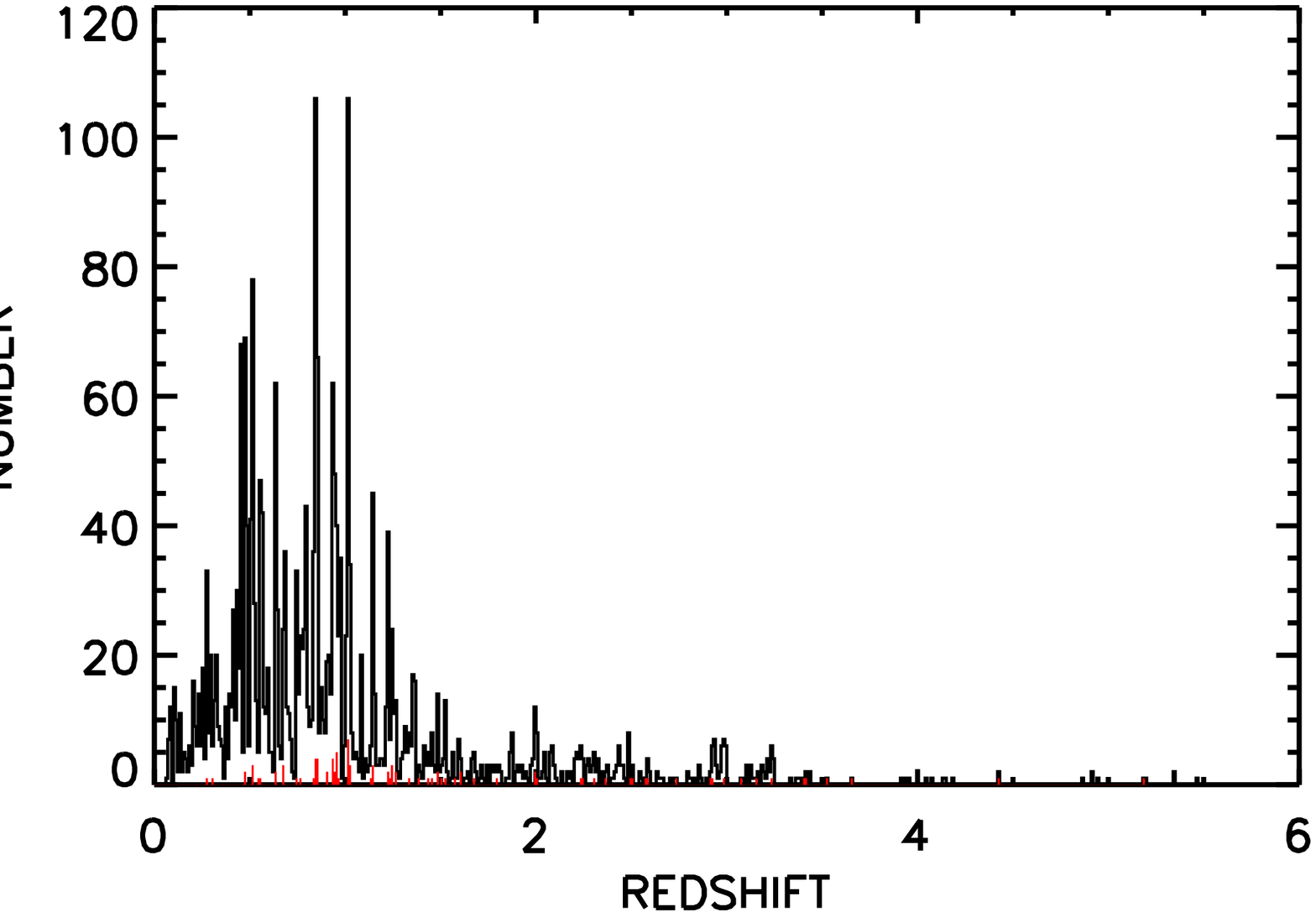,width=3.4in}}
\figcaption[]{Redshift distribution of the sources in 
Table~\ref{tab4}. The bin size is 0.01. 
AGNs are shown in red. 
\label{full_redshift_hist}
}
\end{inlinefigure}

We show the redshift distribution in
Figure~\ref{full_redshift_hist}. Most of the redshifts are 
at $z<1.6$ (2362 sources) where the [OII]~3727~\AA\ line 
still lies in the DEIMOS spectroscopic range, 327 lie
between $z=1.6-3.5$, mostly from the
LRIS observations of Reddy et al.\ (2006), and there
are relatively few objects (only 21) with spectroscopic
redshifts $z>3.5$. In total, 98 of the objects are classified
as containing AGNs that could be significant contributors to the 
galaxy light based on either their rest-frame hard or soft 
X-ray luminosities being above $10^{42}$~ergs~s$^{-1}$. 
We show on the figure the redshift distribution for these
AGNs {\em (red)\/}. 

We also list the spectroscopic redshifts for the $K_{s,{\rm AB}}<24.5$ 
sample as a function of $K_s$ magnitude in Table~\ref{tab1} 
(col.~[12]; there are redshifts for 2596 stars and galaxies in 
this sample), for the 24~$\mu$m sample as a function of 24~$\mu$m 
flux in Table~\ref{tab2} (col.~[12]; the {\em Spitzer\/} area covered 
is larger than the area covered by the optical/NIR catalog, so
213 of the 743 galaxies and stars with spectroscopic identifications 
in this sample lie outside the ACS area), and for the NUV$+$FUV sample 
as a function of NUV magnitude in Table~\ref{tab3} (col~[12]). 

The remaining columns in 
Table~\ref{tab1} are the source of the redshift (col.~[13]), 
the 24~$\mu$m flux in $\mu$Jy (col.~[14]), and the rest-frame 
$2-8$~keV (col.~[15]) and $0.5-2$~keV (col.~[16]) X-ray 
luminosities in units of $10^{40}$~ergs~s$^{-1}$. 
The remaining 
columns in Table~\ref{tab2} and Table~\ref{tab3} are the source 
of the redshift (col.~[13]), the rest-frame $2-8$~keV (col.~[14]) 
and $0.5-2$~keV (col.~[15]) X-ray luminosities in units of 
$10^{40}$~ergs~s$^{-1}$, and the 20~cm flux in $\mu$Jy (col.~[16]).

In Figure~\ref{figz_f24} we show redshift versus 24~$\mu$m flux 
for the $80~\mu$Jy MIR sample. We use red large squares to 
denote AGNs with rest-frame hard or soft X-ray luminosities in 
excess of $10^{42}$~ergs~s$^{-1}$. At the base of the plot we show 
a histogram of the fraction of sources per half magnitude bin 
that remain spectroscopically unidentified. The sample is
essentially complete to 500~$\mu$Jy.

%
%
\begin{inlinefigure}
\vskip 0.4cm
\centerline{\psfig{figure=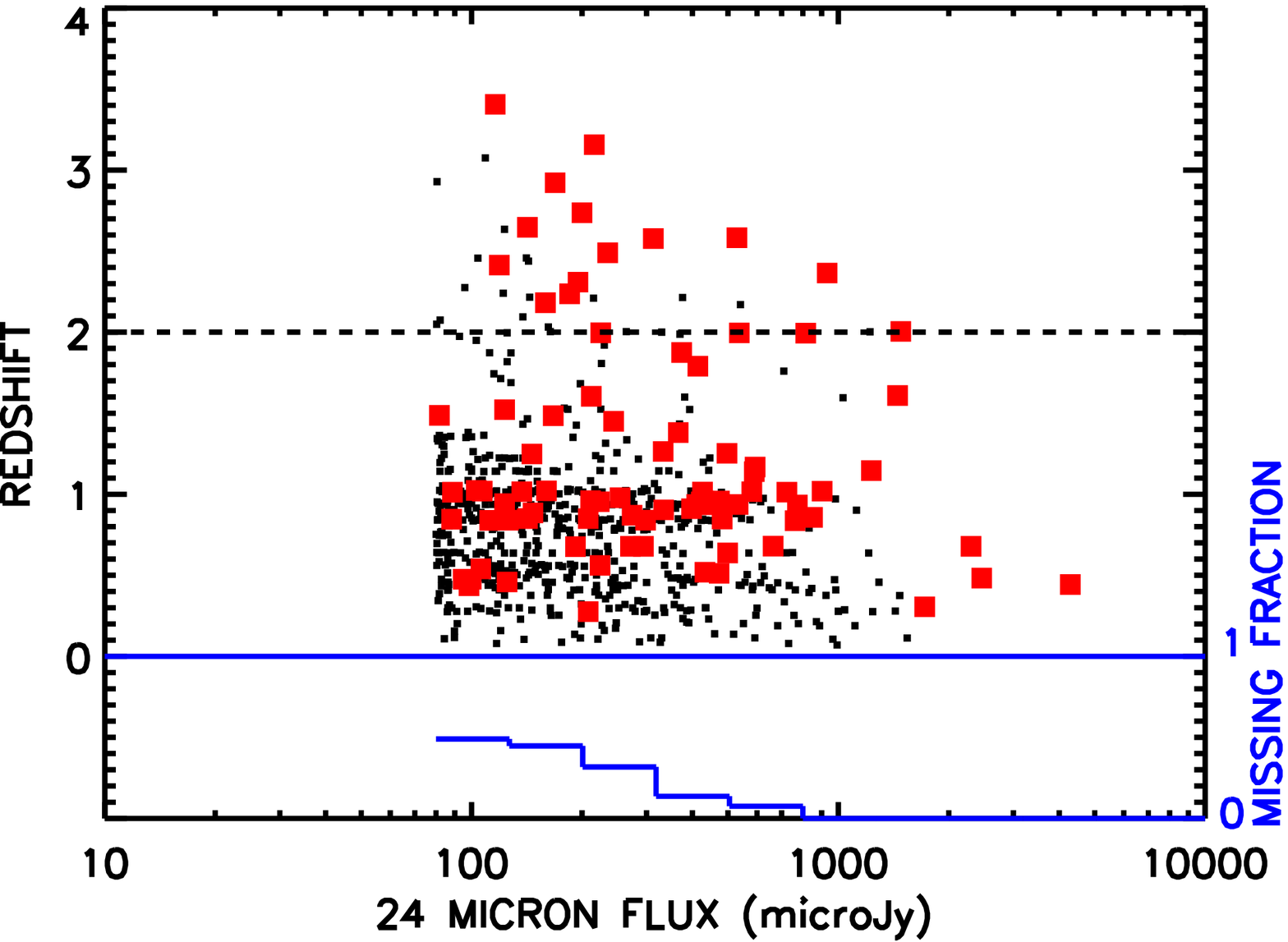,width=3.4in}}
\figcaption[]{Redshift vs. 24~$\mu$m flux for the 80~$\mu$Jy 
MIR sample. AGNs, defined as sources with rest-frame hard or 
soft X-ray luminosities $>10^{42}$~ergs~s$^{-1}$, are denoted 
by red large squares. 
The fraction of unidentified sources per 0.5~mag bin is shown 
in histogram form at the bottom of the plot (label is on 
right-hand y-axis). The dashed horizontal line shows $z=2$.
\label{figz_f24}
}
\end{inlinefigure}

\begin{figure*}
\centerline{\psfig{figure=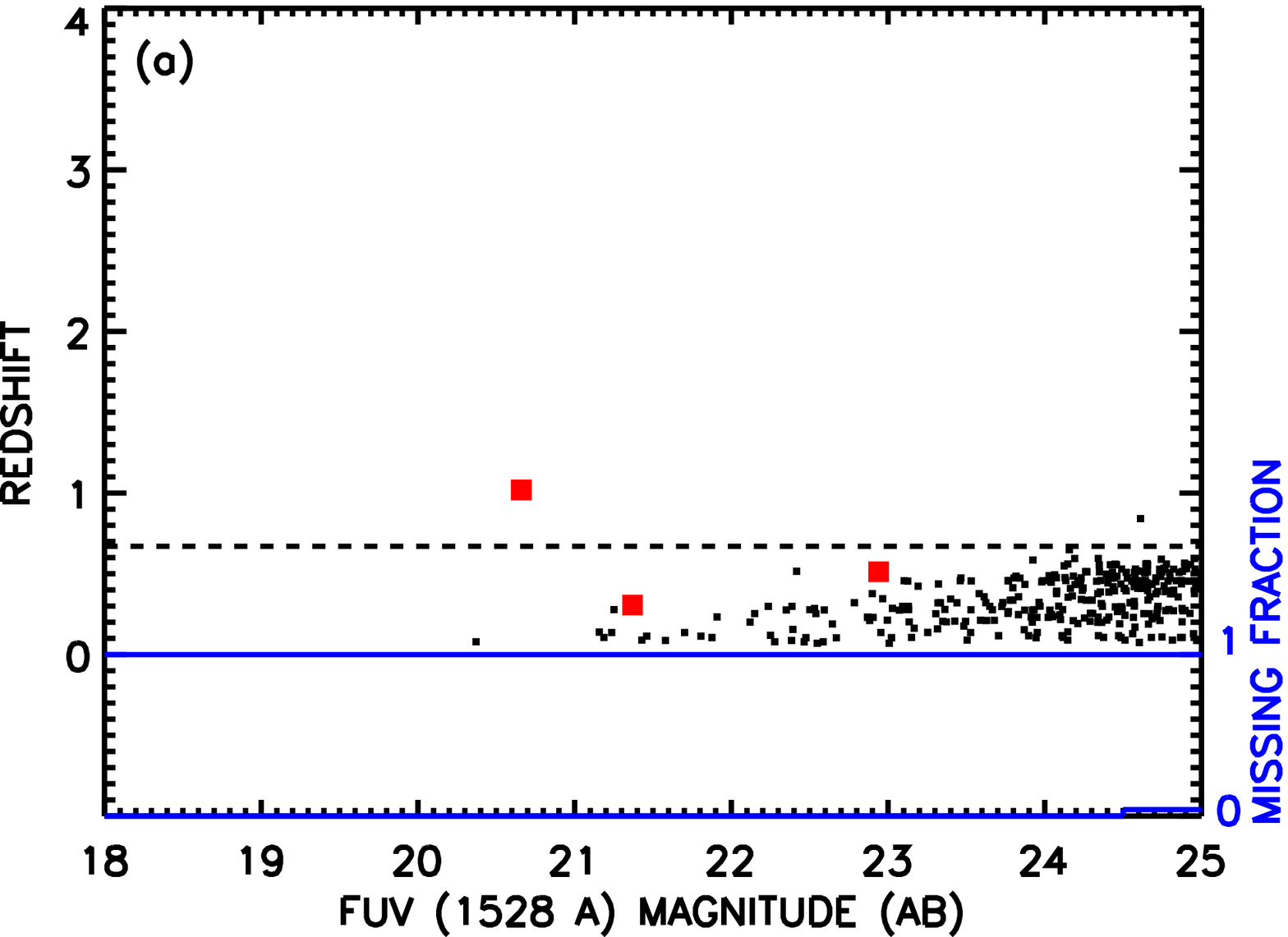,width=3.4in}
\psfig{figure=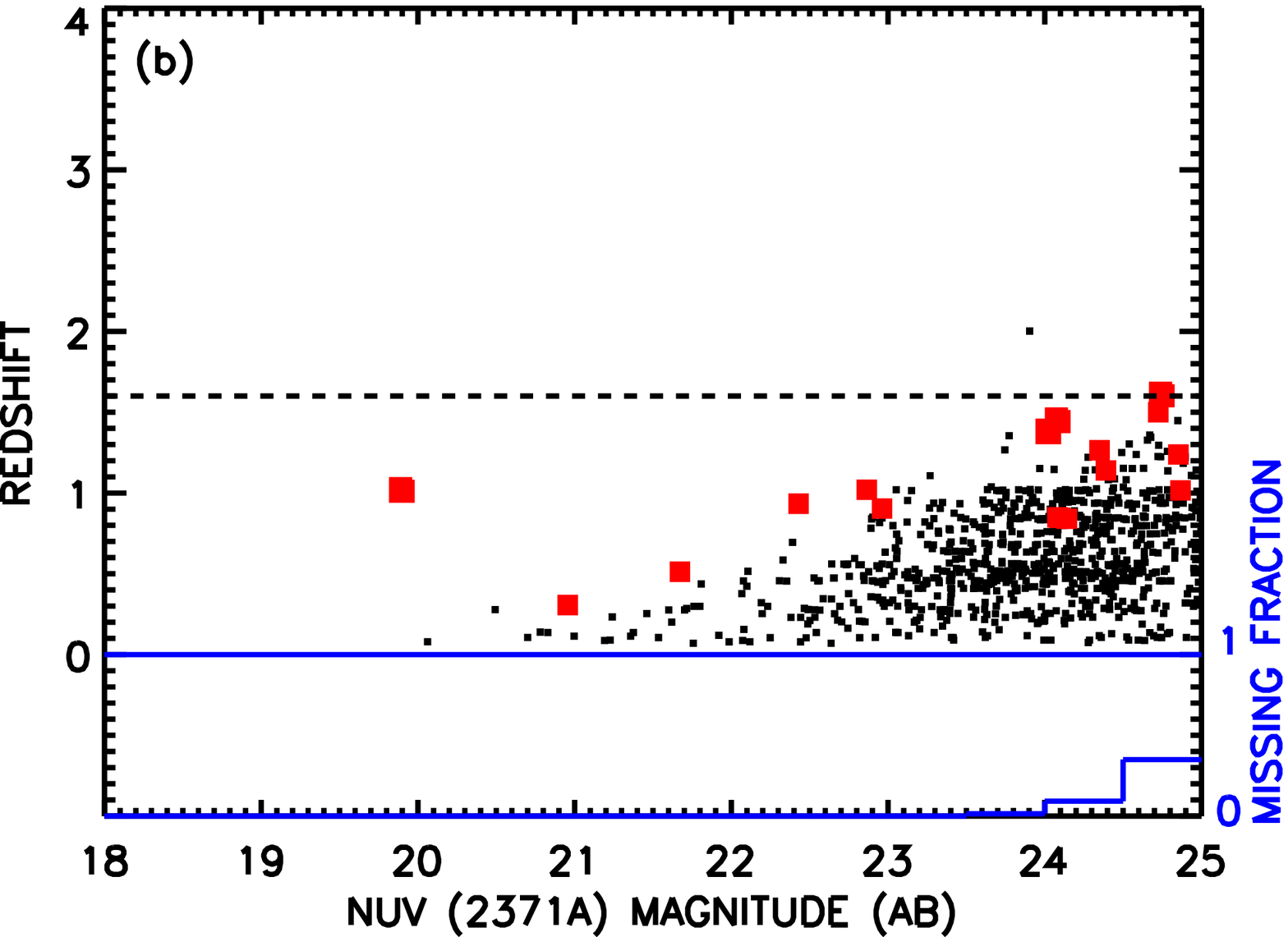,width=3.4in}}
\vskip 0.4cm
\centerline{\psfig{figure=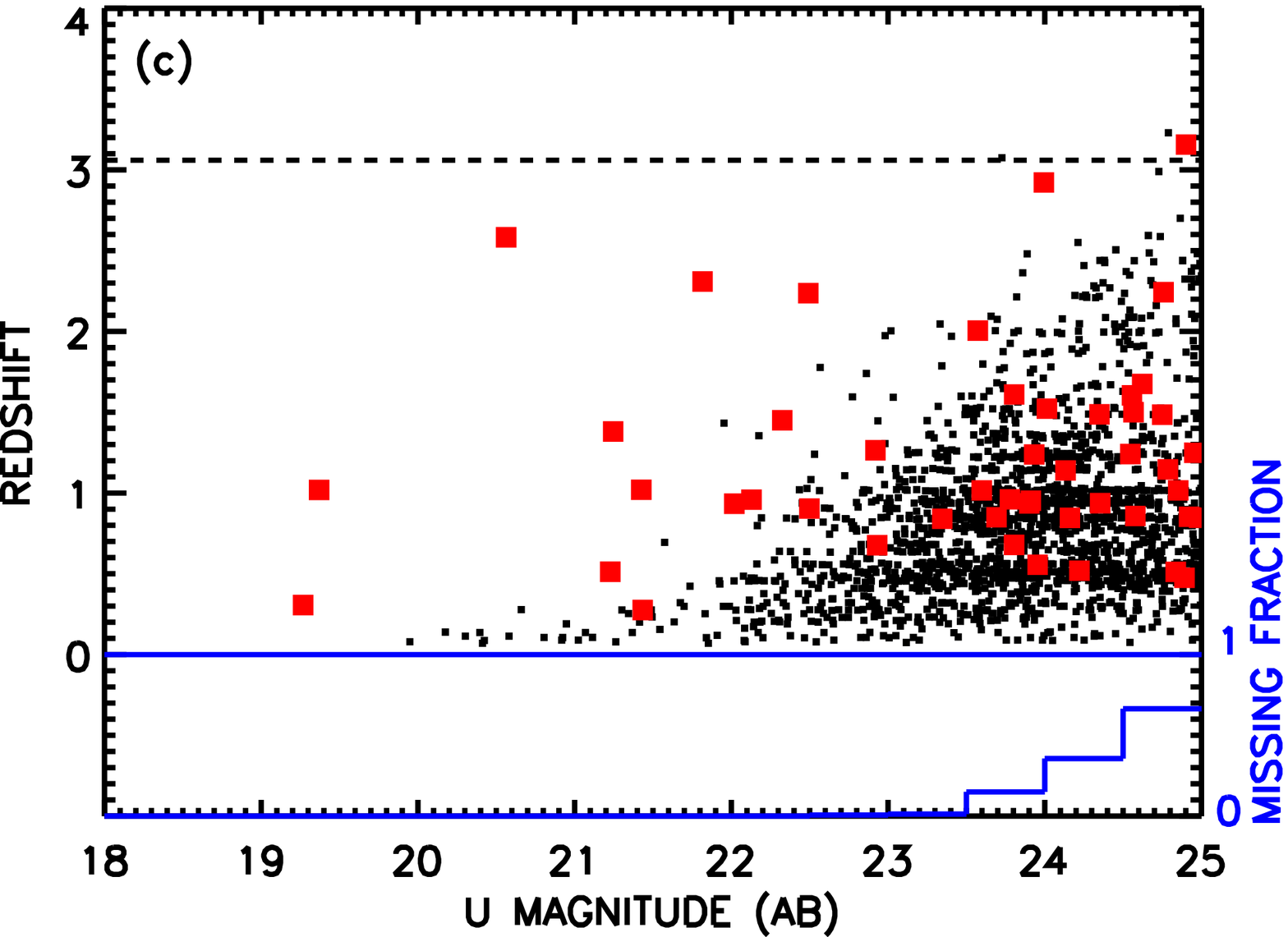,width=3.4in}}
\figcaption[]{Redshift vs. (a) FUV$_{\rm AB}$ magnitude for the
FUV$_{\rm AB}<25$ sample, (b) NUV$_{\rm AB}$ magnitude for the 
NUV$_{\rm AB}<25$ sample, and 
(c) $U_{\rm AB}$ magnitude for the $U_{\rm AB}<25$ sample.
AGNs, defined as sources with rest-frame hard or 
soft X-ray luminosities $>10^{42}$~ergs~s$^{-1}$, are denoted by 
red large squares. 
The dashed horizontal lines show the redshifts at which the 
Lyman continuum break passes through the center of each filter: 
(a) $z=0.67$ for the FUV, (b) $z=1.6$ for the NUV, and (c) $z=3.06$
for the $U$-band.
The fraction of unidentified sources per 0.5~mag bin is shown 
in histogram form at the bottom of each plot (label is on the
right-hand y-axis). 
\label{figz_fuv}
}
\end{figure*}

\begin{figure*}
\centerline{\psfig{figure=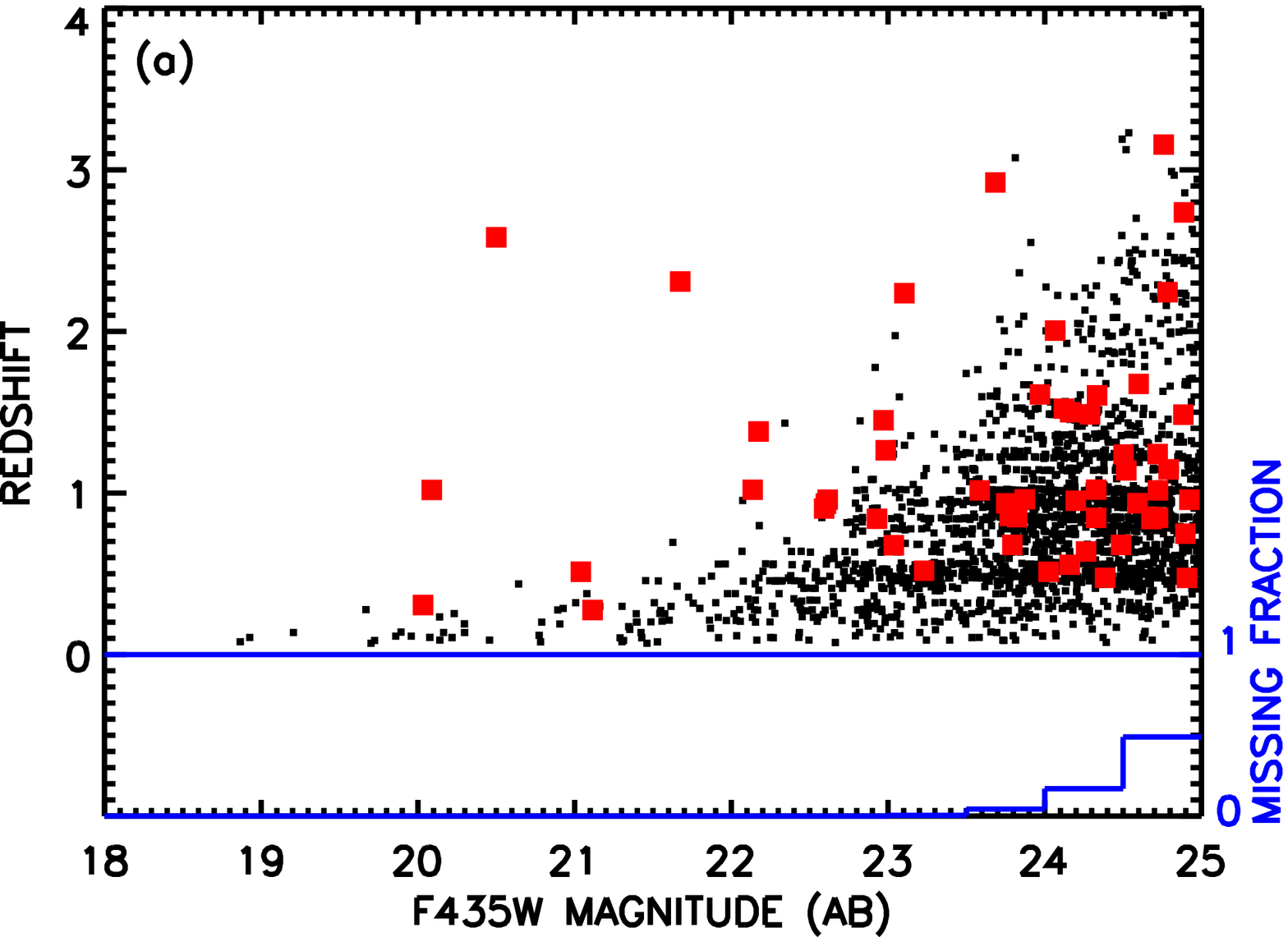,width=3.4in}
\psfig{figure=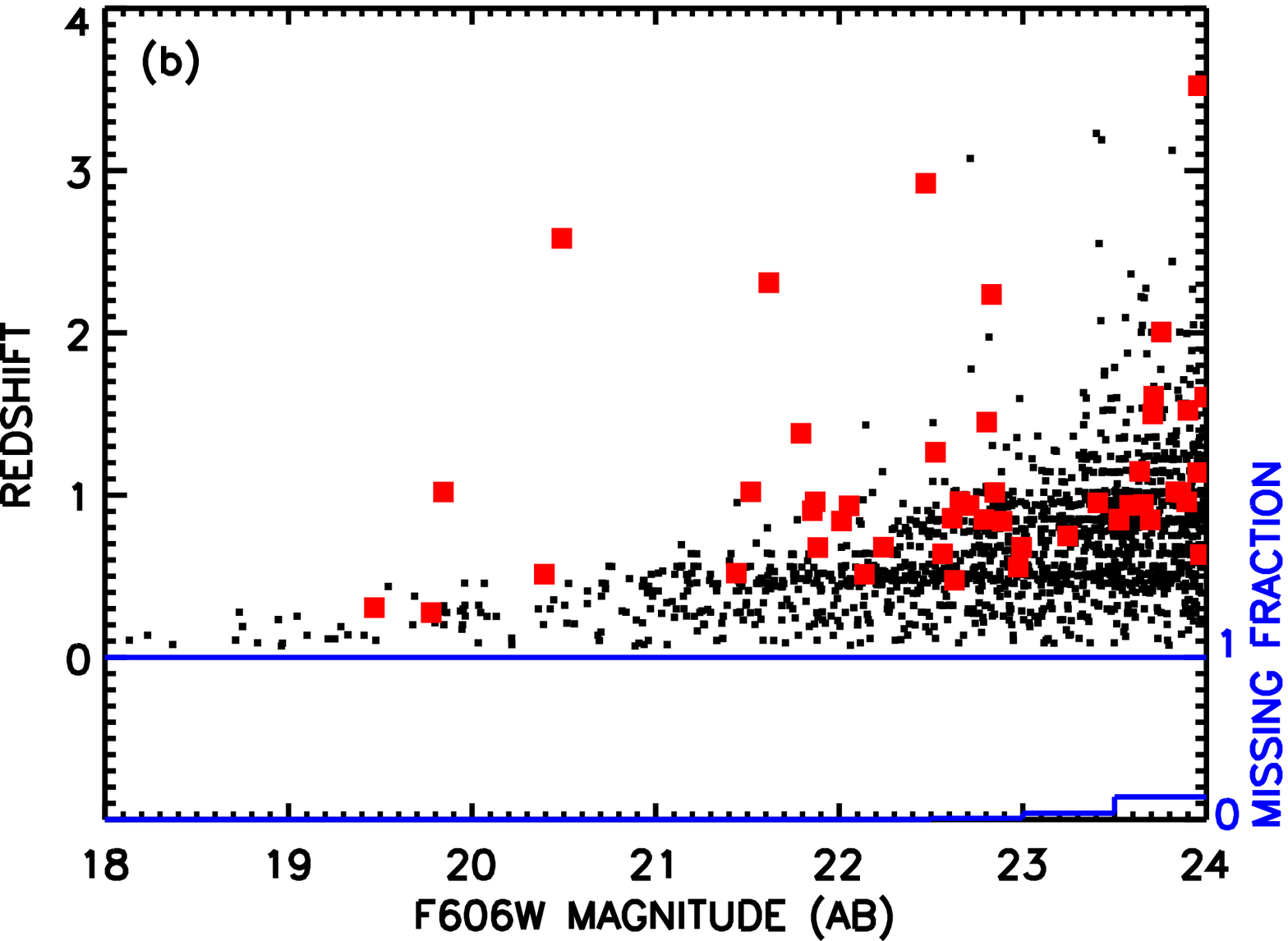,width=3.4in}}
\vskip 0.4cm
\centerline{\psfig{figure=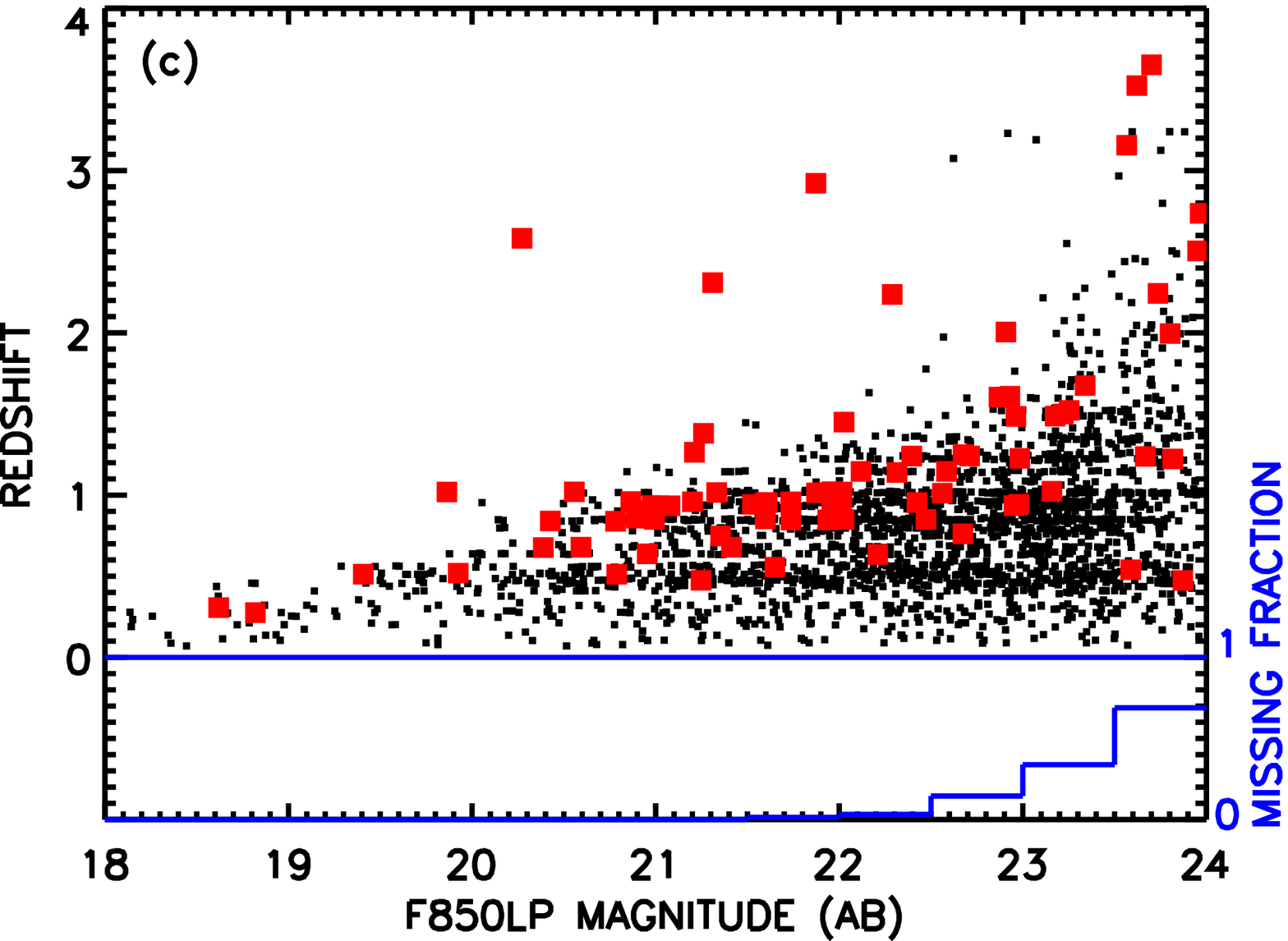,width=3.4in}
\psfig{figure=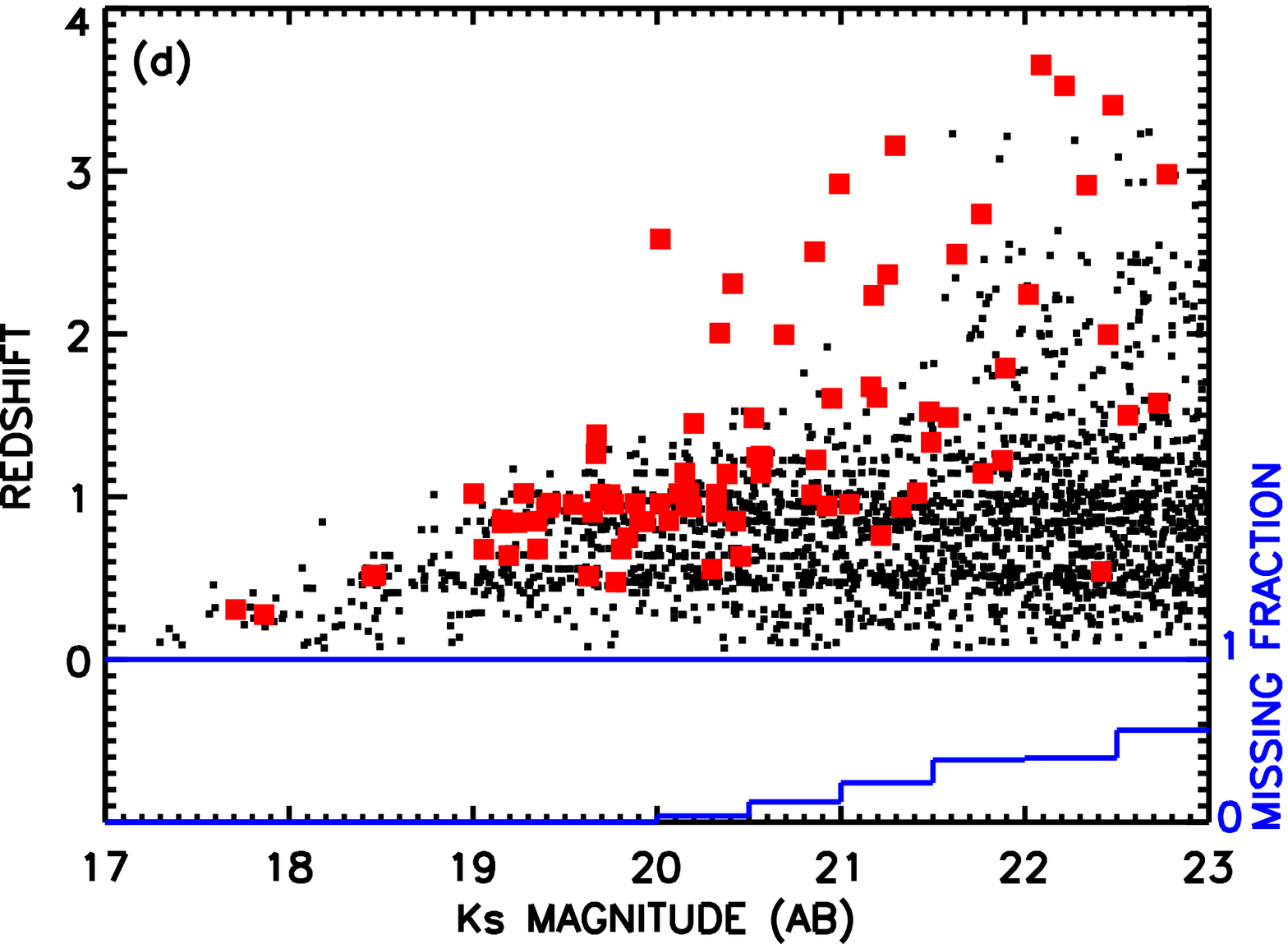,width=3.4in}}
\figcaption[]{
Redshift vs. (a) F435W$_{\rm AB}$ magnitude for the 
F435W$_{\rm AB}< 25$ sample,
(b) F606W$_{\rm AB}$ magnitude for the F606W$_{\rm AB}< 24$ sample,
(c) F850LP$_{\rm AB}$ magnitude for the F850LPW$_{\rm AB}< 24$ sample, 
and (d) $K_{s, {\rm AB}}$ magnitude for the $K_{s, {\rm AB}}< 23$ sample. 
AGNs, defined as sources with rest-frame hard or soft X-ray 
luminosities $>10^{42}$~ergs~s$^{-1}$, are denoted by red large squares. 
The fraction of unidentified sources per 0.5~mag bin is shown in
histogram form at the bottom of each plot (label is on the right-hand
y-axis). 
\label{figzmag}
}
\end{figure*}

In Figure~\ref{figz_fuv} we show redshift-magnitude
diagrams for the (a) FUV$_{\rm AB}<25$, (b) NUV$_{\rm AB}<25$, 
and (c) $U_{\rm AB}<25$ samples.
In each panel we show the redshift at which the Lyman continuum 
break passes through the center of the filter 
({\em dashed horizontal lines\/}; i.e., $z=0.67$, $z=1.6$, 
and $z=3.06$, respectively, for the FUV, NUV, and $U$-band filters).
The sharp cut-offs above these redshifts are clearly seen, 
and, perhaps surprisingly, apply to both the 
AGNs {\em (red large squares)\/} and the galaxies, with
very few sources in either category lying above the cut-offs. 
We shall return to this point when we consider LBG selection in 
\S\ref{secLBG}. 
In Figure~\ref{figzmag} we show redshift-magnitude
diagrams for the (a) F435W$_{\rm AB}<25$, (b) F606W$_{\rm AB}<24$, 
(c) F850LP$_{\rm AB}<24$, and (d) $K_{s, {\rm AB}}<23$ samples.

%
%

%
%

\section{Color Selection Techniques}
\label{seccolor}

We use the spectroscopic sample to test three color
selection techniques that can be used to pick out galaxies in a given 
redshift range: the LBG dropout technique, the $BzK$ color selection, 
and the H$^-$ method in the MIR. For each method
there are two quantities of interest. 
(1) If one were to use the method to select galaxies in the given
redshift range for spectroscopic follow-up or for
direct analysis, then the first quantity of interest is the fraction of objects
which lie outside the redshift range that are found by the selection.
We will call this the selection contamination.
(2) The second quantity of interest is the fraction of objects
which lie in the redshift range that are actually found by the 
color selection. We will call this the selection completeness.

In general the color selections are tuned to provide
a reasonable balance between selection contamination and
selection completeness and can be adjusted to optimize one
or the other. Thus, if one were using the color selection
to choose objects for spectroscopic follow-up, a high level
of contamination might be acceptable in order to make sure
the sample contained most of the objects in the redshift
range. On the other hand, if the primary goal were to
choose a set of objects in the redshift range based on
colors, then one might want to minimize contamination even 
if one missed a number of objects in the redshift range.

\subsection{The LBG Selection}
\label{secLBG}

In Figure~\ref{figlbg} we show the $2.7\le z \le 3.4$ 
LBG color-color selection {\em (blue solid lines)\/} for our 
98.4\% spectroscopically complete 
F606W$_{\rm AB}<23.5$ sample. This sample consists
of 1197 galaxies and stars, only 18 of which
are not identified.  Our selection is
\begin{eqnarray}
(U - {\rm F606W\/})_{\rm AB} > 1.2 \\
\quad {\rm for\/} \quad ({\rm F606W\/} -
{\rm F850LP\/})_{\rm AB} \le 0.25 \nonumber \,,
\end{eqnarray}
and
\begin{eqnarray}
(U - {\rm F606W\/})_{\rm AB} > ({\rm F606W\/} - {\rm F850LP\/})_{\rm AB}
\times 2.4 + 0.6 \\ 
\quad {\rm for\/} \quad 0.25 < ({\rm F606W\/} - 
{\rm F850LP\/})_{\rm AB} < 0.8 \nonumber \,.
\end{eqnarray}

Using the SExtractor stellarity index, as measured in the ACS F850LP 
image, we have divided the sample into extended ($<0.5$;
Fig.~\ref{figlbg}a) and compact ($>0.5$; Fig.~\ref{figlbg}b)
sources {\em (black diamonds for $z\le 2.7$ galaxies; 
red upside-down triangles for $z>2.7$ galaxies; purple stars for
spectroscopically identified stars; green triangles for
unidentified sources).\/} We denote the AGNs
(rest-frame hard or soft X-ray luminosities $>10^{42}$~ergs~s$^{-1}$)
with large squares {\em (blue for $z\le 2.7$; red for $z>2.7$)\/}. 
We enclose in larger black squares those AGNs classified as 
broad-line AGNs, and we enclose in larger black diamonds those 
AGNs with high-excitation emission lines (C~IV or [Ne~V]).

The LBG selection works well for the one $z>2.7$ galaxy
(high selection completeness). This object
is cleanly picked out at these magnitudes (Fig.~\ref{figlbg}a). 
Additionally, only a small number of other sources cross the 
boundary in either Figure~\ref{figlbg}a or Figure~\ref{figlbg}b
(modest selection contamination).
However, the three $z>2.7$ AGNs are not picked out by the LBG selection,
though they do lie close to the boundary. This is reasonable for the 
two compact $z>2.7$ AGNs in Figure~\ref{figlbg}b. Both show AGN 
signatures in their optical spectra, and the dominant AGN 
light would not be expected to have a strong Lyman break. 
However, the $z>2.7$ AGN in Figure~\ref{figlbg}a
shows no AGN signatures in its optical spectrum,
and it is dominated by the extended galaxy light. It would not be 
classified as an AGN based on its optical spectrum, which shows 
strong UV absorption lines and very weak Ly$\alpha$ emission. 
It appears that the AGN is weakening the break signature and 
dropping the object from the LBG selection while not dominating 
the longer wavelength light.

%
%
\begin{inlinefigure}
\vskip 0.8cm
\centerline{\psfig{figure=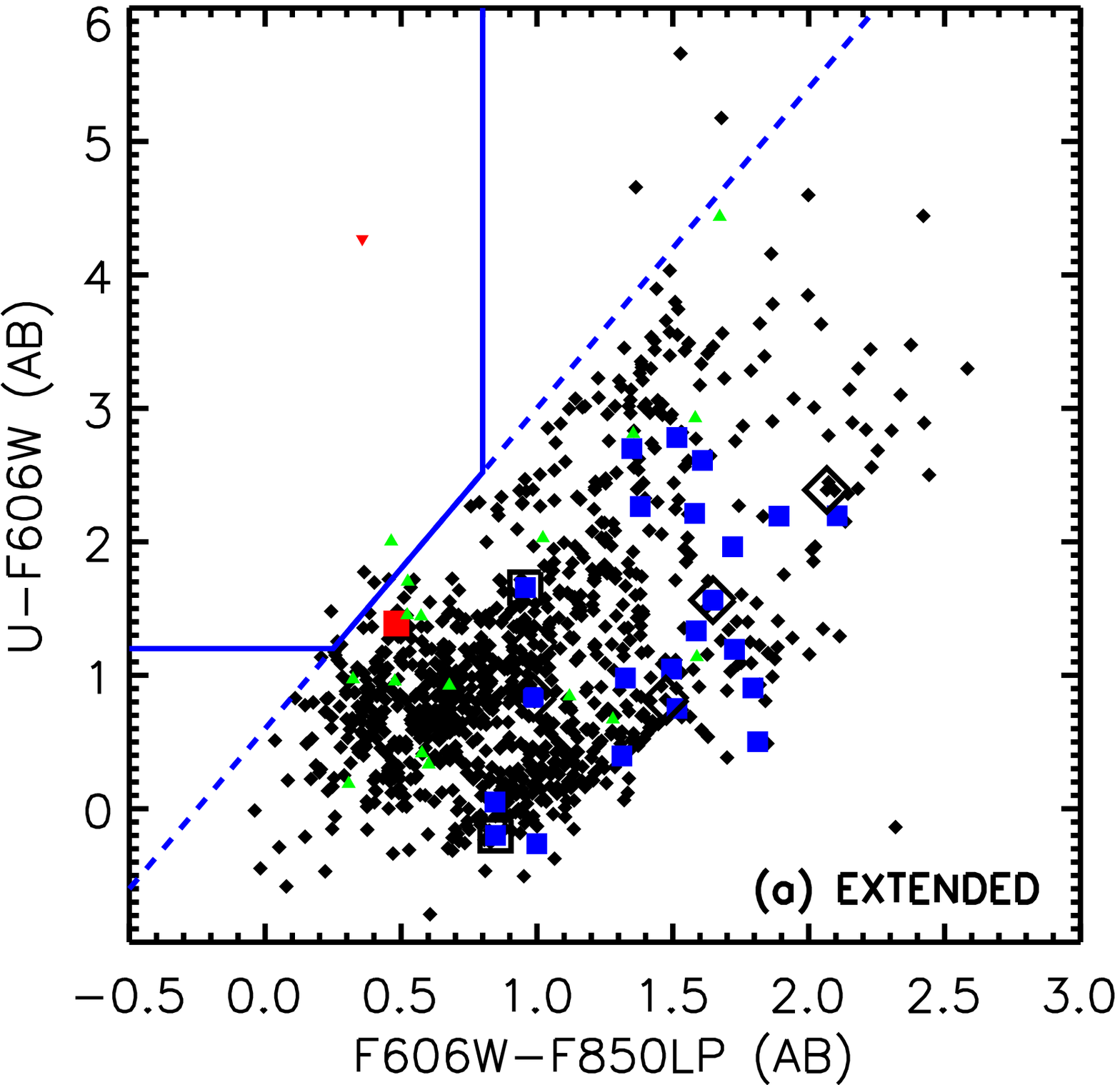,width=3.2in}}
\vskip 0.8cm
\centerline{\psfig{figure=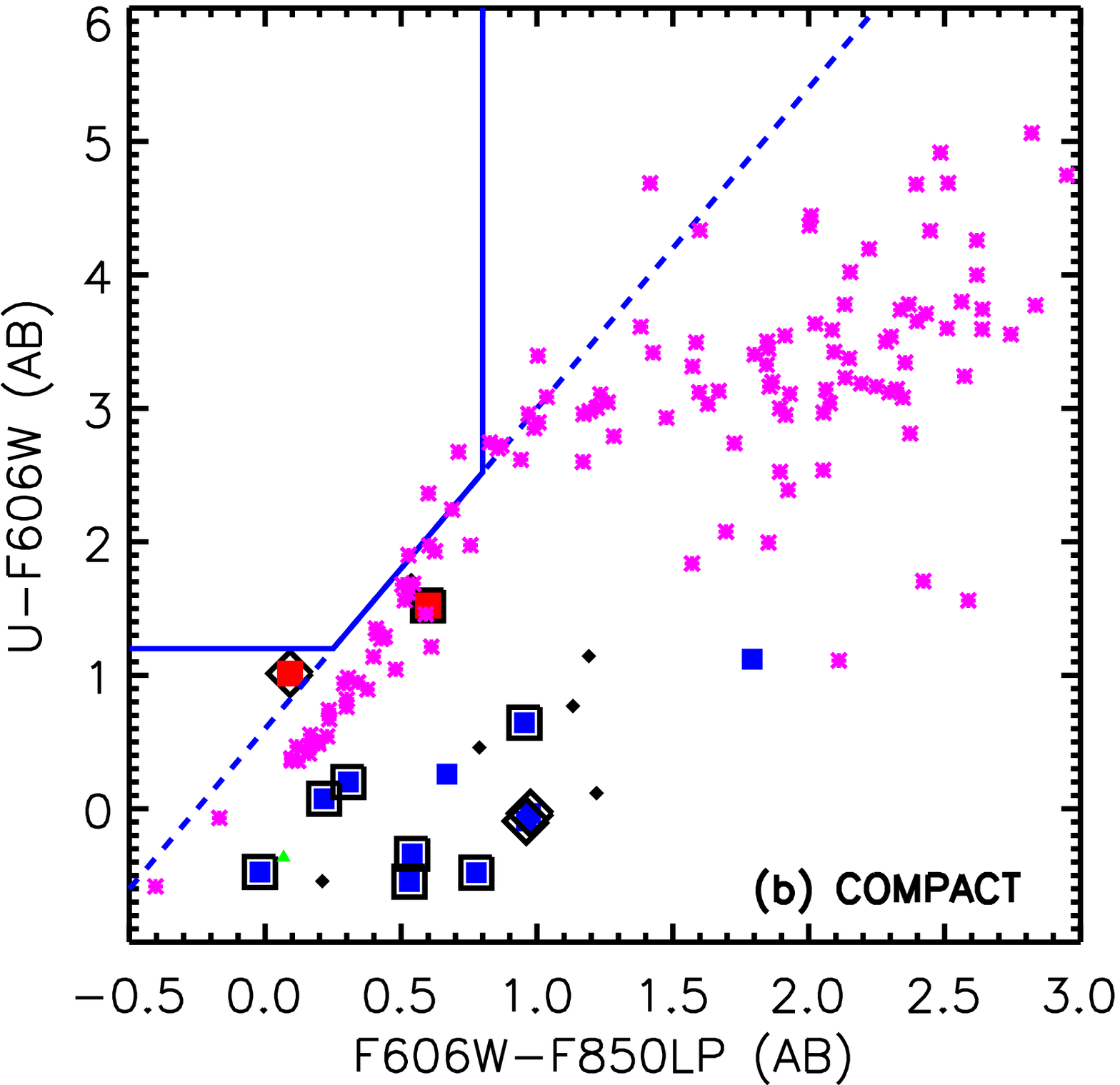,width=3.2in}}
\figcaption[]{
$(U - {\rm F606W\/})_{\rm AB}$ vs. 
$({\rm F606W\/} - {\rm F850LP\/})_{\rm AB}$ 
color-color diagram for our nearly spectroscopically
complete ${\rm F606W\/}_{\rm AB}< 23.5$ sample {\em (black diamonds
for $z\le 2.7$ galaxies; red upside-down triangles for $z>2.7$ galaxies;
purple stars for spectroscopically identified stars;
green triangles for unidentified sources)\/}, 
divided into (a) extended and (b) compact sources, depending on 
whether the SExtractor stellarity index in the ACS F850LP 
image is less than or greater than 0.5, respectively. 
All of the spectroscopically identified stars fall in (b). 
The large squares are AGNs {\em (blue for $z\le 2.7$; red for $z>2.7$)\/}. 
Those AGNs classified as broad-line AGNs are enclosed in larger black 
squares, and those with high-excitation emission lines 
(C~IV or [Ne~V]) are enclosed in larger black diamonds.
The blue solid lines mark where LBGs with $2.7\le z \le 3.4$ would
be searched for in a color-color pre-selection. 
\label{figlbg}
}
\end{inlinefigure}

We show the break selection in a different way in Figure~\ref{figubreak}, 
where we plot a $U$ break parameter versus redshift in three magnitude 
intervals for every ${\rm F850LP\/}_{\rm AB}\ge 19$ 
source with a measured spectroscopic redshift. We define the $U$ break 
parameter to be
\begin{equation}
(U - {\rm F606W\/})_{\rm AB}-({\rm F606W\/}-{\rm F850LP\/})_{\rm AB}
\times 2.4 - 0.6 \,.
\end{equation}
This is just a measure of how far 
above or below the diagonal selection limit 
{\em (blue dashed line in Fig.~\ref{figlbg})\/}
each galaxy lies. We also show the colors
of a few Bruzual \& Charlot (2003) model
tracks, illustrating that the break parameter passes through
zero at $z=2.7$ for a wide range of galaxy types.

We have divided the sources into three magnitude ranges:
(a) $19\le {\rm F606W}_{\rm AB}<22.75$, where all but one of 
the galaxies (with F606W$_{\rm AB}=22.69$) are identified,
(b) $22.75\le {\rm F606W}_{\rm AB}<23.5$, where the galaxy 
identifications are 97\% complete, and 
(c) $23.5\le {\rm F606W}_{\rm AB}<24.5$, where a 
substantial number of $z>2.7$ galaxies enter the sample.
The galaxies follow a remarkably well-defined track in
all the panels. In Figure~\ref{figubreak}a only the two AGNs 
have $z>2.7$ and, as we have discussed above, both lie close 
to the zero $U$ break parameter. In Figure~\ref{figubreak}b 
we see the appearance of the first $z>2.7$ galaxy, which has a 
magnitude of F606W$_{\rm AB}=23.4$. This galaxy lies above the 
selection limit, while the extended galaxy$+$AGN discussed above, 
which also appears in this figure, does not. 

In Figure~\ref{figubreak}c the $z=2.7-3.4$ sources become more 
common. Twelve lie above the selection limit and five lie below, of 
which only two lie fairly far below, including one further extended 
source with a weak AGN. Of particular note is the relatively small 
number of sources in the lower-right quadrant in all three panels.
This is not consistent with the work of Le F{\`e}vre et al.\ (2005)
and Paltani et al.\ (2007), who claim to have discovered 
a significantly larger galaxy population at high redshifts 
($z\sim 3$) than had previously been identified, particularly 
at bright magnitudes. They found that their $2.7\le z\le 3.4$ 
galaxies were mainly located near the boundary used to isolate 
LBGs in their $u-g$, $g-r$ color-color diagram, so they 
adopted a very conservative diagonal limit for their selection
criterion at the expense of more contamination from low-redshift
galaxies. We find no evidence for a strong population of this
type of galaxy.

\begin{figure*}
\centerline{\psfig{figure=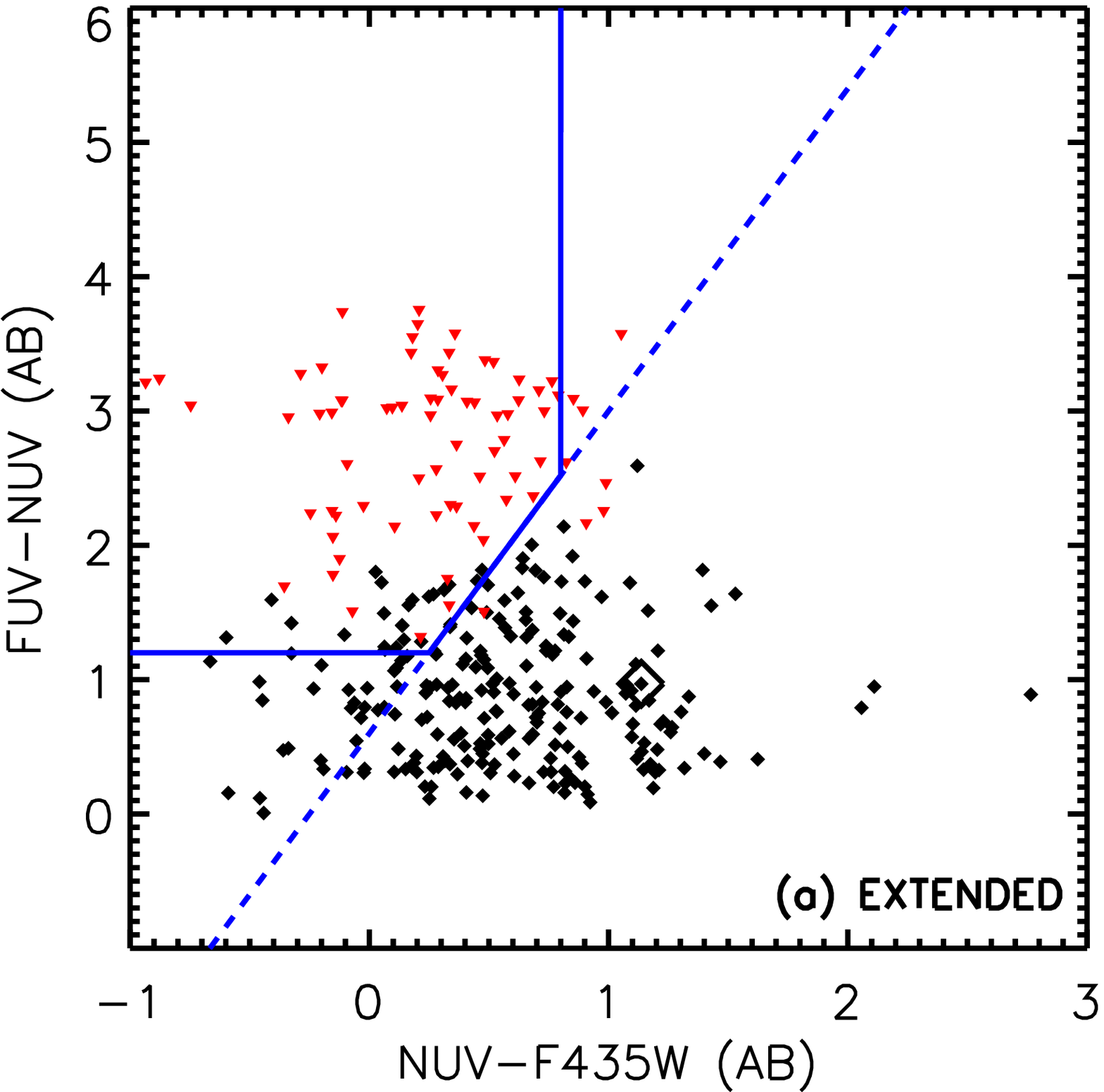,width=3.4in}
\psfig{figure=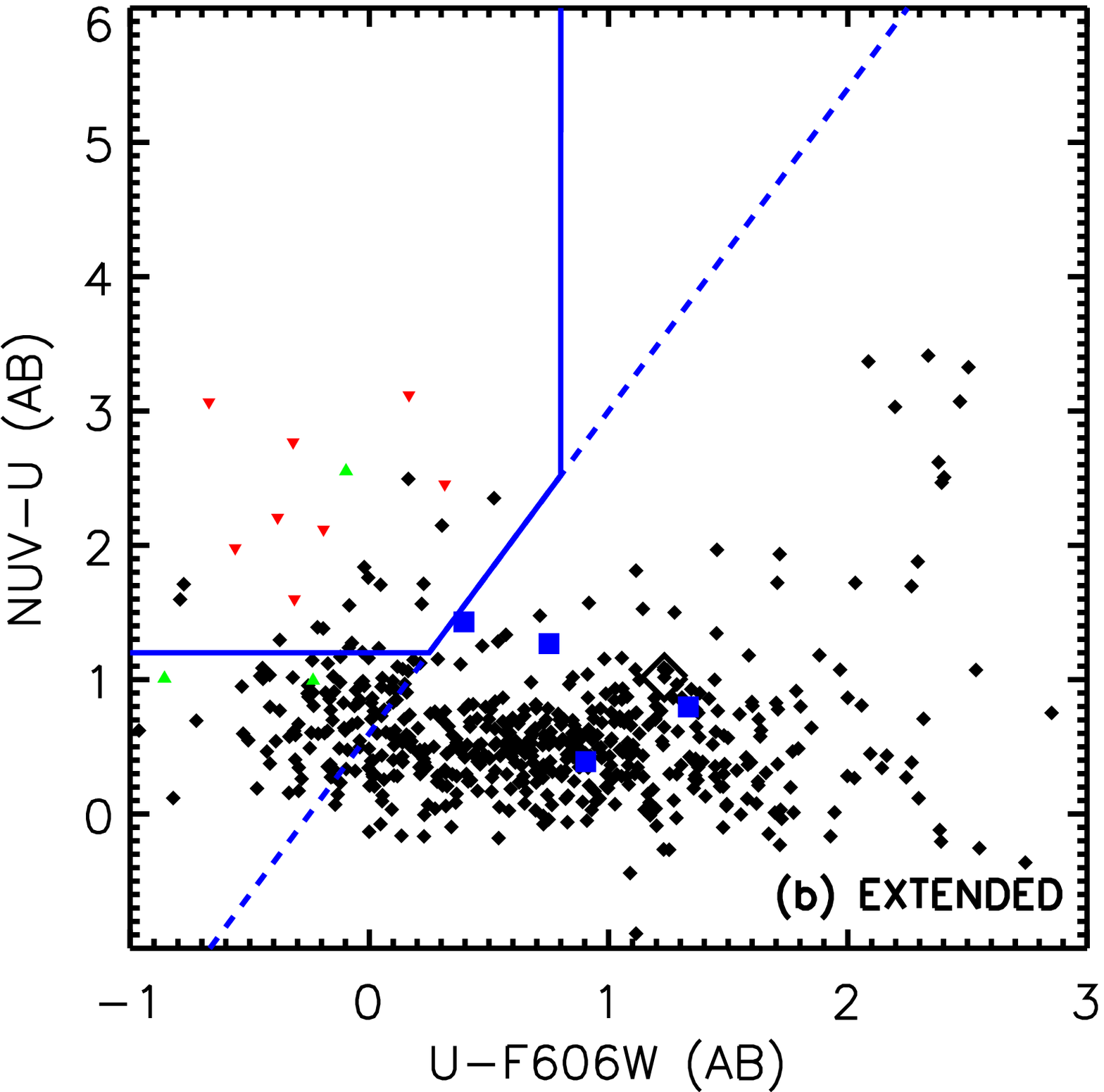,width=3.4in}}
\vskip 0.8cm
\centerline{\psfig{figure=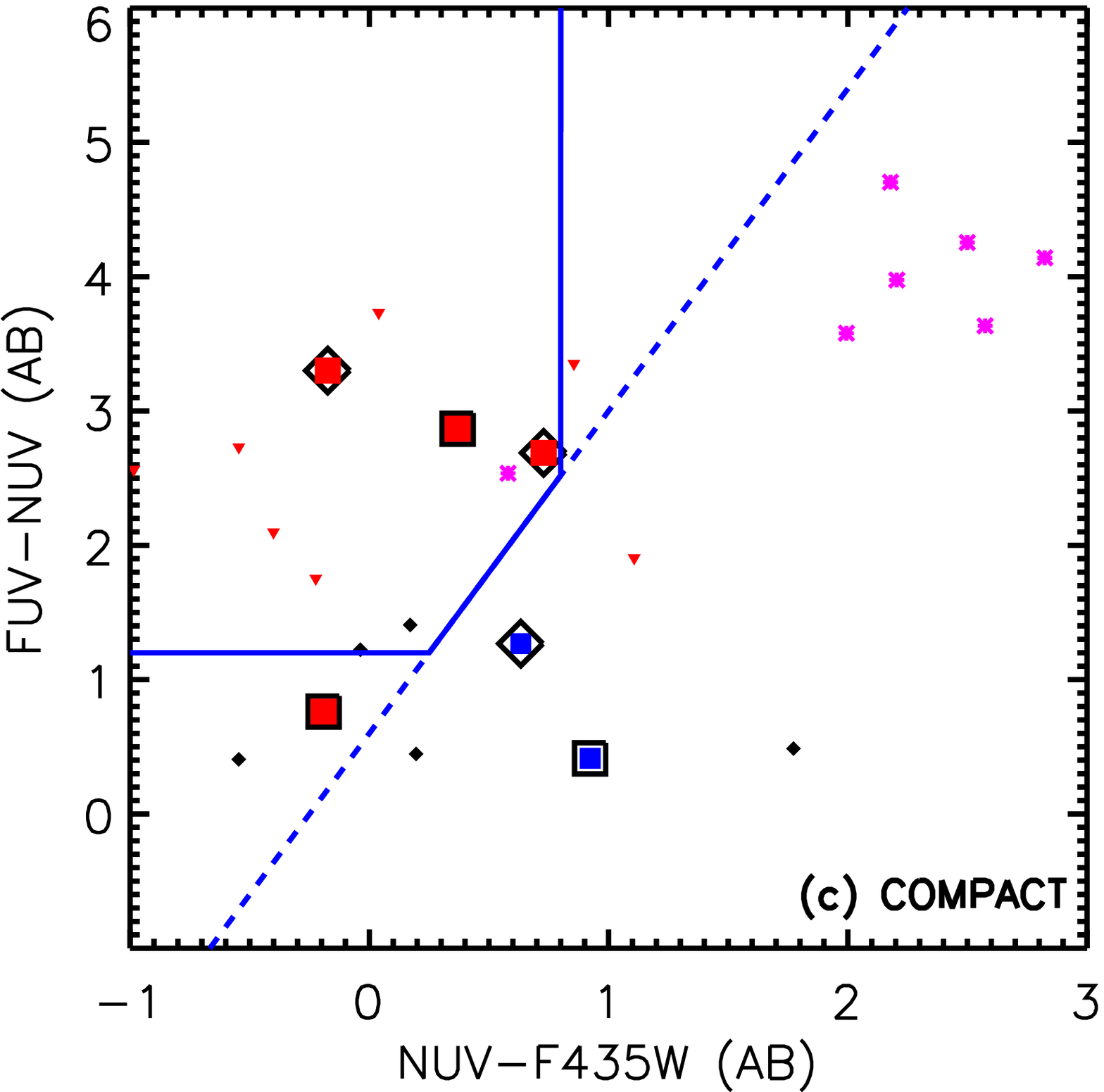,width=3.4in}
\psfig{figure=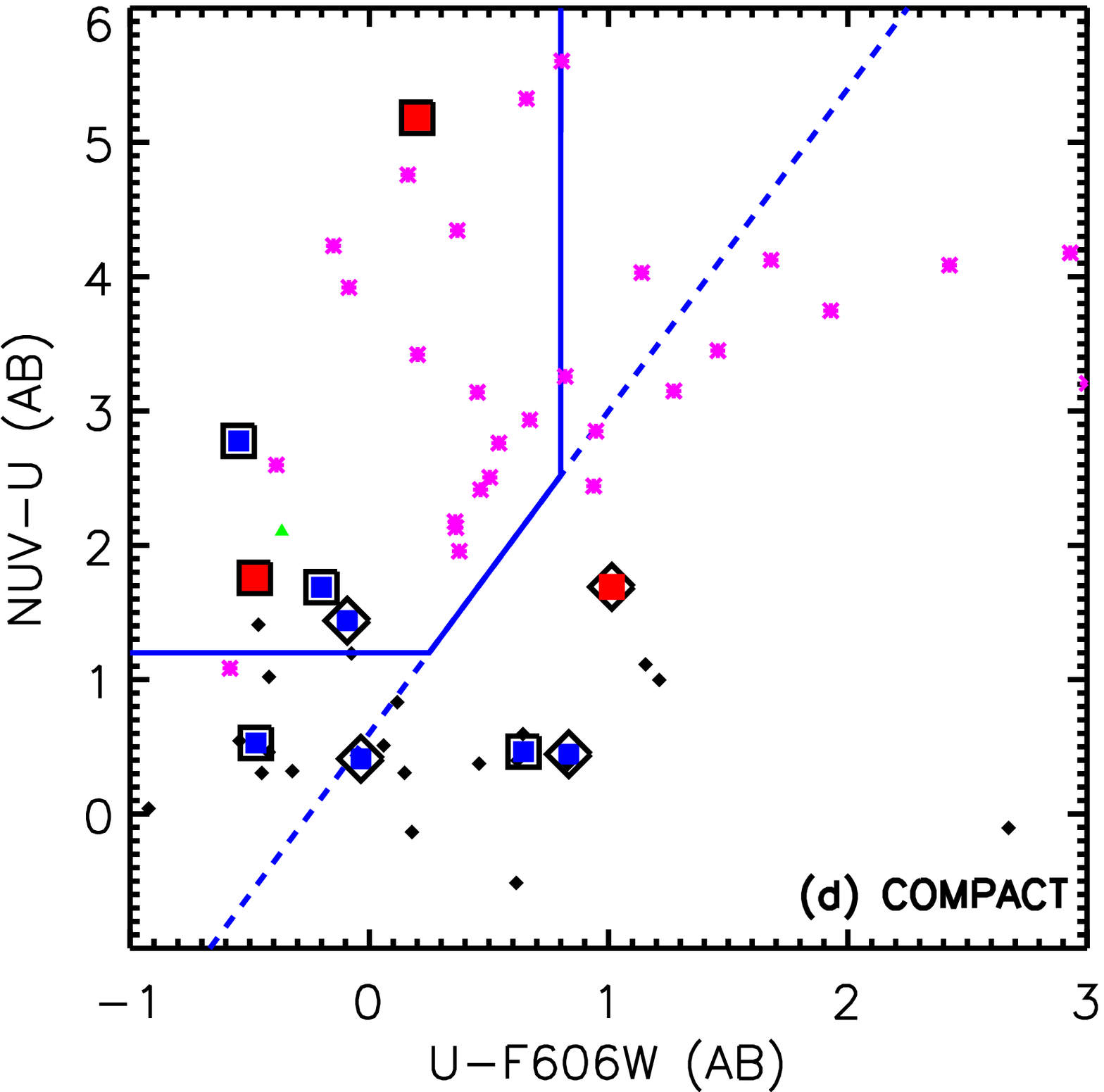,width=3.4in}}
\figcaption[]{
(a) Extended and (c) compact (FUV$-$NUV)$_{\rm AB}$ vs. 
(NUV$-$F435W)$_{\rm AB}$ and (b) extended and (d) compact
(NUV$-U)_{\rm AB}$ vs. ($U-$F606W)$_{\rm AB}$ color-color
diagrams for our 100\% spectroscopically complete 
${\rm NUV\/}_{\rm AB}< 23.75$ sample (a and c) 
and our 97\% spectroscopically complete $U_{\rm AB}<23.75$ 
sample (b and d). The extended and compact designations depend 
on whether the SExtractor 
stellarity index in the ACS F850LP image is less than or greater than
0.5, respectively. In (a) and (c) the black diamonds are galaxies
with spectroscopic redshifts $z\le 0.6$, the red upside-down
triangles are galaxies with spectroscopic redshifts $z>0.6$,
the large squares are AGNs {\em (blue for $z\le 0.6$; 
red for $z>0.6$)\/}, and the blue solid lines mark where LBGs with 
$z>0.6$ would be searched for in a color-color pre-selection.
In (b) and (d) the black diamonds are galaxies with spectroscopic
redshifts $z\le 1.4$, the red upside-down triangles are galaxies
with spectroscopic redshifts $z>1.4$,
the large squares are AGNs {\em (blue for $z\le 1.4$; red for $z>1.4$)\/},
the blue solid lines mark where LBGs with $z>1.4$ would be searched 
for in a color-color pre-selection, and the green triangles are the
unidentified sources. All of the spectroscopically identified 
stars {\em (purple stars)\/} fall in the compact panels. 
In all panels the broad-line AGNs are shown enclosed in larger 
black squares, and sources with high-excitation emission lines 
(C~IV or [Ne~V]) are shown enclosed in larger black diamonds. 
\label{figlbgnuv}
}
\end{figure*}

We can perform a similar analysis at lower redshifts using
the GALEX magnitudes. In Figures~\ref{figlbgnuv}a and 
\ref{figlbgnuv}c we show the LBG selection of $0.6\le z<1.4$ galaxies 
{\em (blue solid lines)\/}
using the (FUV$-$NUV$)_{\rm AB}$ vs. 
(NUV$-$F435W$)_{\rm AB}$ color-color diagram. 
In Figures~\ref{figlbgnuv}b and 
\ref{figlbgnuv}d we show the selection of $1.4\le z<2.7$ galaxies 
{\em (blue solid lines)\/}
using the (NUV$-U)_{\rm AB}$ vs. $(U-$F606W$)_{\rm AB}$ 
color-color diagram. In both cases we have used an equivalent
selection to that of the high-redshift LBGs.
For Figures~\ref{figlbgnuv}a and \ref{figlbgnuv}c we include 
all sources with ${\rm NUV\/}_{\rm AB}<23.75$, which is 100\% 
spectroscopically complete.
For Figures~\ref{figlbgnuv}b and \ref{figlbgnuv}d we include
all sources with $U_{\rm AB}<23.75$, which is 97\% 
spectroscopically complete. An ${\rm NUV\/}_{\rm AB}=23.75$ 
magnitude limit at $z=1$ corresponds to an F606W limit of 26 
at $z=3$, so these near-ultraviolet samples extend to much lower 
luminosities than the F606W sample.
The samples have been divided into extended 
($<0.5$; Figs.~\ref{figlbgnuv}a and \ref{figlbgnuv}b) and 
compact ($>0.5$; Figs.~\ref{figlbgnuv}c and \ref{figlbgnuv}d)
using the SExtractor stellarity index, as measured in the
ACS F850LP image.

%
%
\begin{inlinefigure}
\vskip 0.8cm
\centerline{\psfig{figure=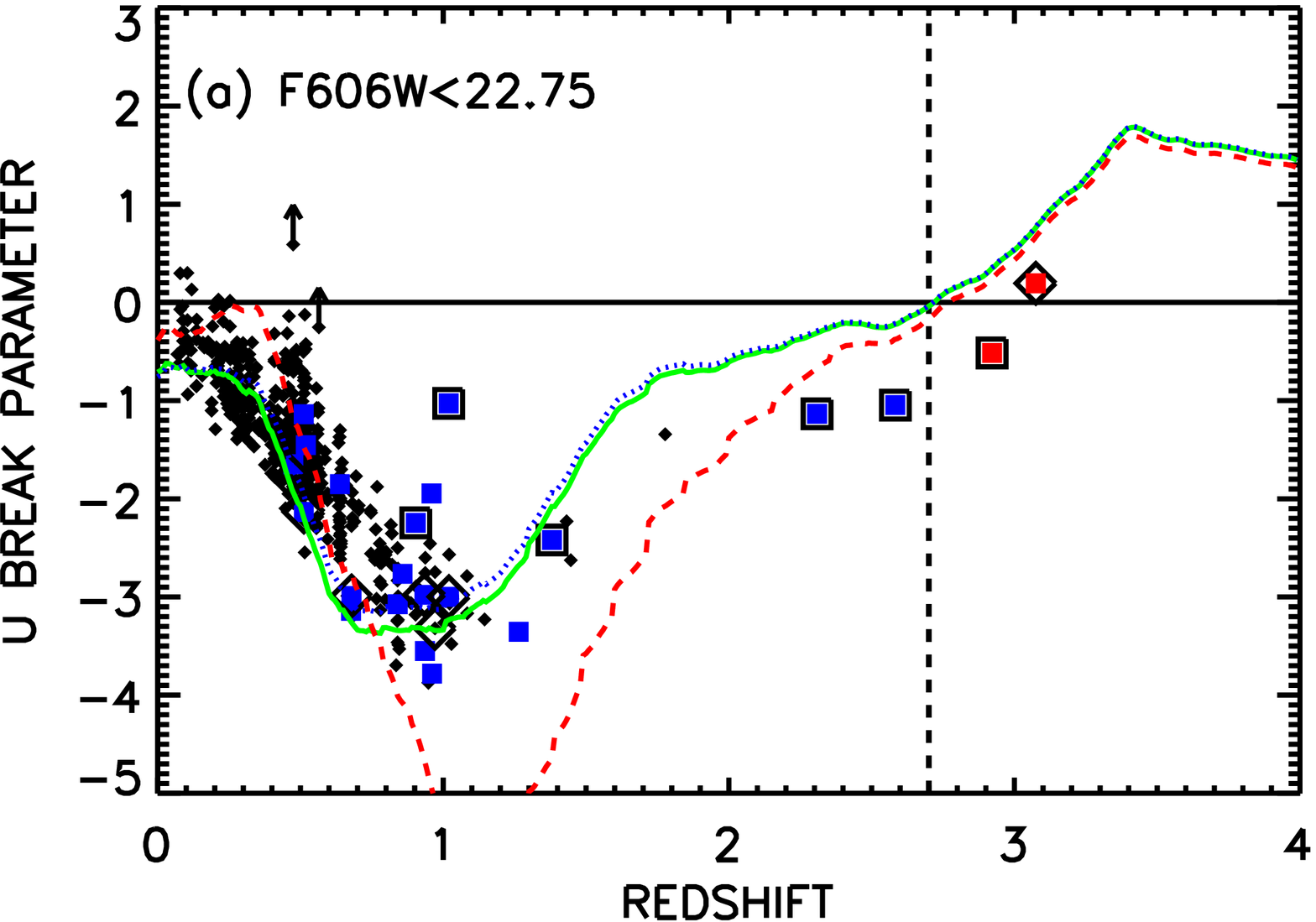,width=3.4in}}
\vskip 0.4cm
\centerline{\psfig{figure=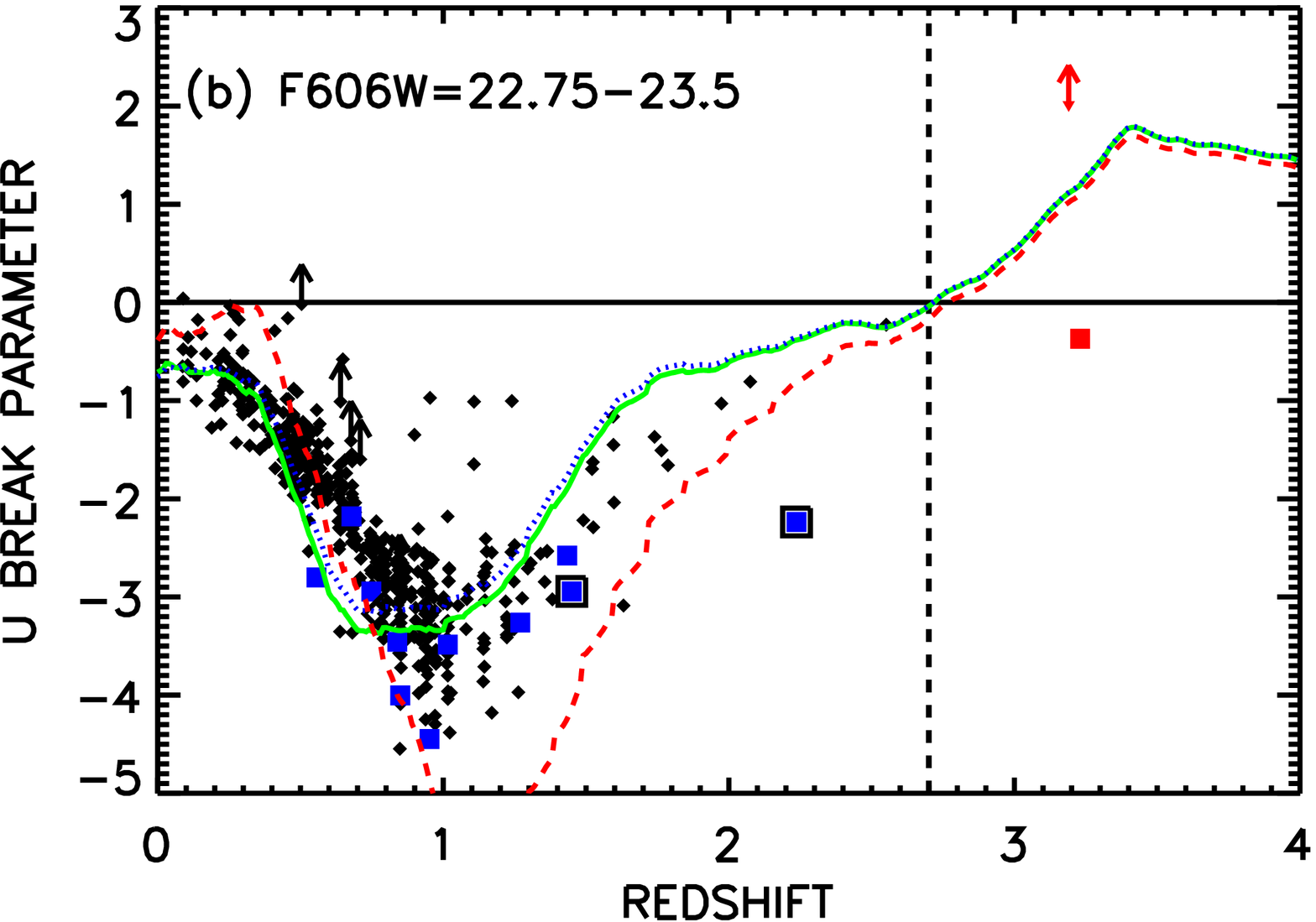,width=3.4in}}
\vskip 0.4cm
\centerline{\psfig{figure=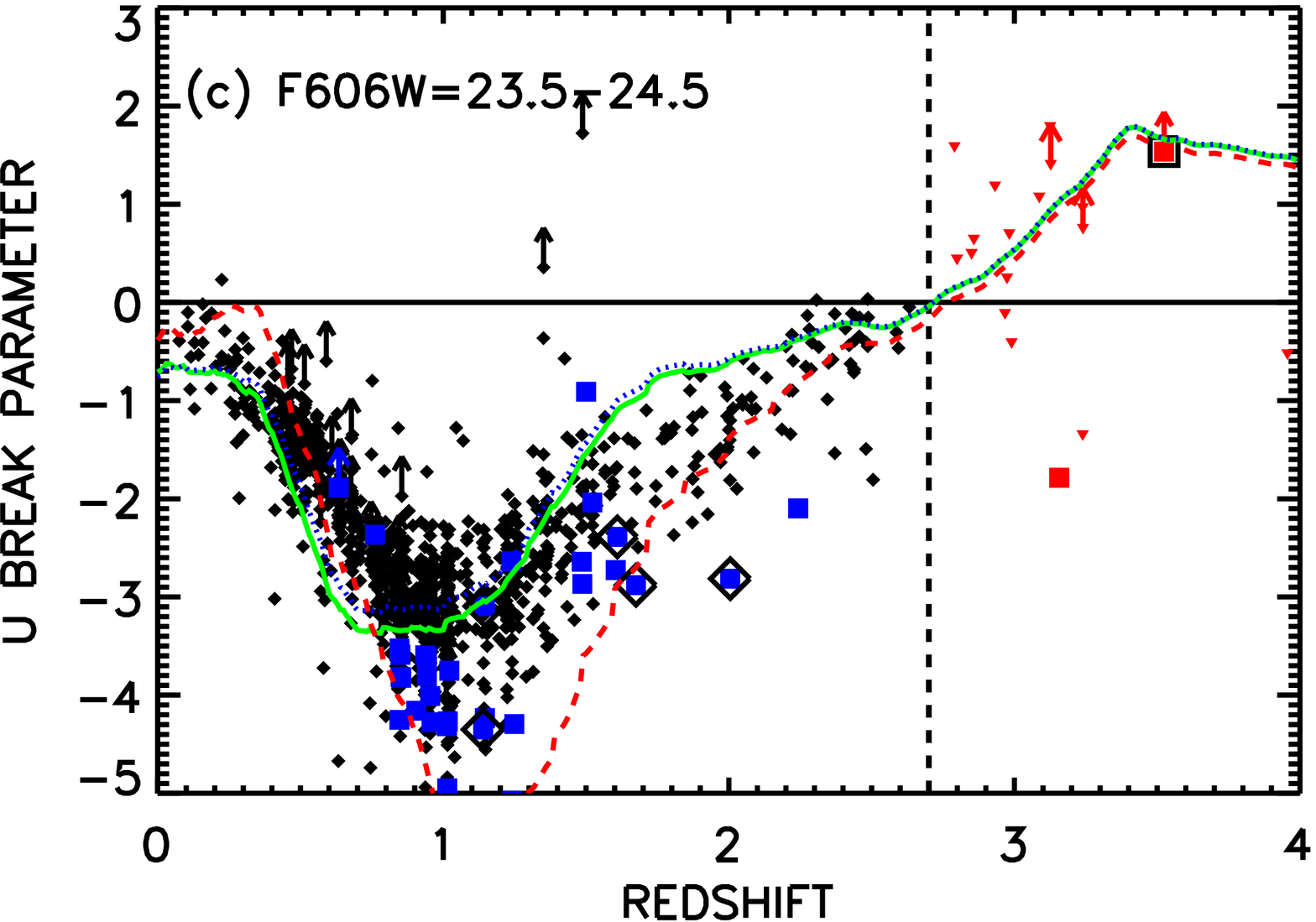,width=3.4in}}
\figcaption[]{
$U$ break parameter 
[$(U-{\rm F606W\/})_{\rm AB}-({\rm F606W\/}-{\rm F850LP\/})_{\rm AB}
\times 2.4 - 0.6$] vs. redshift for every 
${\rm F850LP\/}_{\rm AB}\ge 19$ galaxy with a spectroscopic 
redshift {\em (black for $z\le 2.7$; red for $z>2.7$)\/}. 
The large squares are AGNs {\em (blue
for $z\le 2.7$; red for $z>2.7$)\/}. 
The broad-line AGNs are shown enclosed in larger 
black squares. Sources with high-excitation emission lines 
(C~IV or [Ne~V]) are shown enclosed in larger black diamonds. 
The arrows show $2\sigma$ lower limits. 
(a) $19\le {\rm F606W\/}_{\rm AB}<22.75$,
where every galaxy has a spectroscopic redshift;
(b) $22.75\le {\rm F606W\/}_{\rm AB}<23.5$, where 97\% of the 
galaxies have redshifts and where there is only one $z>2.7$ galaxy;
and (c) $23.5\le {\rm F606W\/}_{\rm AB}<24.5$, where $z>2.7$
sources begin to appear in significant numbers.
The dashed vertical line marks the $z=2.7$ lower limit 
of the LBG redshift selection. The curves
show the expected break parameters from Bruzual \& Charlot
(2003) models for exponentially declining star formation 
rates with decline times of $10^{9}$~yrs {\em (red dashed)\/}, 
$5\times10^{9}$~yrs {\em (green solid)\/}, and 
$10^{10}$~yrs {\em (blue dotted)\/}, solar metallicities,
and ages equal to the age of the universe at that redshift.
\label{figubreak}
}
\end{inlinefigure}

%
%

The color-color diagrams pick out nearly all of the galaxies
above the selection limits ($z>0.6$ or $z>1.4$).
Figures~\ref{figlbgnuv}a and \ref{figlbgnuv}c suggest
that the (NUV$-$F435W)$_{\rm AB}$ constraint could be relaxed
without increasing the selection contamination.
It can be seen from Figures~\ref{figlbg} and \ref{figubreak}
that the vertical constraint can also be relaxed at higher redshifts, 
except as a way to minimize the star contamination problem when the
LBG selection is applied without a compactness
measurement. (Figure~\ref{figlbg}b shows how
the star track edges into the selection region; some
of these can be eliminated by the 
[F606W$-$F850LP]$_{\rm AB}<0.8$ constraint.) 
From Figure~\ref{figlbgnuv}d we see that star contamination can be 
substantial if we use the LBG method to select $z\sim 1$ galaxies 
without compactness information.

The color selections in Figure~\ref{figlbgnuv} are also surprisingly 
good at picking out AGNs, including many of the broad-line AGNs, 
suggesting that some of these have strong Lyman breaks also.
However, as with the high-redshift sample, there are
AGNs which would be missed by the color selection.
Cowie et al.\ (2008) discusses the implications of this 
for the production of the metagalactic ionizing flux.

We may use the 100\% complete NUV$_{\rm AB}<23.75$ 
sample of Figures~\ref{figlbgnuv}a 
and \ref{figlbgnuv}c to assess the selection completeness
and the selection contamination.
Of the 86 galaxies in that sample with $z>0.6$,
79 are selected by the cuts 
\begin{eqnarray}
({\rm FUV - NUV})_{\rm AB} > 1.2 \,,
\end{eqnarray}
and
\begin{eqnarray}
({\rm FUV - NUV})_{\rm AB} > ({\rm NUV} - U)_{\rm AB}\times 2.4 + 0.6 \,.
\end{eqnarray}
This gives a selection completeness of 92\%. Note that
one of the seven omitted sources is an X-ray luminous broad-line
quasar, and one is an AGN with high-excitation lines.
Since the above cuts pick out 116 sources, the
selection contamination by sources with $z\le0.6$ is 32\%.

We conclude that while there may be a very small number of
unusual sources where AGN contamination (or some other cause)
moves the source outside the selection region, more than 90\% of 
galaxies in the desired redshift range will be picked out by the 
LBG selection. While we have based our analysis 
on the low-redshift sample, the selection completeness should only 
be higher at high-redshifts, where the effects of the Ly$\alpha$
forest accentuate the LBG selection.

\subsection{The BzK Selection}
\label{secK}

Daddi et al.\ (2004) proposed a NIR selection technique for
finding high-redshift galaxies defined by
\begin{equation}
BzK \equiv (z-K_s)_{\rm AB} - (B-z)_{\rm AB} \,.
\end{equation}
With this relation, $z>1.4$ star-forming galaxies can be 
identified using the criterion $BzK \ge -0.2$, while $z>1.4$ 
old stellar systems can be isolated using the criteria 
$BzK<-0.2$ and $(z-K_s)_{\rm AB}>2.5$. Daddi et al.\ (2004) 
applied the technique to the K20 ($K_{s, {\rm AB}}<21.8$) survey 
(Cimatti et al.\ 2002). This survey includes a 32~arcmin$^2$ 
region in the GOODS-S, where 328 of the 347 sources have 
spectroscopic redshifts, and a 19~arcmin$^2$ area centered on 
the QSO~0055-269 at $z=3.656$, where 176 of the 198 sources
have spectroscopic redshifts. They also utilized photometric 
redshifts, which make up a sizeable fraction of their $z>1.4$ 
sample. Daddi et al.\ (2004) found that their $BzK$
selection criteria did very well at selecting $z>1.4$ galaxies;
moreover, only about 12\% of their final sample were lower redshift 
``interlopers''. Some of this contamination was due to AGNs.
They also checked their 
criteria on two deeper samples, a spectroscopic sample to 
$K_{s, {\rm AB}}=22.4$ from the Gemini Deep Deep Survey 
(GDDS; Abraham et al.\ 2004) and a photometric sample to 
$K_{s, {\rm AB}}=23.8$ from ISAAC imaging of the same GOODS-S
region covered by the K20 survey. These samples supported 
the continued validity of the method for magnitudes fainter than 
$K_{s, {\rm AB}}=21.8$, though the GDDS sample of $z>1.4$ galaxies
was small, and the ISAAC sample did not have spectroscopic
redshifts. (Popesso et al.\ 2008 have since increased the number 
of spectroscopic redshifts in the GOODS-S field using VIMOS. 
However, it does not appear that they are using a $K_s$ selected 
sample in their Figure~10 $BzK$ diagram.)

Because it is important to understand how well the $BzK$ selection 
works, especially before applying the method to extremely large 
samples, such as those being generated by the UKIRT Infrared Deep 
Sky Survey (UKIDSS; Lane et al.\ 2007),
several other groups have also tried to test its reliability. 
In most of these cases the authors have
compared the locations of the sources in the $BzK$ diagram 
with their photometric redshift estimates
(e.g., Kong et al.\ 2006; Grazian et al.\ 2007; Quadri et al.\ 2007).
However, the difficulty with this approach is that the colors used
to construct the $BzK$ diagram and the photometric redshift
estimates are not independent.

\subsubsection{Our Nearly Spectroscopically Complete Sample}

We can investigate the reliability of the $BzK$ method using
our nearly complete, large spectroscopic sample 
in the GOODS-N region. The area covered by our sample is 
substantially larger than that of the
K20 survey (145~arcmin$^2$ versus 51~arcmin$^2$), though
the $K_{s, {\rm AB}}$ limit to which we are complete to the same 
level as K20 is shallower ($K_{s, {\rm AB}}=21.4$ versus 
$K_{s, {\rm AB}}=21.8$).

In Figure~\ref{figbzzk21} we restrict our sample to the magnitude
range $15<K_{s, {\rm AB}}\le 21$ (the lower limit
is to avoid saturation problems, which would lead to incorrect
colors). Out of the 964 sources in this sample, only 45 have
not been spectroscopically identified {\em (green triangles)\/}.
We use red symbols {\em (upside-down triangles for galaxies and
red large squares for AGNs)\/} to show the $z>1.4$ sources, and 
we use black diamonds for galaxies and blue large 
squares for AGNs to show the $z\le 1.4$ sources. We denote the 
spectroscopically identified stars with purple star symbols, and 
we use purple large open squares to show sources whose SExtractor 
stellarity index, as measured in the {\rm F850LP\/} image, is $>0.5$ 
(i.e., compact). This parameter identifies all of the spectroscopic 
stars and many of the AGNs as compact sources.
We see a very tight sequence for the $z\le1.4$ galaxies. We 
also see that the $BzK$ criterion {\em (diagonal line)\/} 
does a reasonable job of separating most of the sources into the 
correct redshift regimes. However, we note that to this magnitude 
limit only one spectroscopically identified galaxy at
$z>1.4$ lies in the passive galaxies region of 
the diagram (to the right of the diagonal line and above the
horizontal line). 

There are 15 spectroscopically identified
$z>1.4$ sources to the left of the diagonal line and only three 
(one of which lies in the $z>1.4$
passive galaxy region) to the right of the diagonal line.
Thus, the $BzK$ selection of $z>1.4$ star-forming galaxies 
(i.e., excluding the $z>1.4$ source in the passive galaxy region)
has a high selection completeness (88\%).

%
%
\begin{inlinefigure}
\vskip 0.8cm
\centerline{\psfig{figure=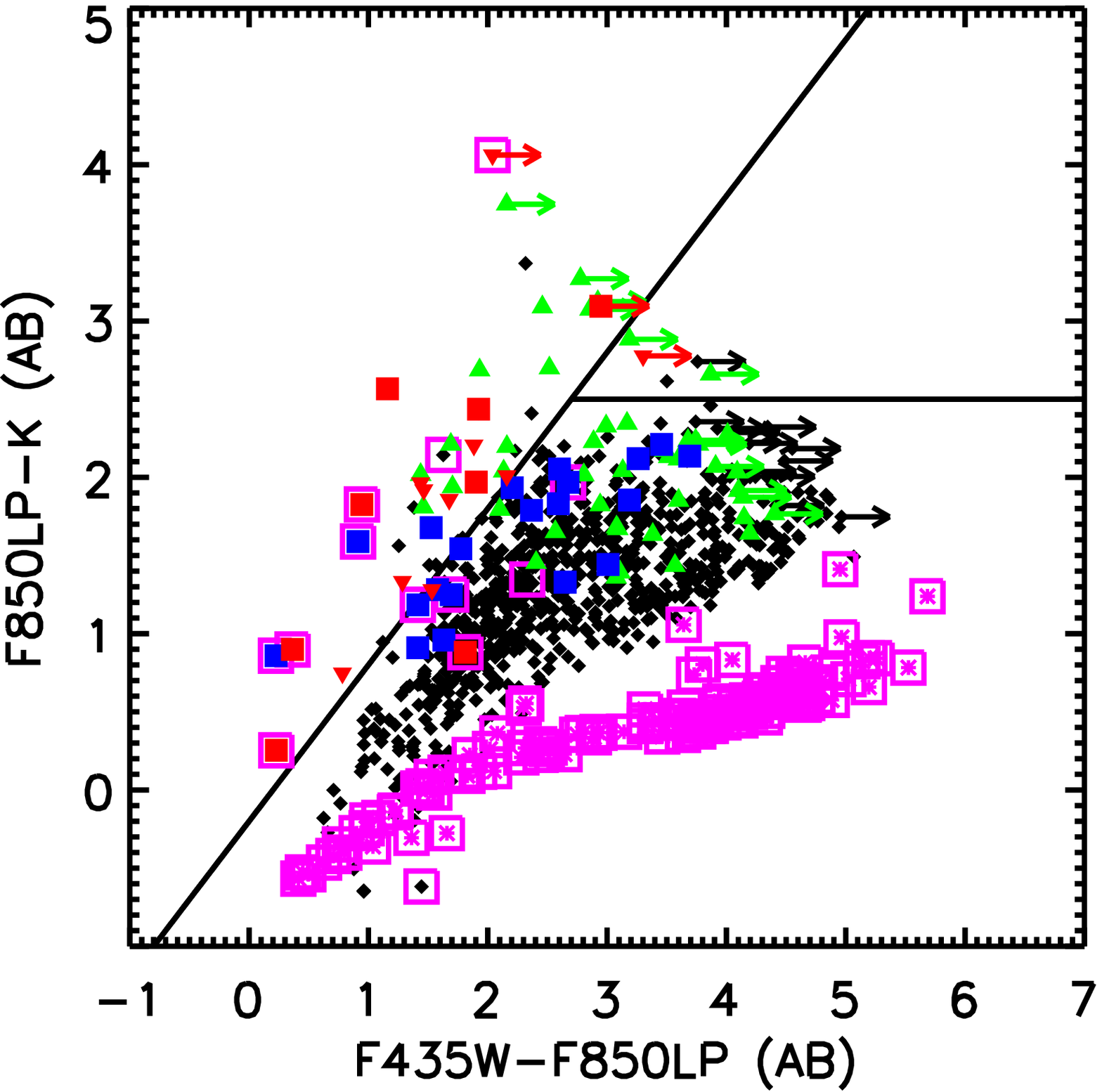,width=3.4in}}
\figcaption[]{
$({\rm F850LP\/}-K_s)_{\rm AB}$ vs. (F435W-F850LP)$_{\rm AB}$ 
for the $15<K_{s,{\rm AB}}\le 21$ sample.
The $z\le 1.4$ galaxies are denoted by black
diamonds, and the $z>1.4$ galaxies are denoted by red
upside-down triangles. AGNs are denoted by
large squares {\em (blue for $z\le 1.4$; red for $z>1.4$)\/}. 
The unidentified sources
are denoted by green triangles. Sources with only $2\sigma$ lower
limits in F435W are denoted by right-pointing arrows.
Spectroscopically identified stars are denoted by purple star symbols.
The purple large open squares show sources whose SExtractor
stellarity index measured on the F850LP image is $>0.5$.
The black diagonal line shows the $BzK\ge -0.2$ selection of
$z>1.4$ star-forming galaxies. This line combined with
the horizontal line shows the ${\rm F850LP\/}-K_s>2.5$
selection of old galaxies at $z>1.4$. Both were proposed by
Daddi et al.\ (2004).
\label{figbzzk21}
}
\end{inlinefigure}

We can also estimate the minimum and maximum selection 
contamination by taking into account the 45 spectroscopically
unidentified sources in the sample. 13 of these lie in the
$z>1.4$ star-forming region, 2 in the $z>1.4$ passive galaxies
region, and the remaining 30 in the $z\le 1.4$ region.

We first consider the $z\le 1.4$ region of the diagram. Here 
there are two spectroscopically identified $z>1.4$ sources
(interlopers) and 768 spectroscopically identified $z\le 1.4$ 
sources. If we assume that all 30 of the unidentified sources 
in the $z\le 1.4$ region do in fact have $z\le 1.4$, then we obtain 
a minimum selection contamination of 0.25\%. If we instead assume 
that all 30 of the unidentified sources in the $z\le 1.4$ 
region actually have $z>1.4$ (interlopers), then we obtain 
a maximum selection contamination of 4\%.

We next consider the $z>1.4$ star-forming region of the diagram
(i.e., we are not considering the $z>1.4$ passive region).
Here there are 14 spectroscopically identified $z\le 1.4$ 
sources (interlopers) and 15 spectroscopically identified $z>1.4$ 
sources. If we assume that all 13 of the unidentified sources
in the $z>1.4$ star-forming region do in fact have $z>1.4$, 
then we obtain a minimum selection contamination
of 33\%. If we instead assume that all 13 of the unidentified 
sources in the $z>1.4$ star-forming region actually 
have $z\le 1.4$ (interlopers), 
then we obtain a maximum selection contamination of 64\%. 

We conclude that at these magnitudes the $BzK$ selection 
of $z>1.4$ star-forming galaxies has a high selection 
completeness but a substantial selection contamination. 
Thus, $BzK$ is effective at picking out high-redshift 
candidates, but spectroscopic follow-up is crucial to weed 
out the substantial number of interlopers.

\subsubsection{Our Fainter Magnitude Sample}

It is important to try to push the $BzK$ analysis to fainter 
magnitudes, even with the increasing spectroscopic incompleteness.
Reddy et al.\ (2005) tested the $BzK$ selection on a fairly 
substantial, albeit not uniformly selected, high-redshift 
spectroscopic sample. They spectroscopically observed a
$z\sim 2$ sample in the GOODS-N field
selected using the observed $U_nG\cal{R}$ 
colors to a limiting AB magnitude of $\cal{R}$=25.5.
These $z\sim 2$ galaxies were selected to be actively
star-forming galaxies with the same range in UV properties
and extinctions as the LBGs at $z\sim 3$. Adelberger et al.\
(2004) and Steidel et al.\ (2004) named sources selected
in this way ``BX" ($2.0\le z\le 2.6$) and ``BM" ($1.5\le z\le 2.0$)
galaxies.

%
%
\begin{inlinefigure}
\vskip 0.8cm
\centerline{\psfig{figure=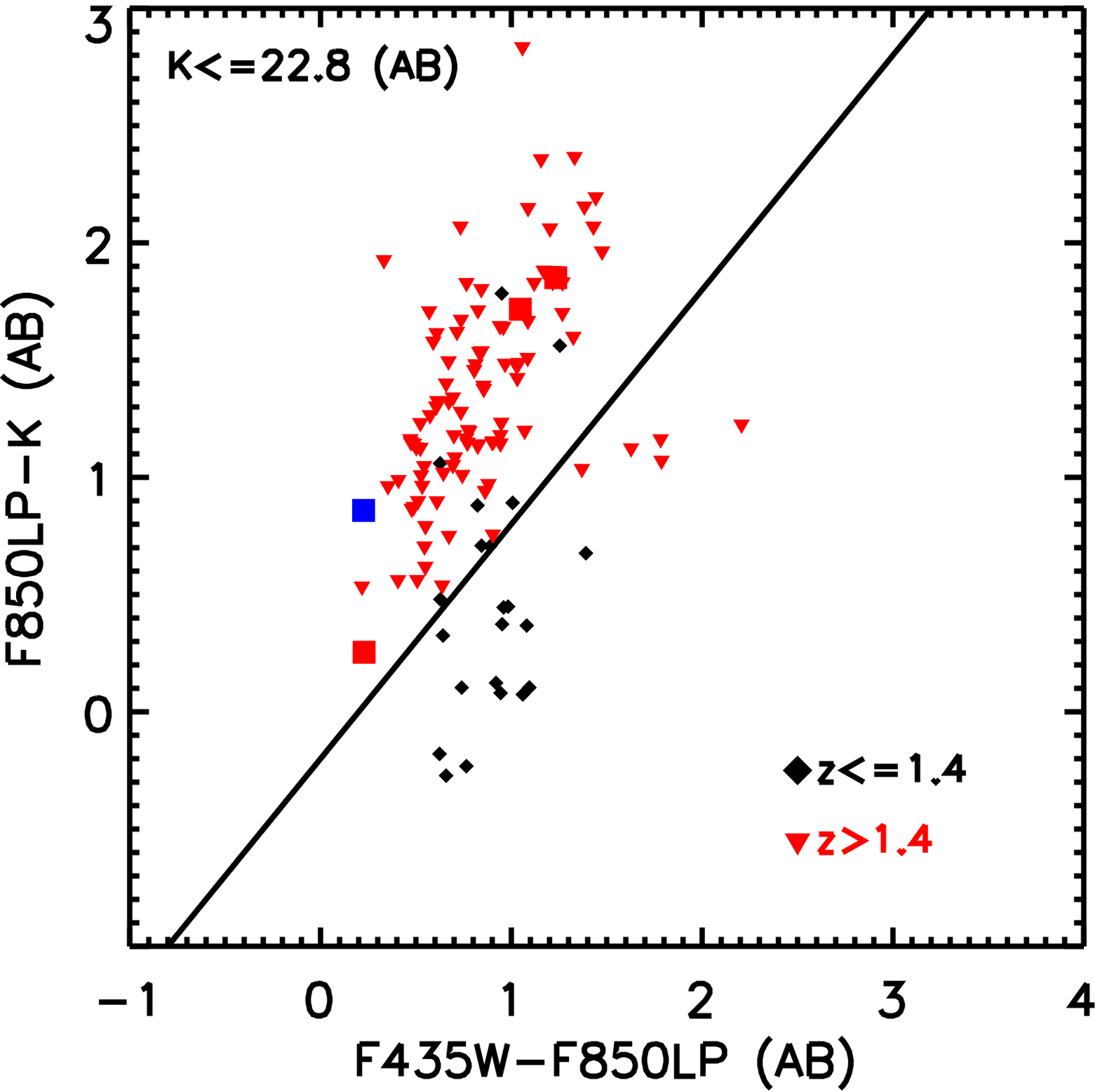,width=3.4in}}
\figcaption[]{
$({\rm F850LP\/}-K)_{s,{\rm AB}}$ vs. (F435W-F850LP)$_{\rm AB}$ 
for the $15< K_{s,{\rm AB}}\le 22.8$ sources with redshifts 
from Reddy et al.\ (2005). The $z\le 1.4$ galaxies are denoted 
by black diamonds, and the $z>1.4$ galaxies are denoted by
red upside-down triangles. AGNs are denoted by large squares
{\em (blue for $z\le 1.4$; red for $z>1.4$)\/}.
The diagonal line shows the $BzK\ge -0.2$ selection of
$z>1.4$ star-forming galaxies proposed by Daddi et al.\ (2004).
\label{figreddy}
}
\end{inlinefigure}

Reddy et al.\ (2005) then used a moderately deep $K_s$ image 
($5\sigma$ limit of $K_{s,{\rm AB}}\sim 23.8$) covering
$\sim 72.3$~arcmin$^2$ to put their galaxies on a
$K_{s, {\rm Vega}}\le 21$ ($K_{s,{\rm AB}}\le 22.8$) 
$BzK$ diagram (see their Fig.~12).
For consistency, in Figure~\ref{figreddy} we replot their 
spectroscopic sample on a $BzK$ diagram using our deeper imaging data
but restricting to their range $15< K_{s,{\rm AB}}\le 22.8$.
We have coded the data points so the $z\le 1.4$ galaxies 
{\em (black diamonds)\/} can be differentiated from the $z>1.4$ 
galaxies {\em (red upside-down triangles)\/}. 
We denote the AGNs (rest-frame hard or soft X-ray luminosities
$>10^{42}$~ergs~s$^{-1}$) with large squares 
{\em (blue for $z\le 1.4$; red for $z>1.4$)\/}.
For this fainter spectroscopic 
sample the $BzK$ selection still seems to do a reasonably 
good job of separating the high-redshift galaxies from the low-redshift 
galaxies, with only a small amount of contamination in either direction
(five $z>1.4$ sources in the low-redshift region;
nine $z\le 1.4$ sources in the high-redshift region, including one AGN). 
The $BzK$ $z>1.4$ star-forming galaxy selection completeness is 91\%
for this sample.

Reddy et al.\ (2005) claim that the $BzK$ selection completeness 
decreases slightly at fainter magnitudes, missing
about 20\% of the $K_{s,{\rm AB}}>22.8$ BX/BM galaxies with spectroscopic
redshifts $1.4<z<2.6$. They point out that a significant fraction 
of these have colors that place them within $\lesssim0.2$~mag of
the selection window, which is comparable to the photometric
uncertainties. Reddy et al.\ (2005) also found that 11\% of their 
BX/BM {\em candidates\/} with $K_s$ data were undetected to 
$K_{s,{\rm AB}}=24.3$ ($3\sigma$). Thus, they conclude that
high-redshift sources may be 
missed from the $BzK$ selection because of photometric scatter or 
because of insufficiently deep $K_s$ data.

Although we are unable to investigate these issues with a 
nearly spectroscopically complete sample, we can use
just our spectroscopically identified sample to study
the selection completeness and selection contamination
with increasing magnitude. In Figure~\ref{figbzzk} we plot 
two redshift samples separately: (a) $z\le 1.4$ sources only 
and (b) $z>1.4$ sources only. For each
we show three magnitude intervals 
($15 < K_{s,{\rm AB}} \le 21$ - {\em green upside-down triangles\/}; 
$21 < K_{s,{\rm AB}} \le 23$ - {\em red triangles\/};
$23 < K_{s,{\rm AB}} \le 24$ - {\em black diamonds\/})
(note that the sources heavily overlap).

With increasing magnitude there is an increasing build-up of
sources along the $BzK$ boundary for both redshift intervals.
However, in contrast to Reddy et al.\ (2005)'s conclusion that 
a substantial fraction of $z>1.4$ sources are missed by the $BzK$
selection, we find that the selection completeness is 
extremely good. In total, 254 of the 267 sources with $1.4<z<2.8$ 
(95\%) are found with the $BzK$ selection. For the faintest magnitude
interval $23< K_{s,{\rm AB}}\le 24$, 69 of the 74 sources with 
$1.4<z<2.8$ (93\%) are found. The 
different conclusion is probably a consequence of the deeper $K_s$ 
image used here and the corresponding increase in precision 
in the determination of the $BzK$ parameter.

%
%
\begin{inlinefigure}
\vskip 0.8cm
\centerline{\psfig{figure=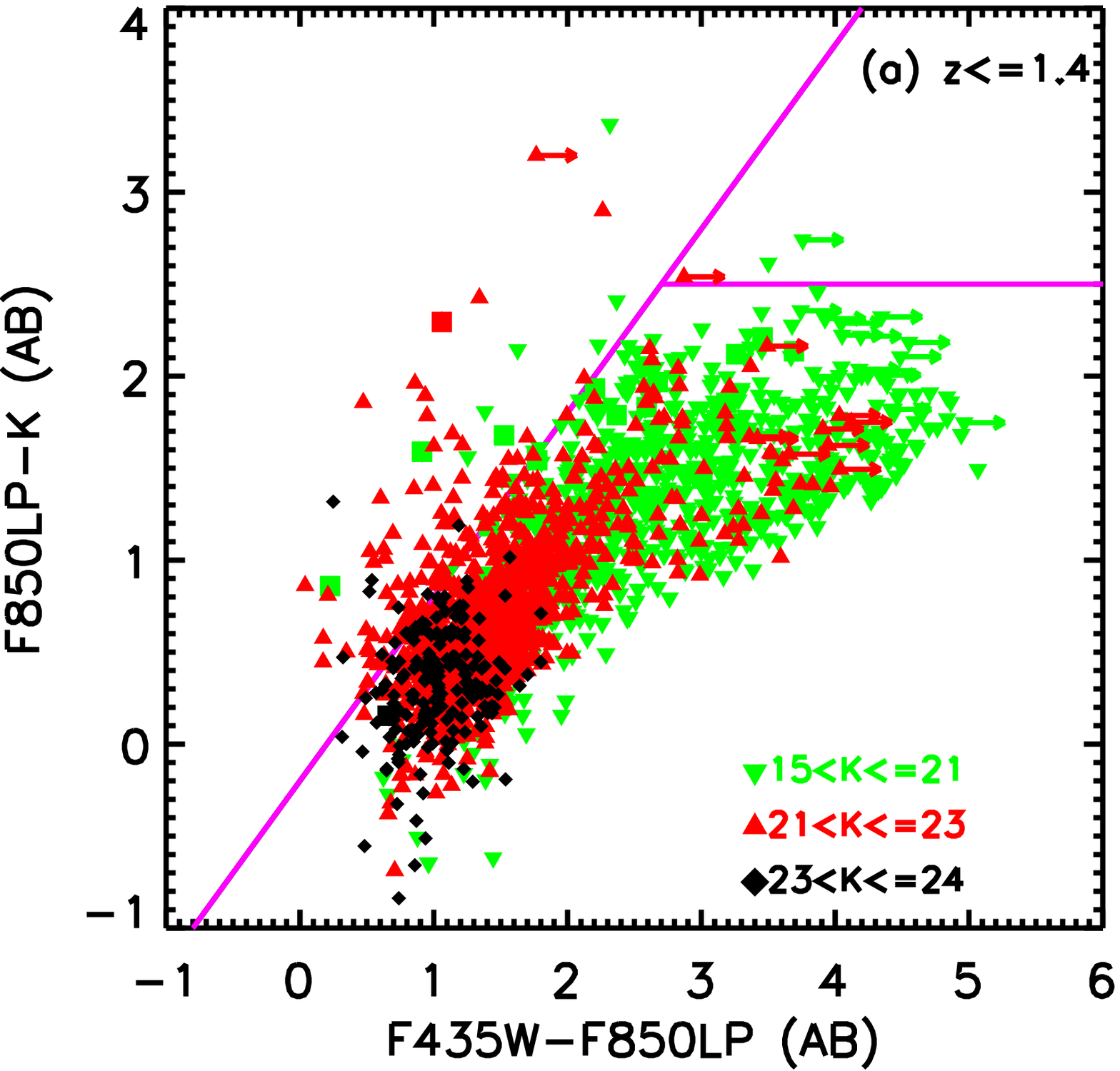,width=3.4in}}
\vskip 0.8cm
\centerline{\psfig{figure=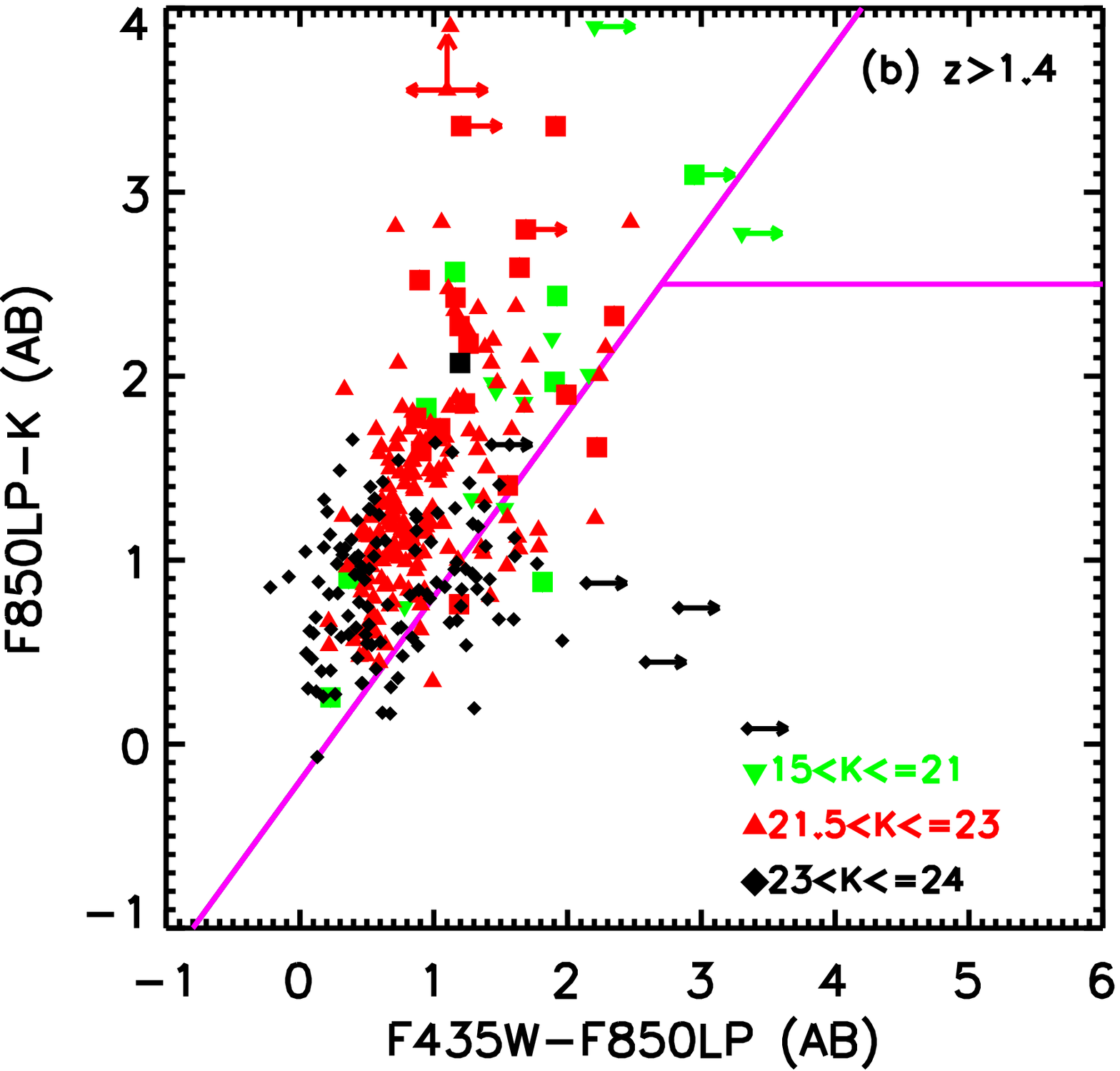,width=3.4in}}
\figcaption[]{
$({\rm F850LP}-K_s)_{\rm AB}$ vs. (F435W-F850LP)$_{\rm AB}$ for all 
of the spectroscopically identified $15<K_{s,{\rm AB}}\le 24$ 
sources with redshifts (a) $z\le 1.4$ and (b) $z>1.4$.
AGNs are denoted by large squares.  
Sources with $({\rm F850LP}-K_s)_{\rm AB}$ colors redder than 3.9
are shown at that value. 
The sources are color and symbol coded by magnitude range 
(see figure legend). The diagonal line shows the $BzK\ge -0.2$ 
selection of $z>1.4$ star-forming galaxies. This line combined
with the horizontal line shows the $BzK\ge -0.2$,
$({\rm F850LP\/}-K_s)_{\rm AB}>2.5$
selection of passive galaxies at $z>1.4$. 
Both were proposed by Daddi et al.\ (2004).
\label{figbzzk}
}
\end{inlinefigure}

%
%
\begin{inlinefigure}
\vskip 0.8cm
\centerline{\psfig{figure=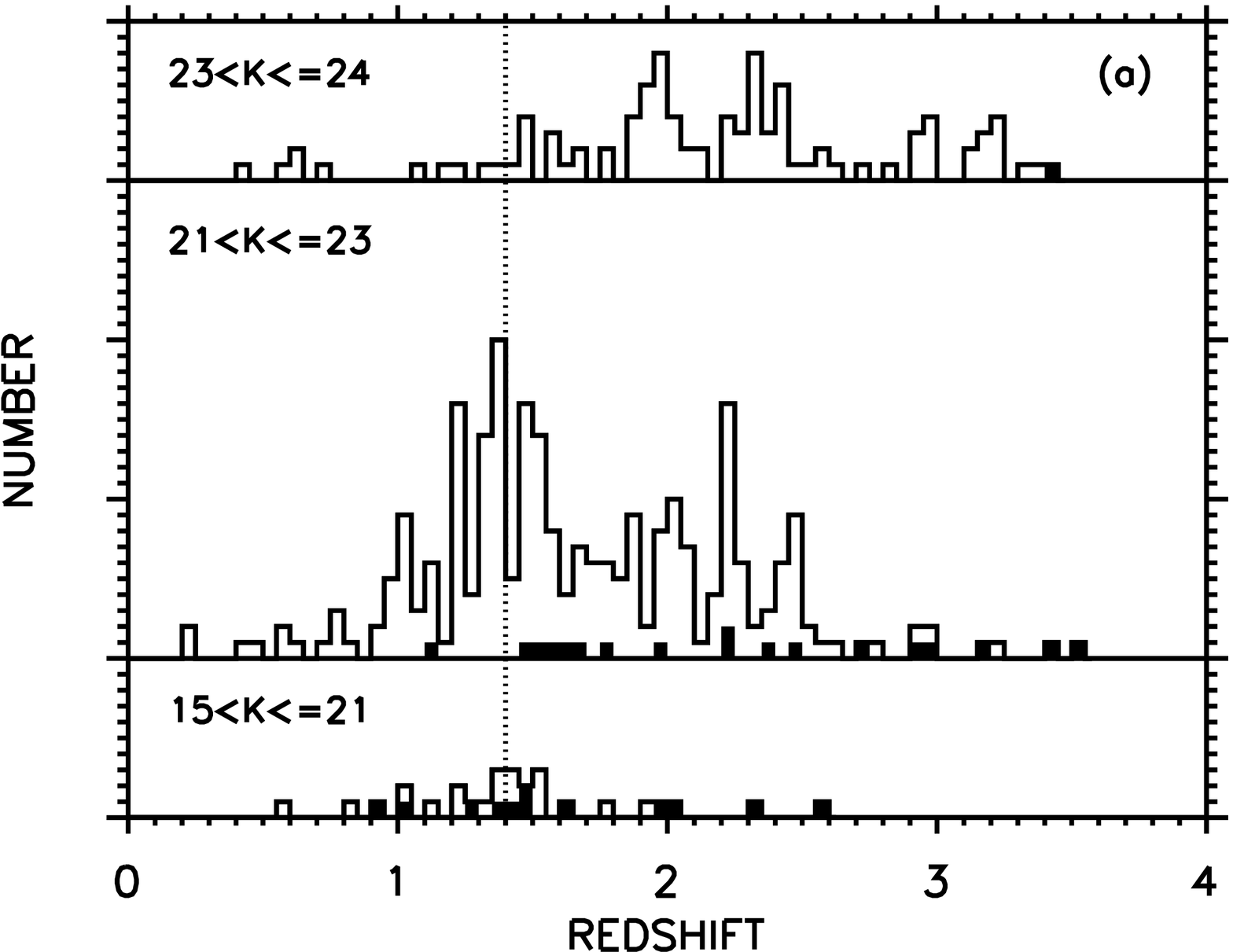,width=3.4in}}
\vskip 0.8cm
\centerline{\psfig{figure=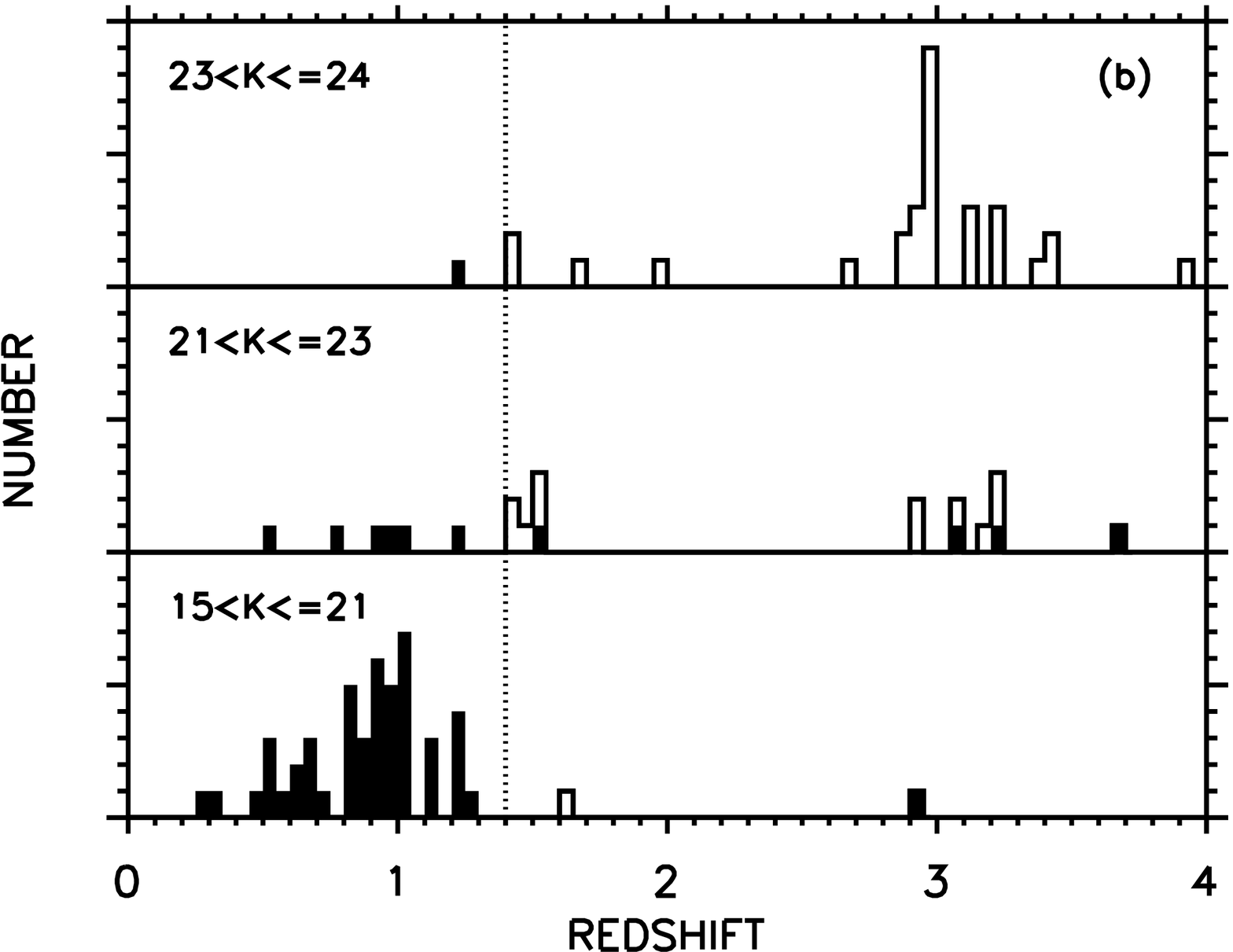,width=3.4in}}
\figcaption[]{
Redshift histograms by magnitude interval (see figure labels) for the 
spectroscopically identified $15<K_{s,{\rm AB}}\le 24$ sources. 
AGNs are denoted by solid shading. The dotted vertical lines show 
$z=1.4$. (a) $BzK\ge -0.2$, the selection of $z>1.4$ star-forming
galaxies proposed by Daddi et al.\ (2004).
(b) $BzK<-0.2$ and $({\rm F850LP}-K_s)_{\rm AB}\le 2.5$ (the
latter constraint to exclude passive galaxies at $z>1.4$).
Here only AGNs are shown at $z\le 1.4$ due to the 
very large numbers of galaxies at $z\le 1.4$.
Each tick mark on the y-axis represents one source in both panels.
\label{fighist}
}
\end{inlinefigure}

To quantitatively assess what fraction of the spectroscopically
identified sources are interlopers on either side of the $BzK$ 
boundary, in Figure~\ref{fighist} we plot redshift histograms of 
(a) sources that satisfy the $BzK$ $z>1.4$ star-forming galaxy 
criterion and (b) sources that do not. Both figures are divided 
into the same $K_{s, \rm AB}$ magnitude intervals used in 
Figure~\ref{figbzzk}. AGNs are denoted by solid shading. Since 
there are such a large number of
galaxies at $z\le 1.4$, in Figure~\ref{fighist}b we 
only show the AGNs at $z\le 1.4$. 
In (a) the sources to the left of the vertical line can be
thought of as interlopers in the $z>1.4$ $BzK$ star-forming
region. The contamination is substantial, about 
35\% in the $21<K_{s, {\rm AB}}\le 23$ interval. 
In (b) the sources to the right of the vertical line can be 
thought of as interlopers in the $z\le 1.4$ region
(or as $z>1.4$ sources missed by the $BzK$ star-forming
galaxy selection).

This is consistent with our results from using our
highly spectroscopically complete sample at $15<K_{s, {\rm AB}}\le 21$.
A $BzK$ selection picks out nearly all the $z>1.4$ sources 
but with a substantial contamination by lower redshift
objects. However, it must be remembered that the spectroscopically 
unidentified sources at these fainter magnitudes may contain 
substantial biases and that the number of unidentified sources 
in these fainter magnitude ranges is very large:
1016 unidentified sources (versus 1331 identified) at 
$21<K_{s, {\rm AB}}\le 23$ and 2083 unidentified sources
(versus 311 identified) at $23<K_{s, {\rm AB}}\le 24$.

\subsection{H$^-$}
\label{sechminus}

The minimum at 1.6~$\mu$m in the opacity of the H$^-$ 
ion present in the atmospheres of cool stars (e.g., John 1988) 
is seen as a spectral ``bump'' in the SEDs of all but
the youngest ($\sim 1$~Myr) composite stellar populations
(see Fig.~1 of Sawicki 2002).
Because of its strength and near universality,
Simpson \& Eisenhardt (1999) and Sawicki (2002)
proposed that the 1.6~$\mu$m bump could be used to obtain 
photometric redshift measurements for galaxies. 
However, in using MIR colors as a redshift indicator, 
one needs to be concerned about the effects of AGN 
contamination. 

Since optically selected luminous AGNs often have red power 
law SEDs (e.g., Neugebauer et al.\ 1979; Elvis et al.\ 1994), 
several groups have used this characteristic to try
to select AGNs in deep surveys using IRAC data 
(Alonso-Herrero et al.\ 2006; Donley et al.\ 2007).
Other authors (Lacy et al.\ 2004; Stern et al.\ 2005;
Hatziminaoglou et al.\ 2005; Sajina et al.\ 2005) have 
proposed that AGNs might be found in a different portion 
of MIR color-color space than star-forming galaxies. Both of 
these approaches have been suggested as ways to locate AGNs 
that may be missed by other techniques. 

In Figure~\ref{hminus} we show the 
$(m_{3.6~\mu{\rm m}}-m_{4.5~\mu{\rm m}})_{\rm AB}$ versus 
$(m_{5.8~\mu{\rm m}}-m_{8.0~\mu{\rm m}})_{\rm AB}$ color-color 
diagram for all the spectroscopically identified sources 
in Table~\ref{tab1} with $m_{8.0~\mu{\rm m}, {\rm AB}}<22$
and (a) a rest-frame hard or soft X-ray luminosity
$<10^{42}$~ergs~s$^{-1}$ and a radio power
$<10^{31}$~ergs~s$^{-1}$~Hz$^{-1}$,
(b) a rest-frame hard or soft 
X-ray luminosity $\ge 10^{42}$~ergs~s$^{-1}$,
(c) a rest-frame hard or soft X-ray luminosity
$<10^{42}$~ergs~s$^{-1}$ and a radio power
$\ge 10^{31}$~ergs~s$^{-1}$~Hz$^{-1}$.
We apply the $m_{8.0~\mu{\rm m}, {\rm AB}}<22$ limit
because at fainter magnitudes the signal to noise
ratio at 8$~\mu$m is poor and the colors would not be reliable. 
The $K_{\rm s,AB}<24.5$ sample should include all of the
$m_{8.0~\mu{\rm m}, {\rm AB}}<22$ sources, since sources
are fainter at 8$~\mu$m. The spectroscopic completeness of
the $m_{8.0~\mu{\rm m}, {\rm AB}}$ sample is 94\% 
at $m_{8.0~\mu{\rm m}, {\rm AB}}=21$ and 79\% at
$m_{8.0~\mu{\rm m}, {\rm AB}}=22$.
An ultraluminous infrared galaxy 
(ULIRG; $L_{\rm FIR}\ge 4\times 10^{45}$~ergs~s$^{-1}$) 
that follows the FIR-radio correlation (Condon 1992) would 
have a radio power of about 
$4.8\times 10^{30}$~ergs~s$^{-1}$~Hz$^{-1}$ (Barger et al.\ 2007).
Thus, it is likely that any source with a radio power 
$\ge10^{31}$~ergs~s$^{-1}$~Hz$^{-1}$ is a radio AGN, or at
least a very unusual source, which is why we show these
sources separately in (c). We will refer to these as
high radio power sources hereafter, but their colors are
clearly very similar to those of the rest of the sample.
In (b) we enclose in black large squares any source with 
a rest-frame hard or soft X-ray luminosity
$\ge 10^{44}$~ergs~s$^{-1}$ (X-ray quasar luminosity).

We mark with the green dashed lines the region considered 
to contain broad-line AGNs by Stern et al.\ (2005; note that their
Fig.~1 shows a few broad-line AGNs and most of the narrow-line 
AGNs lying outside this region). 
We can see from our data that this color-color plot is quite poor 
at picking out lower luminosity AGNs 
(see also Barmby et al.\ 2006; Treister et al.\ 2006;
Donley et al.\ 2007; Cardamone et al.\ 2008), while also
missing one of the more luminous X-ray AGNs and
six of the high radio power sources.
It cleanly selects just a few additional AGNs that are 
not already identified as AGNs based on their X-ray properties, 
though there are a number of sources on the boundaries
(see Figure~\ref{hminus}a).
Thus, it appears that the AGNs are not significantly contaminating 
the colors of most of the galaxies, and hence the colors can 
be used as a redshift measure. 

In Figure~\ref{hminus} we divide the galaxies into $z\le 1.3$ 
{\em (black diamonds)\/} and $z>1.3$ {\em (red upside-down triangles)\/}.
These two broad redshift categories occupy fairly distinct
regions of color-color space, which can be separated by the 
condition $(m_{3.6~\mu{\rm m}}-m_{4.5~\mu{\rm m}})_{\rm AB} = 
0.24\times (m_{5.8~\mu{\rm m}}-m_{8.0~\mu{\rm m}})_{\rm AB}$
{\em (red solid line)\/}. We hereafter refer to this relation
as the H$^-$ selection.

The H$^-$ selection is quite efficient at picking out high-redshift 
galaxies and AGNs, with 84 of the 94 sources with 
$m_{8.0~\mu{\rm m}, {\rm AB}}<22$ and $z>1.3$ satisfying it. This corresponds
to a selection completeness of just under 90\%. However,
there is substantial selection contamination: 53 $z<1.3$
sources also satisfy the H$^-$ selection, giving a 39\% selection
contamination. Thus, as with the $BzK$ selection, the H$^-$ selection
is a very effective way of finding the high-redshift galaxies,
but in order to create a clean sample, spectroscopic
follow-up is needed. (Note that Papovich 2008 suggests that an
efficient way to reduce the contamination due to $z<1.0$ interlopers
is to apply an apparent magnitude cut.)

If we require the sources 
to satisfy both the $BzK$ selection and the H$^{-}$ selection,
then we can substantially reduce the selection
contamination, but at the expense of reducing the
selection completeness to about 80\%. However,
if we instead require the sources to satisfy one or the other
selection, then we can find almost all of the $z>1.4$ 
sources (63/64 with $m_{8.0~\mu{\rm m}, {\rm AB}}<22$) at the expense of 
a high selection contamination (about 55\%). This may be the 
best way to generate a highly complete sample at these redshifts. 

Lacy et al.\ (2004) suggest that interspersed colors,
which are a measure of the curvature of the SED, may
be more optimal in selecting AGNs. 
In Figure~\ref{hminus_2} we show the 
$(m_{3.6~\mu{\rm m}}-m_{5.8~\mu{\rm m}})_{\rm AB}$ versus 
$(m_{4.5~\mu{\rm m}}-m_{8.0~\mu{\rm m}})_{\rm AB}$ color-color diagram
for all the sources in Table~\ref{tab1} with $m_{8.0~\mu{\rm m}, {\rm AB}}<22$ 
and (a) a rest-frame hard or soft X-ray luminosity
$<10^{42}$~ergs~s$^{-1}$ and a radio power
$<10^{31}$~ergs~s$^{-1}$~Hz$^{-1}$,
(b) a rest-frame hard or soft 
X-ray luminosity $\ge 10^{42}$~ergs~s$^{-1}$,
(c) a rest-frame hard or soft X-ray luminosity 
$<10^{42}$~ergs~s$^{-1}$ and a radio power
$\ge 10^{31}$~ergs~s$^{-1}$~Hz$^{-1}$.
In (b) we again enclose in black large squares the sources 
with a rest-frame hard or soft X-ray luminosity
$\ge 10^{44}$~ergs~s$^{-1}$ (X-ray quasar luminosity).

We mark with green dashed lines the region considered to 
contain AGNs by Lacy et al.\ (2004).
This selection is better than the Stern et al.\ (2005) 
selection in choosing luminous AGNs, but again it misses 
some of the lower luminosity sources, as well as three (one is
just outside the boundary) of the high radio power sources.  
However, the galaxy contamination is severe. 

In Figure~\ref{hminus_2} we
divide the galaxies into $z\le 1.6$ {\em (black diamonds)\/}
and $z>1.6$ {\em (red upside-down triangles)\/}. These two broad
redshift categories again fall into fairly distinct regions of 
color-color space, which can be separated by the condition
$(m_{3.6~\mu{\rm m}}-m_{5.8~\mu{\rm m}})_{\rm AB} = 
0.24\times (m_{4.5~\mu{\rm m}}-m_{8.0~\mu{\rm m}})_{\rm AB}$
{\em (red dashed line)\/}.
We hereafter refer to this relation as the IRAC color selection.

The IRAC color selection has a high selection completeness,
selecting 47 of the 50 galaxies and AGNs with $m_{8.0~\mu{\rm m}}
< 22$ and $z>1.6$. However, once again the contamination by lower
redshift sources is substantial (about 40\%), and spectroscopic
follow-up is necessary to provide a clean sample. As with
the H$^-$ selection, requiring the sources to satisfy either the 
$BzK$ selection or the IRAC color selection can provide an extremely 
complete selection (49/50 with $m_{8.0~\mu{\rm m},{\rm AB}}<22$ 
and $z>1.6$) at the expense of a high selection contamination (about 56\%).

%
%
\begin{inlinefigure}
\centerline{\psfig{figure=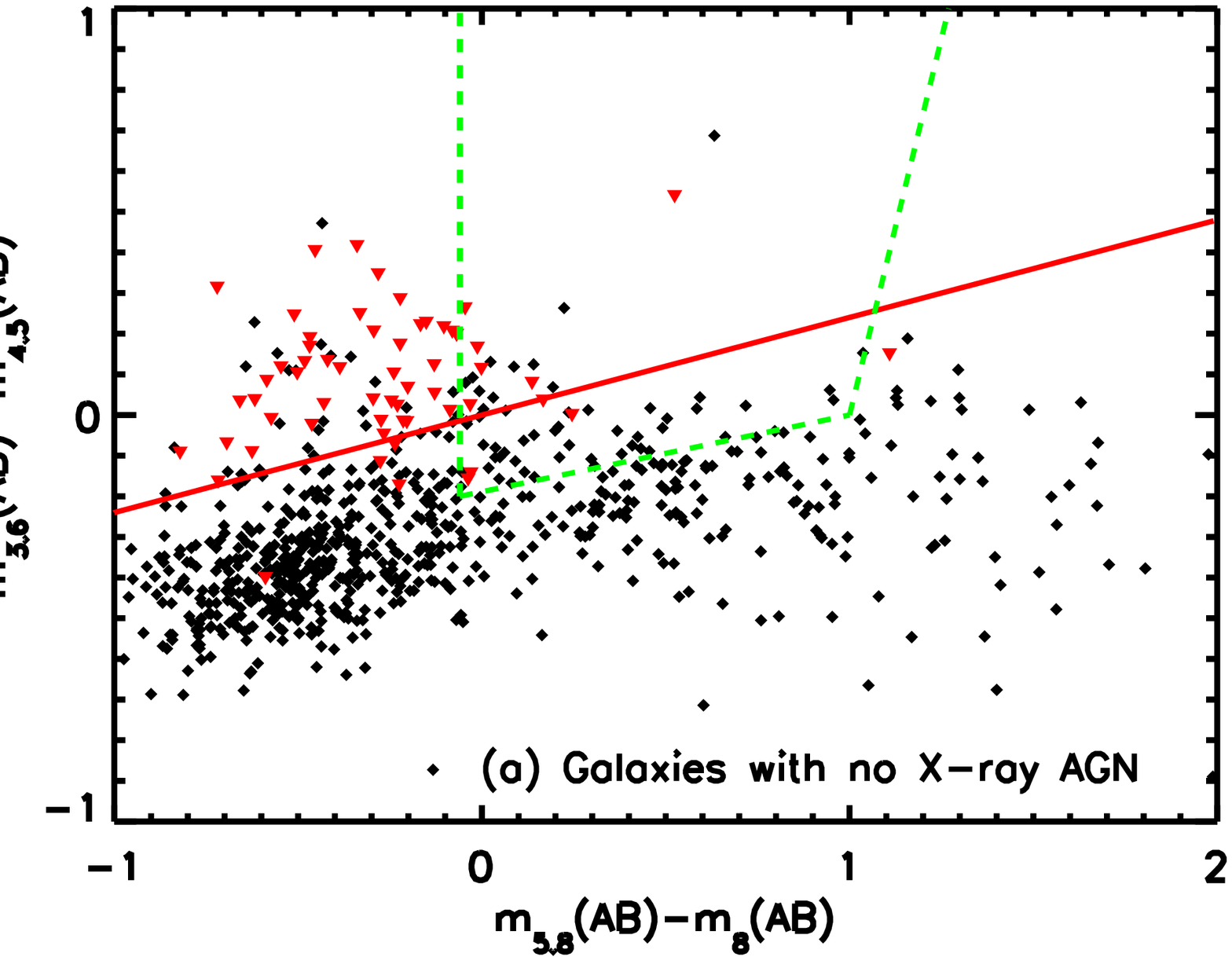,width=3.4in}}
\vskip 0.2cm
\centerline{\psfig{figure=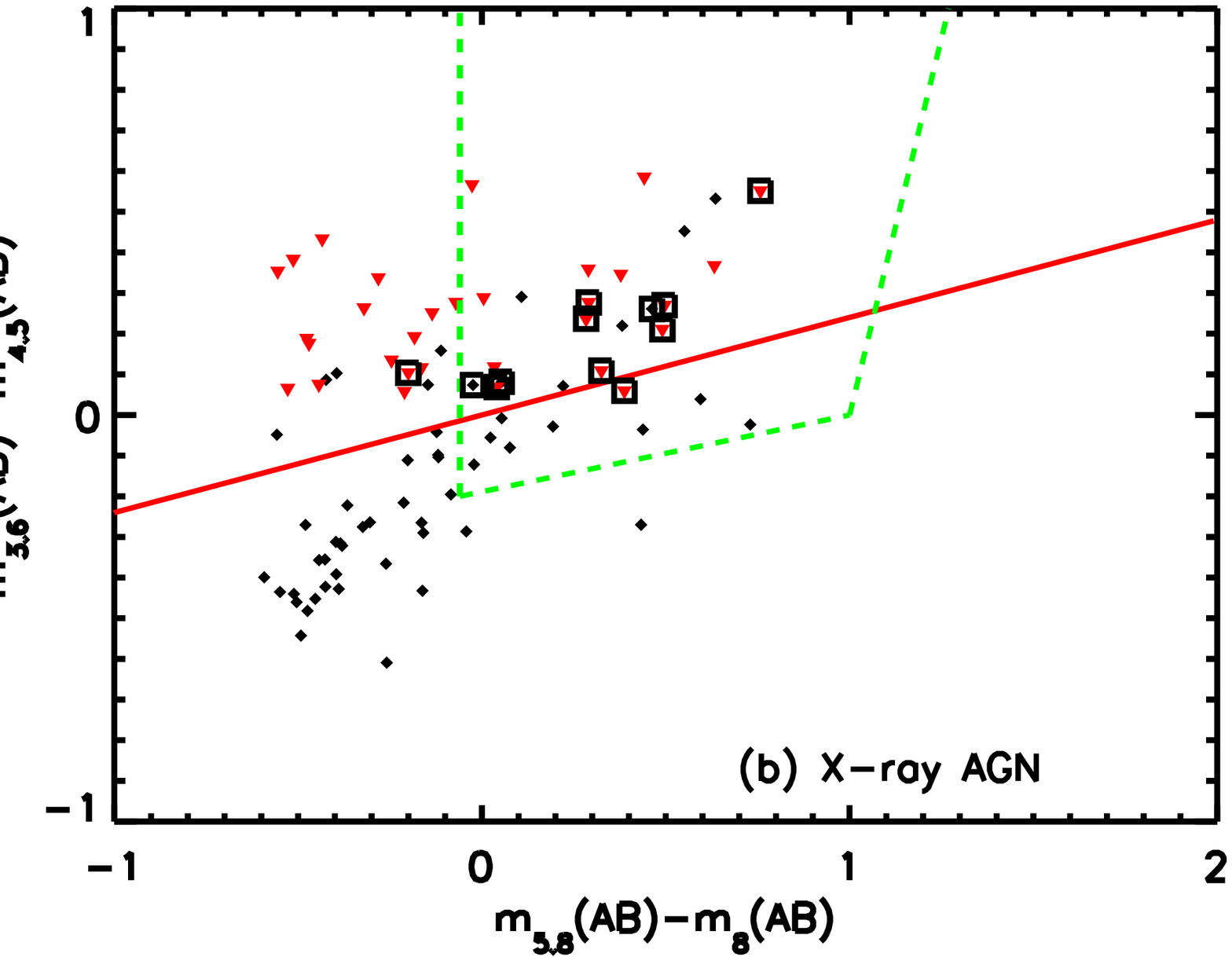,width=3.4in}}
\vskip 0.2cm
\centerline{\psfig{figure=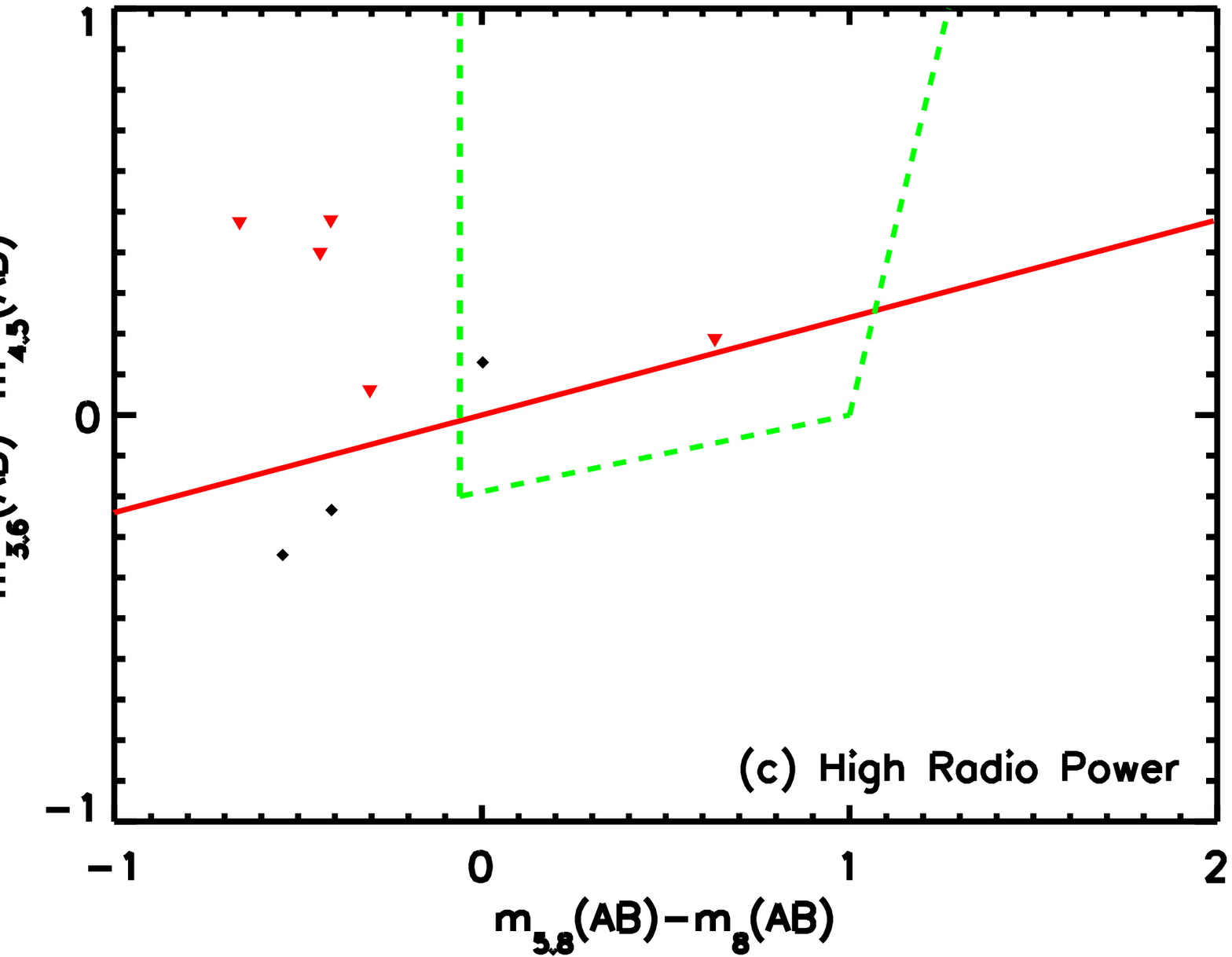,width=3.4in}}
\figcaption[]{
$(m_{3.6~\mu{\rm m}}-m_{4.5~\mu{\rm m}})_{\rm AB}$ vs. 
$(m_{5.8~\mu{\rm m}}-m_{8.0~\mu{\rm m}})_{\rm AB}$ for all 
of the spectroscopically identified galaxies
with $m_{8.0~\mu{\rm m, AB}}<22$ and 
(a) a rest-frame hard or soft X-ray luminosity
$<10^{42}$~ergs~s$^{-1}$ 
and a radio power $<10^{31}$~ergs~s$^{-1}$~Hz$^{-1}$,
(b) a rest-frame hard or soft X-ray luminosity 
$\ge 10^{42}$~ergs~s$^{-1}$, and (c) a rest-frame hard or 
soft X-ray luminosity $<10^{42}$~ergs~s$^{-1}$  
and a radio power $\ge10^{31}$~ergs~s$^{-1}$~Hz$^{-1}$.
Sources with $z\le 1.3$ ($z>1.3$) are 
shown as black diamonds (red upside-down triangles). AGNs with quasar X-ray 
luminosities are enclosed in black large squares.  
The diagonal red solid line shows the condition 
$(m_{3.6~\mu{\rm m}}-m_{4.5~\mu{\rm m}})_{\rm AB} = 
0.24\times (m_{5.8~\mu{\rm m}}-m_{8.0~\mu{\rm m}})_{\rm AB}$, 
which roughly divides the sources in both panels at $z=1.3$. 
The green dashed lines enclose the region where Stern et al.\ (2005) 
suggested that these MIR colors select out most broad-line AGNs.  
\label{hminus}
}
\end{inlinefigure}

%
%
\begin{inlinefigure}
\centerline{\psfig{figure=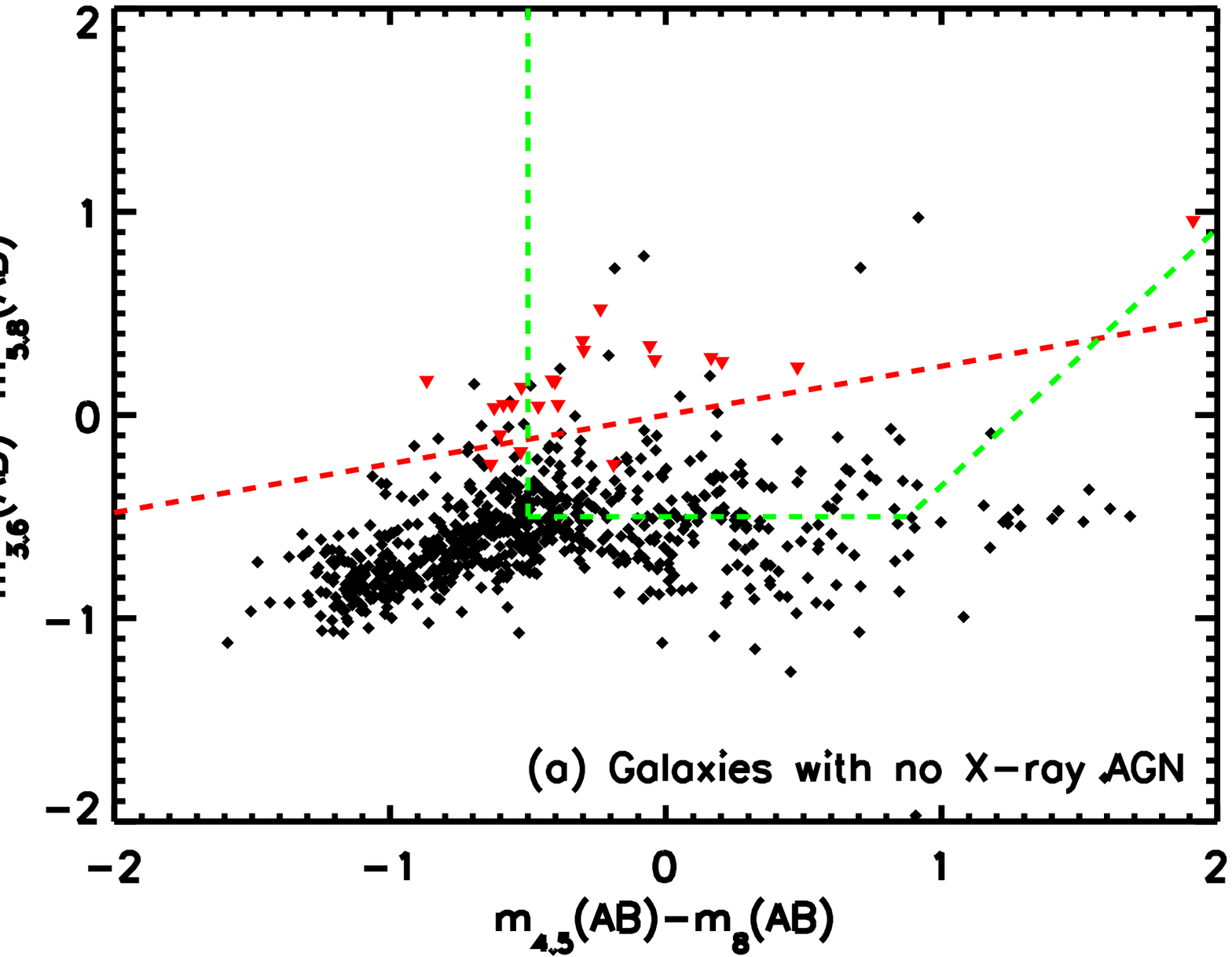,width=3.4in}}
\vskip 0.2cm
\centerline{\psfig{figure=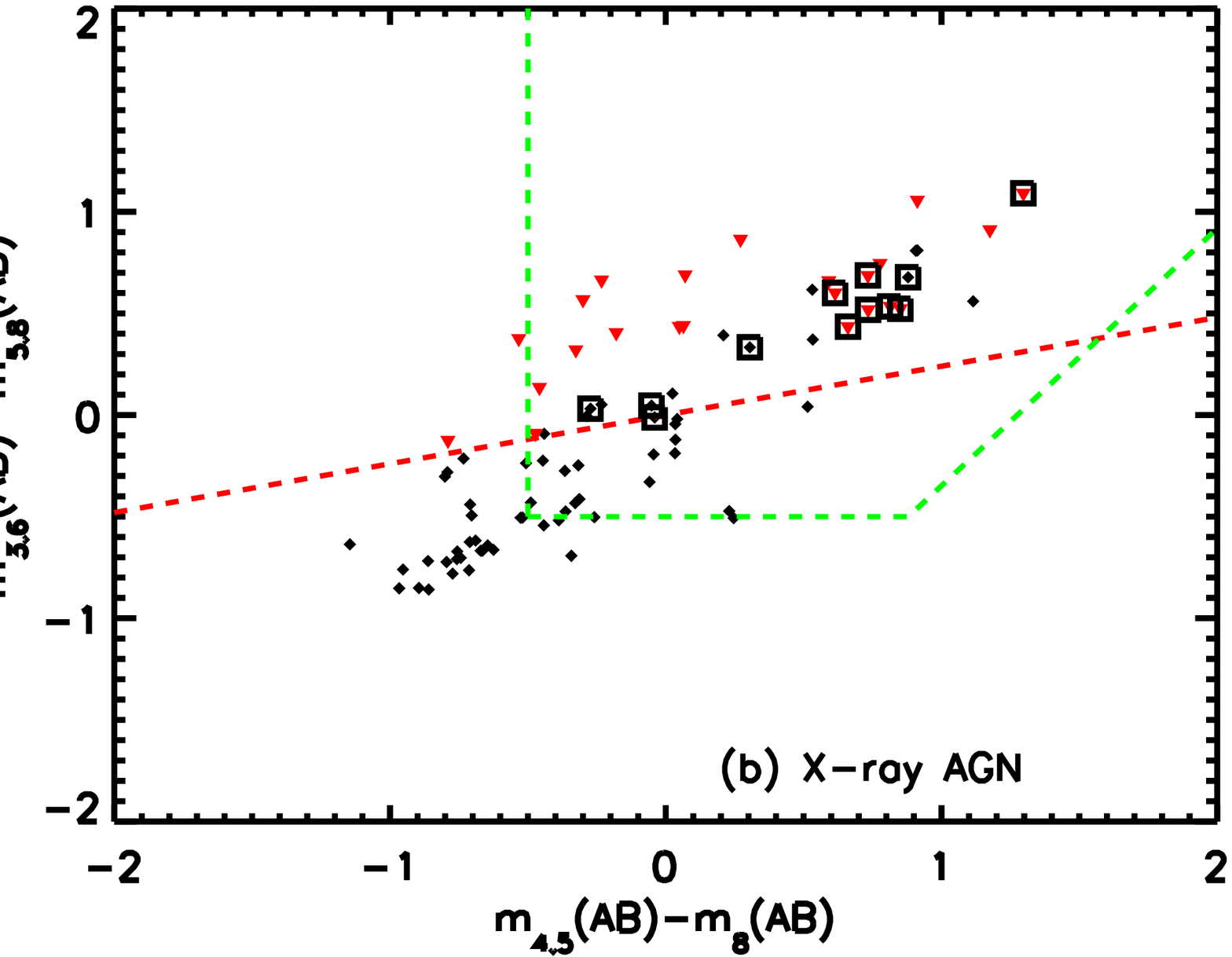,width=3.4in}}
\vskip 0.2cm
\centerline{\psfig{figure=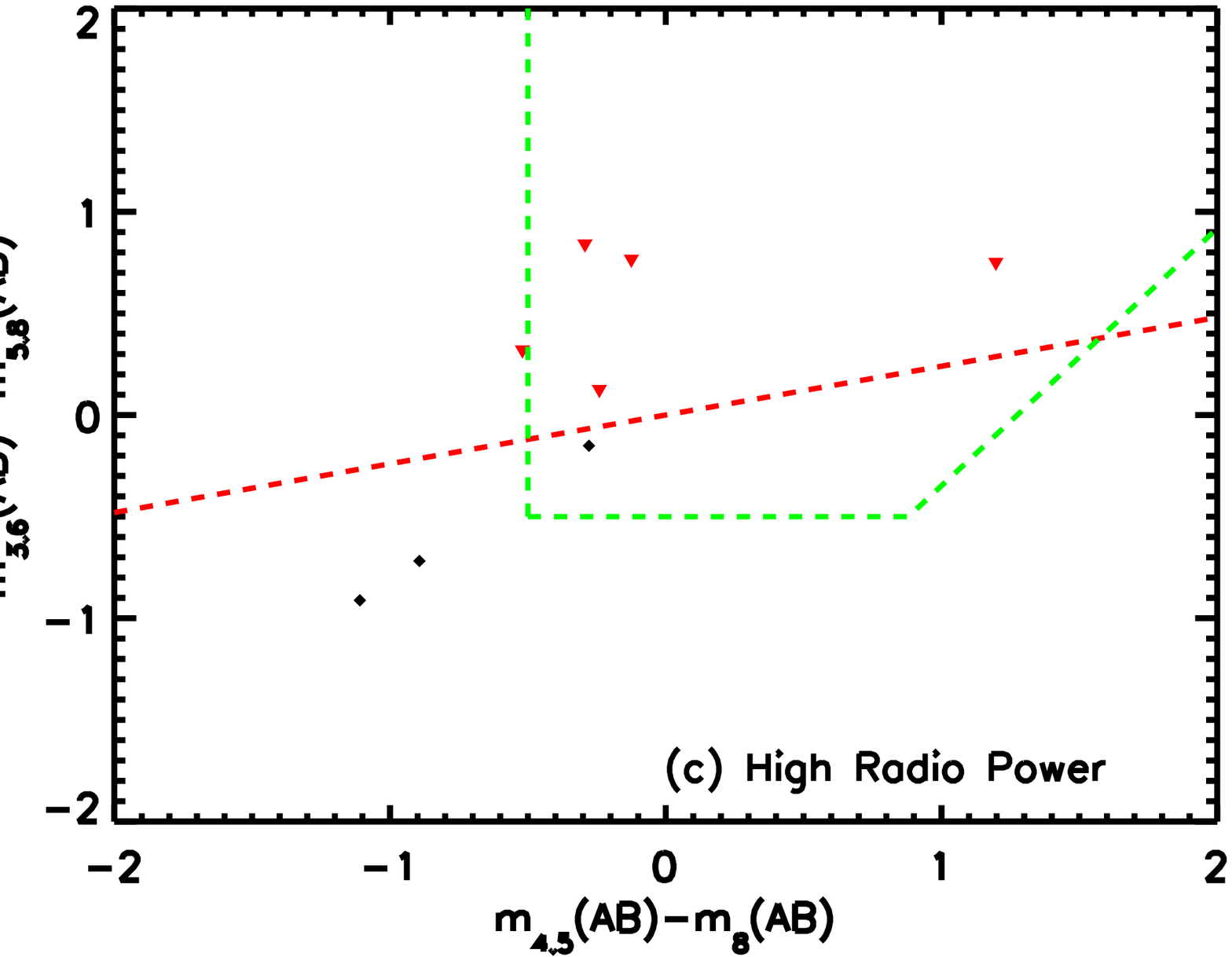,width=3.4in}}
\figcaption[]{
$(m_{3.6~\mu{\rm m}}-m_{5.8~\mu{\rm m}})_{\rm AB}$ vs. 
$(m_{4.5~\mu{\rm m}}-m_{8.0~\mu{\rm m}})_{\rm AB}$ for all 
of the spectroscopically identified galaxies
with $m_{8.0~\mu{\rm m}, {\rm AB}}<22$ and
(a) a rest-frame hard or soft X-ray luminosity
$<10^{42}$~ergs~s$^{-1}$ and a radio power
$<10^{31}$~ergs~s$^{-1}$~Hz$^{-1}$,
(b) a rest-frame hard or soft X-ray luminosity 
$\ge 10^{42}$~ergs~s$^{-1}$, and
(c) a rest-frame hard or soft
X-ray luminosity $<10^{42}$~ergs~s$^{-1}$  
and a radio power $\ge10^{31}$~ergs~s$^{-1}$~Hz$^{-1}$.
Sources with $z\le 1.6$ ($z>1.6$) are shown
as black diamonds (red upside-down triangles). AGNs with quasar
luminosities are enclosed in black large squares. The diagonal red dashed line 
shows the condition $(m_{3.6~\mu{\rm m}}-m_{5.8~\mu{\rm m}})_{\rm AB} =
0.24\times (m_{5.8~\mu{\rm m}}-m_{8.0~\mu{\rm m}})_{\rm AB}$, 
which roughly divides the sources in both panels at $z=1.6$. 
The green dashed lines enclose the region where Lacy et al.\ (2004) 
suggested that these MIR colors select out AGNs.
\label{hminus_2}
}
\end{inlinefigure}

\section{Intergalactic Medium Tomography}
\label{sectom}

One interesting application of this data set is to
investigate the feasibility of one of the major science projects 
proposed for a 30~m telescope: tomography of the IGM. The 
goal of the tomography project is to observe sufficient 
numbers of background sources (galaxies and AGNs) to have enough 
lines of sight to probe the distribution of the intervening gas. 
The Thirty Meter Telescope (TMT) Science Advisory Committee (2007) 
estimated that $R\sim 24.5$ is approximately the apparent 
magnitude at which TMT/WFOS can obtain a spectrum at a 
spectral resolution of 6000 with a signal-to-noise ratio (S/N) 
of about 30 (see their Figure~5-12 for a simulated spectrum of an
$R=24$ galaxy observed with WFOS). The high-resolution
ACS images of the present data set allow us to model this
in more detail by correctly including the slit losses.
Here we do this by measuring the S/N, which can be
obtained with an optimal positioning of each galaxy in the
slit and for a given natural seeing.

%
%
\begin{figure*}
\centerline{\psfig{figure=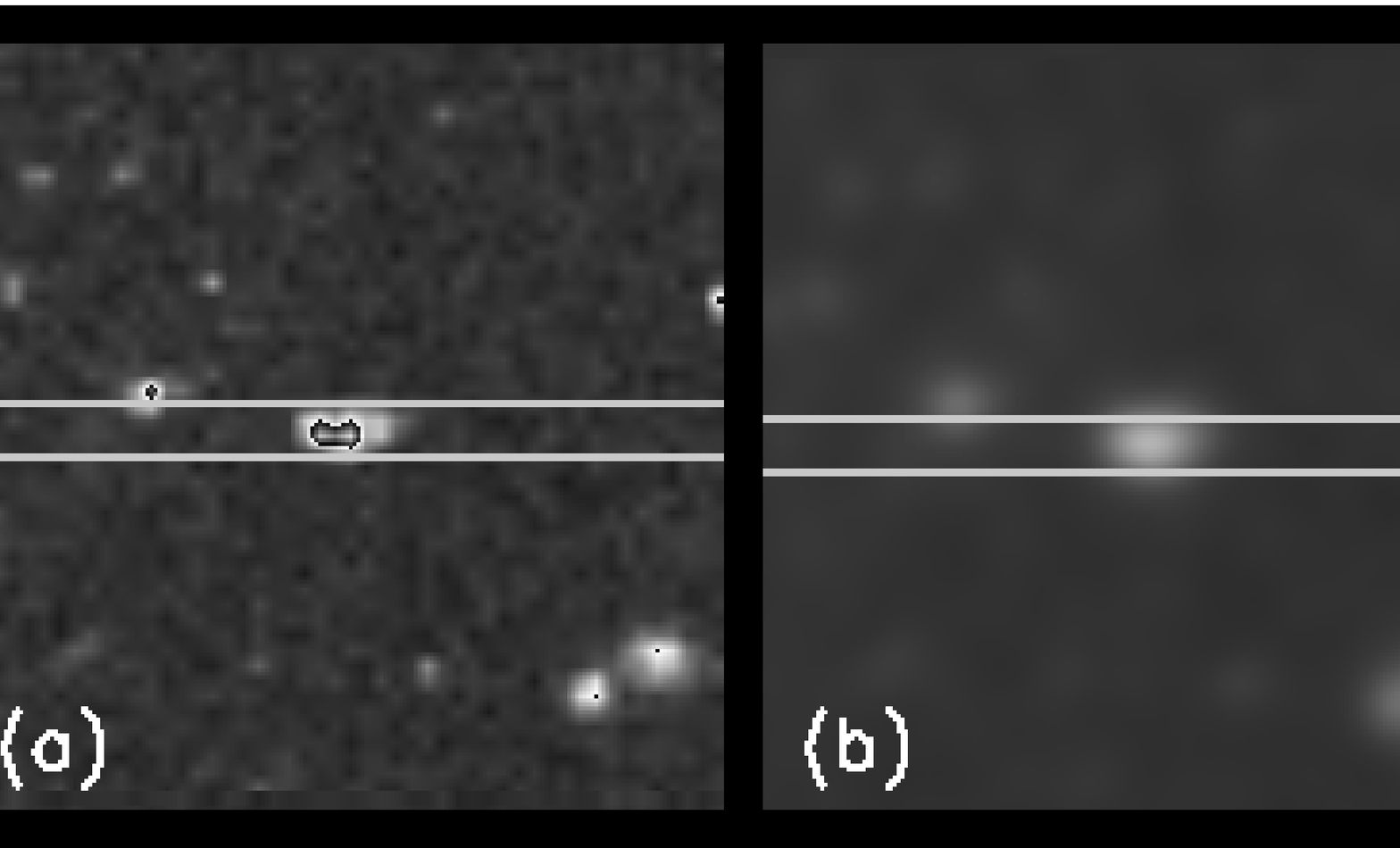,width=3.4in}}
\figcaption[]{A F606W$_{\rm AB}=24$ galaxy at $z=2.05$ 
(a) before and (b) after smoothing to $0.65''$ ground-based 
seeing. The white lines show the position of the $0.75''$ 
wide slit.
\label{image_smooth}
}
\end{figure*}

For the telescope and spectrograph properties, we follow
the TMT Observatory Requirements Document (2007).
We take the telescope's collecting area to be 655~m$^2$.
(The proposed Giant Magellan Telescope has an area of 368~m$^2$,
and the exposure times would be nearly doubled.)
We assume a $0.75''$ wide slit and a spectral resolution of 6000 
for the instrumental parameters. We take a typical sky to collection
efficiency of 25\%, though the peak efficiency could be slightly 
higher than this. We assume a 10~hr integration.

We calculate the S/N at the central wavelength (6060~\AA) of the 
F606W filter. Longer wavelengths are in the complex OH night sky 
spectrum, and hence the calculation of the
S/N becomes extremely wavelength dependent. However, at 6060~\AA\
we can assume a smooth continuum sky brightness. The choice
of sky brightness is not straightforward, since it is somewhat
variable and site dependent. We adopt a continuum AB surface
brightness of 21.4~mags~arcsec$^{-2}$ at this wavelength, which
is based on the mean zenith $V$ (21.7~mags~arcsec$^{-2}$) 
and $R$ (21.1~mags~arcsec$^{-2}$) AB surface brightnesses
at Cerro Paranal (Patat 2003). (The $V$-band sky background at 
Mauna Kea is similar; Krisciunas 1997.) We have also corrected 
for the strong emission line contributions, and we
have assumed that the galaxy is being observed at an average 
airmass of 1.3. The choice of the sky background is probably 
the largest uncertainty in our calculation.

\begin{figure*}
\vskip 0.4cm
\centerline{\psfig{figure=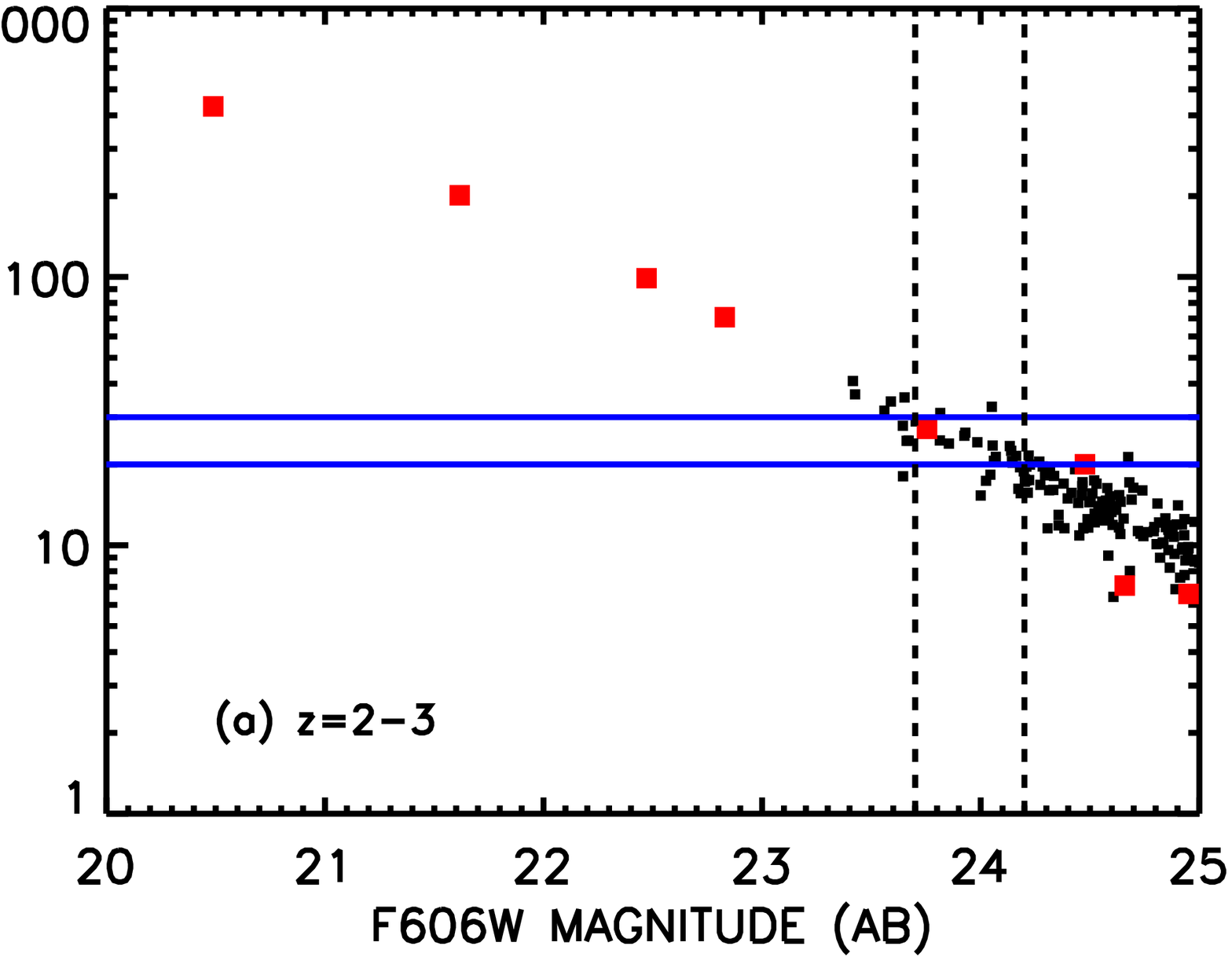,width=3.2in}\hskip 0.8cm \psfig{figure=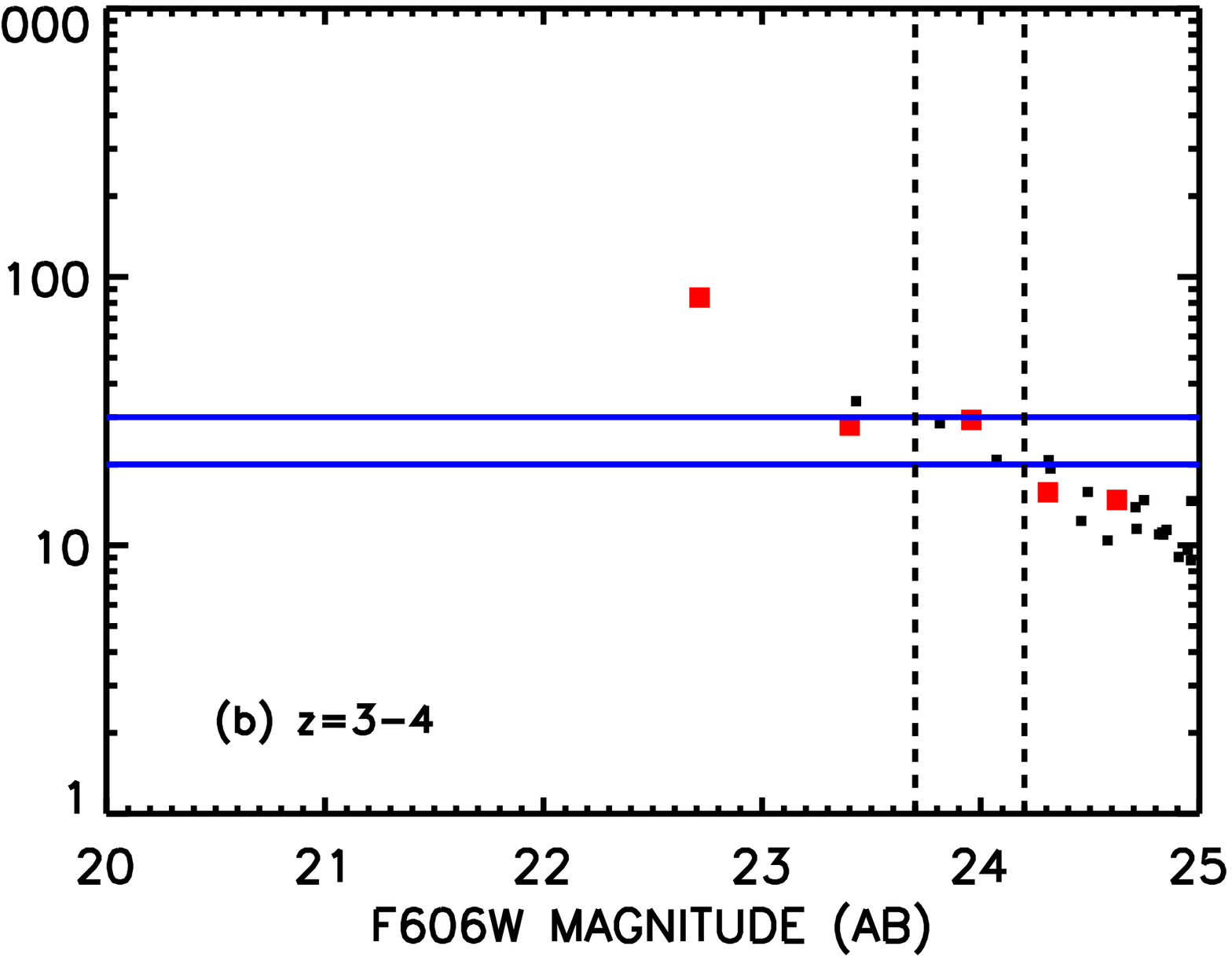,width=3.2in}}
\vskip 0.4cm
\figcaption[]{
Expected S/N per resolution element vs. F606W magnitude for a 
30~m telescope and an exposure time of 10~hr for the sources 
(galaxies and AGNs) in the redshift intervals (a) $z=2-3$ and 
(b) $z=3-4$ in the ACS GOODS-N field. The 
adopted seeing is $0.65''$, the spectral resolution is 6000, and 
the slit width is $0.75''$. AGNs are denoted by red large squares. 
The blue horizontal lines mark S/N per resolution element values of 20 
and 30, and the black vertical dashed lines show the F606W AB 
magnitudes (24.2 and 23.7, respectively) at which these S/N values
are reached for a typical galaxy.  
\label{slit_sn}
}
\end{figure*}

We use the point spread function determined from stars in the 
MOIRCS $K_s$ image with an appropriate scaling to smooth the ACS 
images to the desired ground-based seeing. For our reference 
calculation we adopt a high-quality seeing of $0.65''$, assuming 
that these observations would be prioritized to optimal times.
We show a typical $z=2.05$ galaxy with respect to the $0.75''$
wide slit in Figure~\ref{image_smooth} at (a) the original ACS 
resolution and (b) the ground-based seeing. We place each 
galaxy in an optimal position in the slit and compute the S/N 
which would be obtained in an optimal spectral extraction of the 
light distribution along the slit. Because the noise is sky-dominated, 
the fainter portions of the light profile do not contribute much to 
the S/N. In Figure~\ref{slit_sn} we show the expected S/N per 
resolution element for all the galaxies {\em (black squares)\/} 
and AGNs {\em (red large squares)\/} with 
${\rm F606W\/}_{\rm AB}<25$ and known redshifts in the 
(a) $z=2-3$ and (b) $z=3-4$ ranges versus F606W magnitude.

%
%

\begin{figure*}
\vskip 4cm
\centerline{\psfig{figure=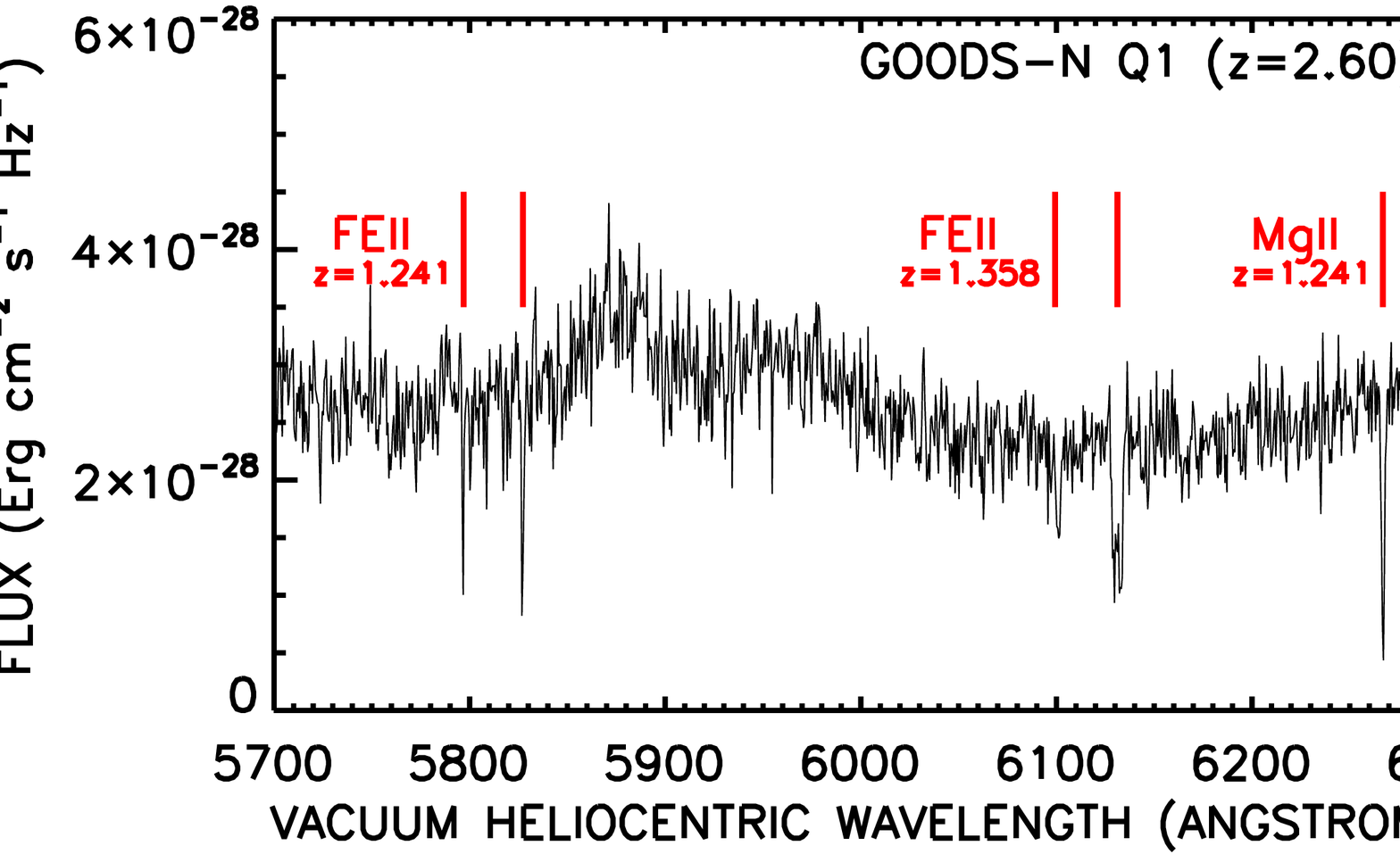,width=3.4in}}
\figcaption[]{Spectrum of the brightest AGN in the $z=2-3$
range in the ACS GOODS-N field (Songaila 2008, in preparation). 
The spectrum is a 1.5~hr exposure with a spectral
resolution of 5400 obtained with a $0.75''$ wide slit under 
$0.9''$ seeing using ESI on Keck~II. The spectrum is shown with 
pixels spaced to Nyquist sample the resolution, and the S/N is 40 
per resolution element. Fe and Mg lines from two absorption systems 
at $z=1.241$ and $z=1.358$ are marked with the red lines. 
\label{showsp}
}
\end{figure*}

We test our S/N calculation using the spectrum of the brightest 
AGN in the $z=2-3$ range in the ACS GOODS-N field, the source with 
${\rm F606W\/}_{\rm AB}=20.5$ in Figure~\ref{slit_sn}a. 
The redshift of this source is $z=2.6$. We use a 1.5~hr 
spectrum obtained with the Echellette Spectrograph and Imager
(ESI; Epps \& Miller 1998; Sheinis et al.\ 2000) on the Keck~II 
telescope from A. Songaila (2008, in preparation). The observation 
was made with a $0.75''$ wide slit under $0.9''$ seeing with some 
windshake. The spectral resolution is 5400. 
We show a segment of the spectrum around 6000~\AA\
in Figure~\ref{showsp}. The measured S/N is 40 in a resolution
element, while our S/N calculation gives a value of 45 for a 
1.5~hr exposure on a 10~m telescope at this seeing. The slight 
degradation may be a consequence 
of windshake effects on the guiding or of imperfections in the 
extraction of the spectrum, but overall the agreement is excellent.

%
%

The choice of S/N for the tomographic project is complex. 
If we simply want to identify kinematic structures, then we can 
tolerate lower S/N spectra, but if we wish to measure column 
densities and abundances, then we need high S/N spectra. In
general the strong lines which can be measured in a low-resolution 
spectrum with low S/N are saturated, and it is not possible to 
derive accurate column densities using them (e.g., Jenkins 1986; 
Prochaska 2007). Prochaska (2007) shows that in order to measure
weak lines where useful information on the column densities can 
be derived, we need to detect rest-frame equivalent widths of 
100~m\AA\ or less. For a $5\sigma$ measurement of a 100~m\AA\
absorption-line system at $z=1$, this translates to a S/N of 25 
per resolution element at a spectral resolution of 6000, which 
seems a reasonable figure of merit. However, the surface density 
of sources is extremely sensitive to the limiting magnitude, 
since we are on the exponential tail of the galaxy distribution.
In Figure~\ref{slit_sn} we show two S/N values
that bracket the above figure of merit:
S/N of 20 and S/N of 30 per resolution element
{\em (blue horizontal lines)\/}. 
These intersect with the galaxy track at magnitudes of 
F606W$_{\rm AB}=24.2$ and 23.7 {\em (vertical dashed lines)\/}, 
respectively, for the $z=2-3$ and $z=3-4$ redshift intervals,
resulting in substantially different surface densities.

To quantify this, in Figure~\ref{fignumab} we show the 
cumulative surface densities of sources (galaxies and AGNs) 
in the ACS GOODS-N field versus F606W$_{\rm AB}$ 
magnitude for the redshift intervals (a) $z=2-3$ and (b) $z=3-4$. 
We denote all sources together by black diamonds with $1\sigma$ 
uncertainties {\em (black solid curves)\/} and AGNs alone by red 
open squares with (in Fig.~\ref{fignumab}a) $1\sigma$ uncertainties 
{\em (red dashed curves)\/}. We obtain maximum
surface densities by putting all of the spectroscopically 
unidentified sources into each redshift interval {\em (blue curves)\/}.

%
%
\begin{inlinefigure}
\centerline{\psfig{figure=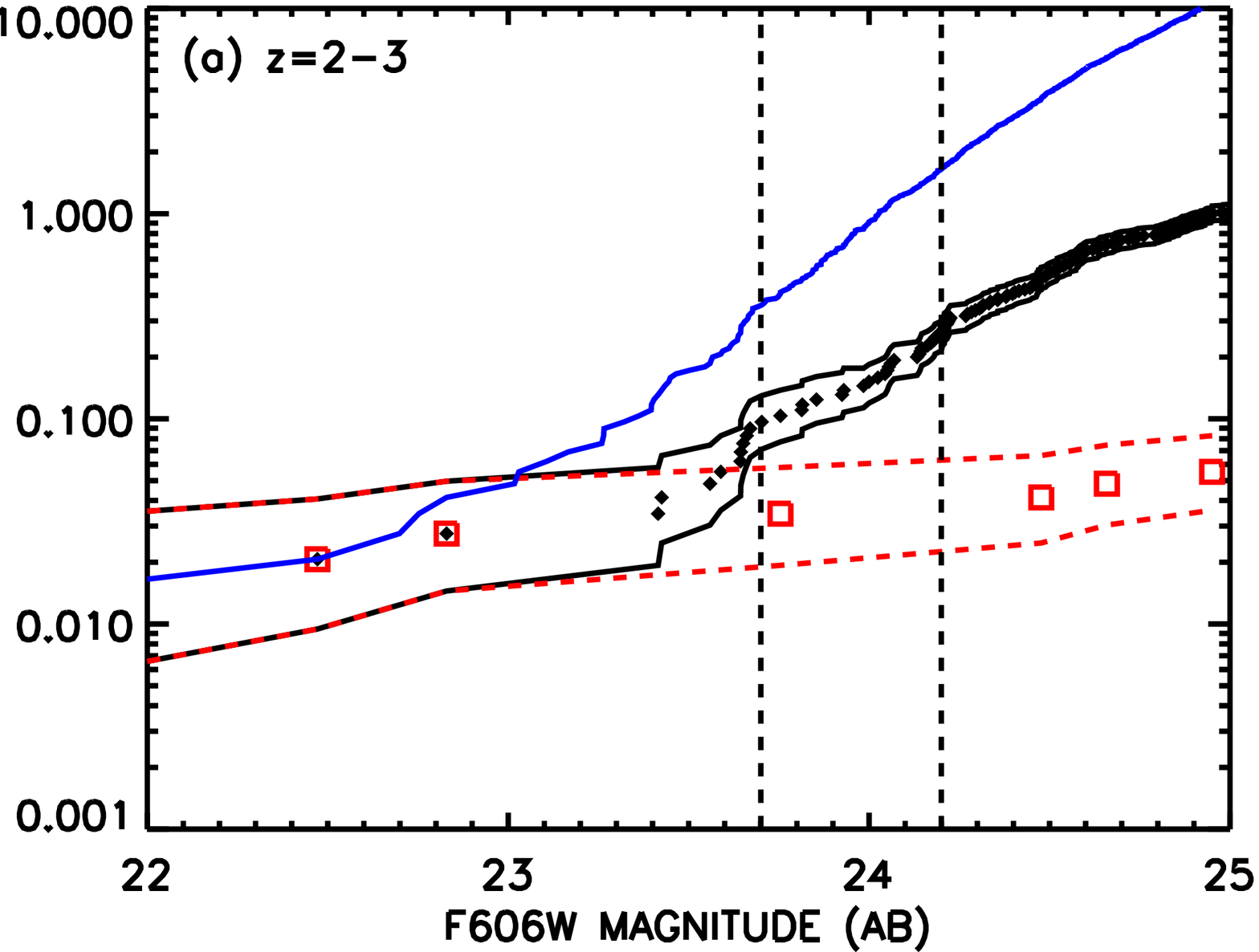,width=3.4in}}
\vskip 0.4cm
\centerline{\psfig{figure=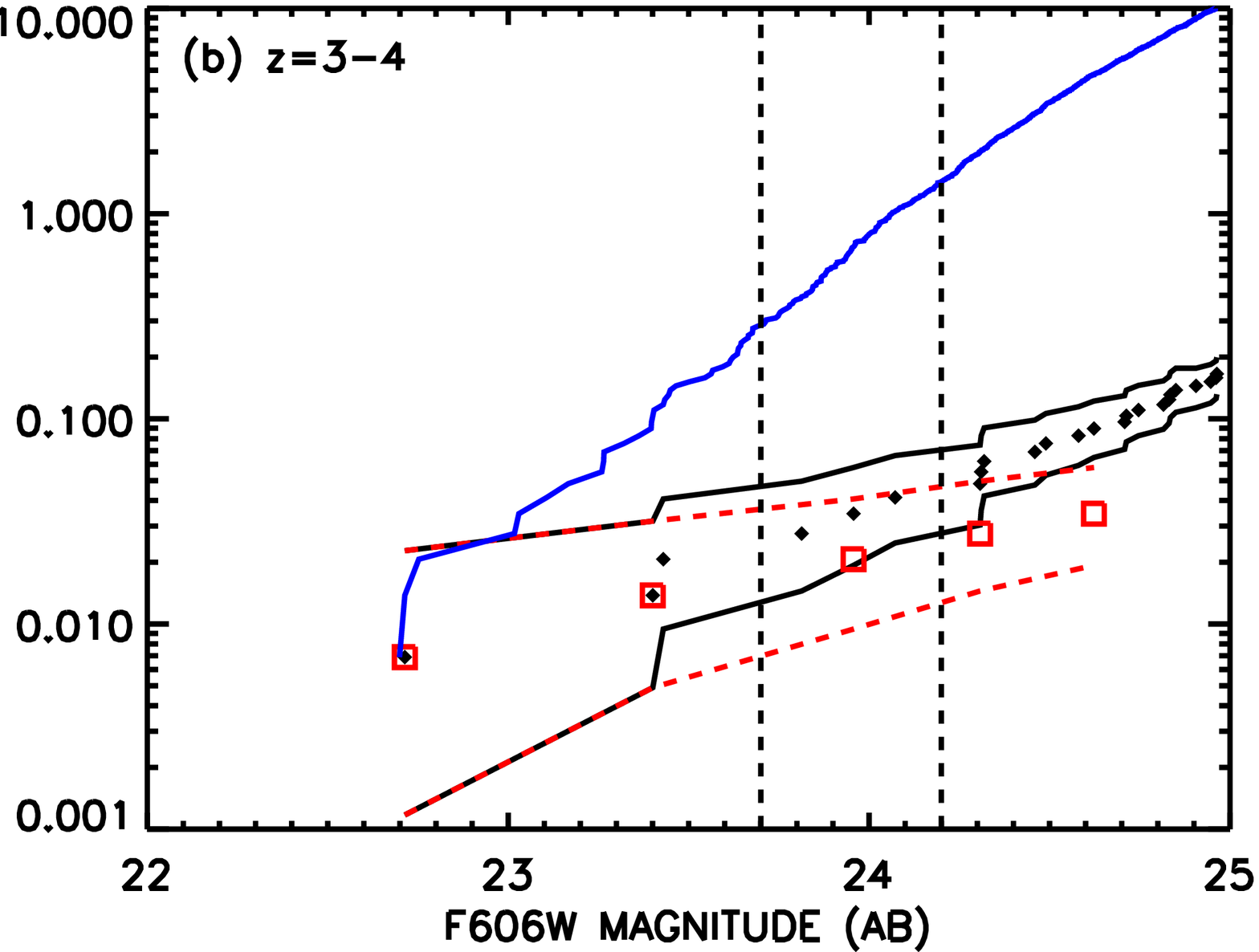,width=3.4in}}
\vskip 0.4cm
\figcaption[]{
Cumulative surface densities of sources (galaxies and AGNs) 
in the ACS GOODS-N field vs. F606W$_{\rm AB}$ magnitude
for the redshift intervals (a) $z=2-3$ and (b) $z=3-4$ 
{\em (black diamonds with black solid curves for the $1\sigma$ 
uncertainties)\/}. AGNs alone are shown with red open squares. 
The red dashed curves in (a) show the $1\sigma$ uncertainties for 
the AGNs. The maximum surface densities for the two 
redshift intervals are obtained by placing all of the spectroscopically 
unidentified sources into each redshift interval {\em (blue curves)\/}.
The black vertical dashed lines show the F606W AB magnitudes (24.2 
and 23.7, respectively, for S/N per resolution element values of 20 and 30).
\label{fignumab}
}
\end{inlinefigure}

We can see from Figure~\ref{fignumab}a that at magnitudes 
significantly brighter than ${\rm F606W}_{\rm AB}=23.7$ the 
cumulative surface densities are still dominated by AGNs, and 
it is only near ${\rm F606W}_{\rm AB}=24$ that galaxies begin 
to dominate. After allowing for the maximum possible correction 
in each redshift interval, the upper bound on the
surface densities is
$\lesssim1.5$~arcmin$^{-2}$ at ${\rm F606W}_{\rm AB}=24.2$ (S/N$=20$)
in both the $z=2-3$ and $z=3-4$ ranges. This is an extreme upper 
limit, since the unidentified galaxies will not all lie in an 
individual redshift range and many may lie below $z=2$. 
At ${\rm F606W}_{\rm AB}=23.7$ (S/N$=30$) the upper bound on the 
surface densities in both the $z=2-3$ and $z=3-4$ ranges is 
$\lesssim0.4$~arcmin$^{-2}$. Again this is an extreme upper
limit, and the true value may be closer to the measured source
value of $0.1$~arcmin$^{-2}$ in the $z=2-3$ range. 
For a 10~m telescope the limiting magnitudes are 1.2~mag
brighter, which means the surface densities that can be 
observed with a 30~m telescope are a factor of 5 to 20 higher 
than those which can be achieved with a 10~m telescope
(e.g., $0.02$~arcmin$^{-2}$ at ${\rm F606W}_{\rm AB}=22.5$
for $z=2-3$ and S/N$=30$). 

We conclude that with a 30~m telescope it will be possible 
to study examples of near neighbors and to map regions at 
arcminute scales. However, tomography at the sub-arcminute 
level is probably still beyond the reach of even an ideal 
30~m telescope.

\section{Summary}
\label{secsummary}

In this paper we presented deep $K_s$-band imaging and the most 
spectroscopically complete redshift sample obtained to date 
of the 145~arcmin$^2$ {\em HST\/} ACS GOODS-N region. We
provided the data in a variety of tables as a function of
magnitude from the NUV to the MIR.
The redshift identifications are greater than 90\%
complete to magnitudes of ${\rm F435W\/}_{\rm AB}=24.5$,
${\rm F850LP\/}_{\rm AB}=23.3$, and $K_{s,{\rm AB}}=21.5$
and to 24~$\mu$m fluxes of 250~$\mu$Jy.

We used the data to analyze various color selection
techniques that have been proposed to find high-redshift
galaxies. The LBG technique still appears to be the most robust
method for providing a high selection completeness with low contamination.
We do not confirm the presence of additional luminous galaxies
not picked out by the LBG selection, as proposed by 
Le F{\`e}vre et al.\ (2005). 
$BzK$, H$^-$, and IRAC color techniques can be highly efficient 
at selecting high-redshift ($z>1.4$, $z>1.3$, and $z>1.6$, 
respectively) galaxies and AGNs but at the expense of a fairly 
high degree of contamination by lower redshift galaxies. Samples 
selected using these techniques therefore need to be 
spectroscopically followed up in order to obtain clean 
high-redshift samples.

Finally, we have used the ACS images to make detailed
calculations of the S/N which can be obtained with spectral
resolution 6000 seeing-limited spectra on a 30~m telescope. 
We find that spectra with S/N$\ge 30$ per resolution element 
in 10~hr integrations can only be obtained for galaxies brighter 
than ${\rm F606W}_{\rm AB}=23.7$.
We have also computed the observed cumulative surface densities 
of galaxies and AGNs as a function of magnitude in the $z=2-3$
and $z=3-4$ ranges, as well as the maximum surface 
densities which could lie in these intervals when unidentified 
sources are included. These surface densities suggest that 
moderate spectral resolution (6000) observations with S/N$\ge 30$ 
on a 30~m telescope will still primarily be of the AGN 
population and that the samples will be sparse with average 
separations greater than 1~arcmin at $z=2-4$.

\acknowledgments
We thank the anonymous referee for a helpful report.
We gratefully acknowledge support from NSF grants
AST 0239425 and AST 0708793 (A.~J.~B.) and 
AST 0407374 and AST 0709356 (L.~L.~C.),
the University of Wisconsin Research Committee with funds 
granted by the Wisconsin Alumni Research Foundation, 
and the David and Lucile Packard Foundation (A.~J.~B.).
We would like to thank Greg Wirth, Jim Lyke, Marc Kassis,
and Grant Hill for their unstinting help with DEIMOS 
through the course of this project. This research used the facilities 
of the Canadian Astronomy Data Centre operated by the National 
Research Council of Canada with the support of the Canadian 
Space Agency. Based in part on data collected at Subaru Telescope 
and obtained from the SMOKA, which is operated by the Astronomy 
Data Center, National Astronomical Observatory of Japan.
Some of the data presented in this paper were obtained from the 
Multimission Archive at the Space Telescope Science Institute (MAST). 
STScI is operated by the Association of Universities for Research in 
Astronomy, Inc., under NASA contract NAS5-26555. Support for MAST 
for non-HST data is provided by the NASA Office of Space Science 
via grant NAG5-7584 and by other grants and contracts.

\clearpage

%
%
\begin{deluxetable}{rccrrrrrrrrrrrrr}
\scriptsize
\tablewidth{0pt}
\tablecaption{ACS GOODS-N Region $K_{s,{\rm AB}}<24.5$ Sample\label{tab1}}
\tablehead{Number & R.A. & Decl. & Flux & Error & $K_s$ & F850LP & F775W & F606W & F435W & $U$ & $z_{spec}$ & source & $f_{24~\mu{\rm m}}$ & $L_{2-8}$ & $L_{0.5-2}$ \\ 
 & (J2000.0) & (J2000.0) & & & & & & & & & & & ($\mu$Jy) & 
\multicolumn{2}{c}{($10^{40}$~ergs~s$^{-1}$)} \\
(1) & (2) & (3) & (4) & (5) & (6) & (7) & (8) & (9) & (10) & (11) & (12) & (13) & (14) & (15) & (16) }
\startdata
       1  &  189.377319  &  62.29675  &  $1905.8$  &  $  0.17   $  &  $ 15.59$
 &  $ 15.95$  &  $ 16.85$  &  $ 17.38$  &  $ 18.03$  &  $ 18.04$  &  star   &  1
 &  \nodata  &  \nodata  &  \nodata  \\ 
       2  &  189.394943  &  62.24295  &  $1893.3$  &  $  0.18   $  &  $ 15.61$
 &  $ 16.16$  &  $ 16.86$  &  $ 17.60$  &  $ 18.48$  &  $ 19.20$  &  star   &   
3                    &  \nodata  &  \nodata  &  \nodata  \\ 
       3  &  189.105804  &  62.23464  &  $1311.5$  &  $  0.17   $  &  $ 15.99$
 &  $ 16.73$  &  $ 17.44$  &  $ 18.75$  &  $ 20.46$  &  $ 21.82$  &  star   &   
3    5                &  \nodata  &  \nodata  &  \nodata  \\ 
       4  &  189.318695  &  62.28870  &  $1145.1$  &  $  0.17   $  &  $ 16.13$
 &  $ 16.93$  &  $ 17.62$  &  $ 19.00$  &  $ 20.73$  &  $ 22.14$  &  star   &  1
 &  \nodata  &  \nodata  &  \nodata  \\ 
       5  &  189.241913  &  62.29114  &  $955.51$  &  $  0.16   $  &  $ 16.32$
 &  $ 16.55$  &  $ 16.99$  &  $ 17.80$  &  $ 18.39$  &  $ 17.71$  &  star   &  1
 &  \nodata  &  \nodata  &  \nodata  \\ 
       6  &  189.069931  &  62.26203  &  $925.89$  &  $  0.16   $  &  $ 16.36$
 &  $ 17.20$  &  $ 17.93$  &  $ 19.39$  &  $ 21.25$  &  $ 22.58$  &  star   &   
3                    &  \nodata  &  \nodata  &  \nodata  \\ 
       7  &  189.436905  &  62.25447  &  $904.58$  &  $  0.16   $  &  $ 16.39$
 &  $ 16.51$  &  $ 17.04$  &  $ 17.73$  &  $ 18.57$  &  $ 17.92$  &  star   &  1
 &  \nodata  &  \nodata  &  \nodata  \\ 
       8  &  189.331253  &  62.21222  &  $858.91$  &  $  0.16   $  &  $ 16.44$
 &  $ 16.97$  &  $ 17.39$  &  $ 18.20$  &  $ 19.27$  &  $ 20.62$  &  star   &  1
3    5                &  \nodata  &  \nodata  &  \nodata  \\ 
       9  &  189.406876  &  62.29158  &  $866.33$  &  $  0.15   $  &  $ 16.44$
 &  $ 16.44$  &  $ 16.87$  &  $ 17.60$  &  $ 17.93$  &  $ 17.45$  &  star   &  1
 &  \nodata  &  \nodata  &  \nodata  \\ 
      10  &  189.171738  &  62.12906  &  $850.35$  &  $  0.16   $  &  $ 16.45$
 &  $ 16.71$  &  $ 17.02$  &  $ 18.05$  &  $ 18.73$  &  $ 18.86$  &  star   &   
3                    &  \nodata  &  \nodata  &  \nodata  \\ 
      11  &  189.324051  &  62.36736  &  $591.90$  &  $  0.18   $  &  $ 16.89$
 &  $ 17.43$  &  $ 18.22$  &  $ 19.76$  &  $ 21.68$  &  $ 22.90$  &  star   &  1
 &  \nodata  &  \nodata  &  \nodata  \\ 
      12  &  189.243301  &  62.15983  &  $558.92$  &  $  0.16   $  &  $ 16.90$
 &  $ 16.89$  &  $ 17.11$  &  $ 17.89$  &  $ 18.44$  &  $ 18.26$  &  star   &  1
5                &  \nodata  &  \nodata  &  \nodata  \\ 
      13  &  189.059051  &  62.22419  &  $539.65$  &  $  0.17   $  &  $ 16.95$
 &  $ 17.42$  &  $ 18.10$  &  $ 19.32$  &  $ 21.04$  &  $ 22.32$  &  star   &   
3                    &  \nodata  &  \nodata  &  \nodata  \\ 
      14  &  189.195038  &  62.27681  &  $490.52$  &  $  0.17   $  &  $ 17.06$
 &  $ 17.42$  &  $ 18.02$  &  $ 18.84$  &  $ 20.22$  &  $ 21.64$  &  star   &  1
5                &  \nodata  &  \nodata  &  \nodata  \\ 
      15  &  189.315536  &  62.33242  &  $480.08$  &  $  0.15   $  &  $ 17.07$
 &  $ 17.57$  &  $ 18.49$  &  $ 19.86$  &  $ 21.77$  &  $ 22.98$  &  star   &  1
 &  \nodata  &  \nodata  &  \nodata  \\ 
      16  &  189.484085  &  62.25370  &  $611.27$  &  $  0.30   $  &  $ 17.09$
 &  $ 17.90$  &  $ 18.15$  &  $ 18.74$  &  $ 20.29$  &  $ 20.94$  &  0.1901  & 
1     3        7            &  \nodata  &  \nodata  &  \nodata  \\ 
      17  &  189.316330  &  62.19973  &  $543.37$  &  $  0.25   $  &  $ 17.29$
 &  $ 17.78$  &  $ 18.08$  &  $ 18.71$  &  $ 20.07$  &  $ 21.11$  &  0.1063  & 
1     3                    &  $ 221$  &  \nodata  &  $   0.17$  \\ 
      18  &  189.287079  &  62.29700  &  $385.62$  &  $  0.16   $  &  $ 17.30$
 &  $ 17.02$  &  $ 17.31$  &  $ 18.09$  &  $ 18.68$  &  $ 17.70$  &  star   &   
3                    &  \nodata  &  \nodata  &  \nodata  \\ 
      19  &  189.481995  &  62.25178  &  $830.05$  &  $  0.26   $  &  $ 17.34$
 &  $ 18.14$  &  $ 18.50$  &  $ 19.28$  &  $ 20.90$  &  $ 22.35$  &  0.1890  &  
2  3                    &  $ 683$  &  \nodata  &  \nodata  \\ 
      20  &  189.243088  &  62.16611  &  $652.10$  &  $  0.32   $  &  $ 17.38$
 &  $ 17.54$  &  $ 17.74$  &  $ 18.23$  &  $ 19.20$  &  $ 20.39$  &  0.1363  & 
1     3    5    7            &  $ 518$  &  \nodata  &  $   0.38$  \\ 
      21  &  189.304703  &  62.26983  &  $359.44$  &  $  0.16   $  &  $ 17.39$
 &  $ 17.48$  &  $ 17.85$  &  $ 18.61$  &  $ 19.31$  &  $ 20.26$  &  star   &  1
 &  \nodata  &  \nodata  &  \nodata  \\ 
      22  &  189.205917  &  62.22967  &  $410.38$  &  $  0.23   $  &  $ 17.41$
 &  $ 17.99$  &  $ 18.26$  &  $ 18.83$  &  $ 20.14$  &  $ 20.91$  &  0.0893  &  
2  3    5                &  \nodata  &  \nodata  &  $  0.09$  \\ 
      23  &  189.046081  &  62.14755  &  $316.06$  &  $  0.14   $  &  $ 17.51$
 &  $ 17.72$  &  $ 18.23$  &  $ 18.97$  &  $ 20.00$  &  $ 21.34$  &  star   &  1
 &  \nodata  &  \nodata  &  \nodata  \\ 
      24  &  189.429108  &  62.30836  &  $310.04$  &  $  0.15   $  &  $ 17.52$
 &  $ 18.10$  &  $ 18.86$  &  $ 20.84$  &  $ 22.98$  &  $ 24.40$  &  star   &  1
 &  \nodata  &  \nodata  &  \nodata  \\ 
      25  &  189.124496  &  62.29128  &  $310.71$  &  $  0.15   $  &  $ 17.54$
 &  $ 17.98$  &  $ 18.40$  &  $ 19.55$  &  $ 21.42$  &  $ 22.87$  &  star   &  1
3                    &  \nodata  &  \nodata  &  \nodata  \\ 
      26  &  189.356873  &  62.28011  &  $307.88$  &  $  0.15   $  &  $ 17.54$
 &  $ 18.08$  &  $ 18.68$  &  $ 20.33$  &  $ 22.24$  &  $ 23.49$  &  star   &   
3                    &  \nodata  &  \nodata  &  \nodata  \\ 
      27  &  189.515167  &  62.28647  &  $408.17$  &  $  0.24   $  &  $ 17.56$
 &  $ 18.82$  &  $ 19.15$  &  $ 19.83$  &  $ 21.15$  &  $ 21.99$  &  0.2790  & 
1     3                    &  $1430$  &  \nodata  &  \nodata  \\ 
      28  &  189.143616  &  62.20361  &  $391.80$  &  $  0.25   $  &  $ 17.58$
 &  $ 18.79$  &  $ 19.15$  &  $ 19.95$  &  $ 21.45$  &  $ 22.27$  &  0.4575  &  
2  3    5                &  $1290$  &  $  20.3$  &  $  18.5$  \\ 
      29  &  189.373825  &  62.32628  &  $448.78$  &  $  0.22   $  &  $ 17.59$
 &  $ 18.61$  &  $ 18.95$  &  $ 19.84$  &  $ 21.75$  &  $ 22.97$  &  0.3177  & 
1   2                      &  \nodata  &  \nodata  &  \nodata  \\ 
      30  &  189.369370  &  62.30920  &  $285.35$  &  $  0.15   $  &  $ 17.64$
 &  $ 18.13$  &  $ 18.70$  &  $ 20.22$  &  $ 22.17$  &  $ 23.65$  &  star   &   
3                    &  \nodata  &  \nodata  &  \nodata
\enddata
\end{deluxetable}

%
%
\pagestyle{empty}
\begin{deluxetable}{ccccrrrrrrrrrrrr}
\scriptsize
\tablewidth{0pt}
\tablecaption{Spitzer GOODS-N Region 24~$\mu$m Sample\label{tab2}}
\tablehead{R.A. & Decl.& R.A. & Decl. & $f_{24}$ & $K_s$ & F850LP & F775W & F606W & F435W & $U$ & $z_{spec}$ & source & $L_{2-8}$ & $L_{0.5-2}$ & $f_{20~{\rm cm}}$ \\ 
(J2000.0) & (J2000.0) & (J2000.0) & (J2000.0) & ($\mu$Jy) & & & & & & & & & \multicolumn{2}{c}{($10^{40}$~ergs~s$^{-1}$)} & ($\mu$Jy) \\
(1) & (2) & (3) & (4) & (5) & (6) & (7) & (8) & (9) & (10) & (11) & (12) & (13) & (14) & (15) & (16)}
\startdata
189.503754  &  62.22667  &  189.503876  &  62.22667  &  $4290$  &  $ 17.28$  & 
$ 18.41$  &  $ 18.72$  &  \nodata  &  $ 19.11$  &  $ 18.59$  &  0.4430  &      
3                    &  $ 3019.$  &  $  3051$  &  $ 190$  \\ 
189.125214  &  62.09511  &  189.124863  &  62.09505  &  $2460$  &  $ 18.86$  & 
$ 20.24$  &  $ 20.53$  &  $ 21.35$  &  $ 25.76$  &  $ 22.64$  &  0.4832  &  1   
3                    &  $ 504.1$  &  \nodata  &  $   82$  \\ 
189.033844  &  62.17664  &  189.033783  &  62.17661  &  $2300$  &  $ 19.35$  & 
$ 20.59$  &  $ 20.98$  &  $ 22.24$  &  $ 23.79$  &  $ 23.83$  &  0.6790  &     2
3        7            &  $ 318.0$  &  $  24.2$  &  $ 217$  \\ 
189.469635  &  62.27450  &  189.469711  &  62.27450  &  $1720$  &  $ 17.70$  & 
$ 18.61$  &  $ 18.94$  &  $ 19.46$  &  $ 20.03$  &  $ 19.23$  &  0.3060  &     2
3                    &  $ 1353.$  &  $  797.$  &  $   79$  \\ 
189.622528  &  62.28028  &  189.622498  &  62.28042  &  $1660$  &  $ 17.64$  & 
$ 18.71$  &  $ 18.98$  &  \nodata  &  $ 20.49$  &  $ 20.51$  &  0.2777  &  1    
3                    &  \nodata  &  \nodata  &  \nodata  \\ 
189.204666  &  62.07744  &  189.204224  &  62.07747  &  $1540$  &  $ 15.42$  & 
$ 16.23$  &  $ 17.09$  &  \nodata  &  $ 18.61$  &  $ 19.30$  &  0.1130  &      
3                    &  \nodata  &  $   2.2$  &  $ 850$  \\ 
189.148239  &  62.24006  &  189.148239  &  62.24000  &  $1480$  &  $ 20.34$  & 
$ 22.90$  &  $ 23.27$  &  $ 23.75$  &  $ 24.06$  &  $ 23.57$  &   2.0050  &     
3          8          &  $ 5510.$  &  $  459.$  &  $   87$  \\ 
188.998795  &  62.26389  &  188.998871  &  62.26386  &  $1470$  &  $ 17.88$  & 
$ 18.62$  &  $ 18.94$  &  $ 19.45$  &  $ 24.02$  &  $ 21.00$  &  0.3750  &      
3                    &  \nodata  &  $   7.5$  &  $ 212$  \\ 
189.207001  &  62.12728  &  189.206909  &  62.12725  &  $1450$  &  $ 21.19$  & 
$ 22.87$  &  $ 23.21$  &  $ 23.64$  &  $ 23.89$  &  $ 23.70$  &   1.6095  &  1  
3                    &  $ 30340$  &  $  6040$  &  $ 307$  \\ 
189.515091  &  62.28647  &  189.515167  &  62.28650  &  $1430$  &  $ 17.56$  & 
$ 18.82$  &  $ 19.15$  &  $ 19.83$  &  $ 21.15$  &  $ 21.99$  &  0.2790  &  1   
3                    &  \nodata  &  \nodata  &  $ 253$  \\ 
189.143600  &  62.20364  &  189.143661  &  62.20359  &  $1290$  &  $ 17.57$  & 
$ 18.79$  &  $ 19.15$  &  $ 19.95$  &  $ 21.45$  &  $ 22.20$  &  0.4575  &  1  
2  3    5                &  $  20.7$  &  $  18.9$  &  $ 233$  \\ 
189.051635  &  62.19453  &  189.051940  &  62.19458  &  $1240$  &  $ 18.05$  & 
$ 18.96$  &  $ 19.24$  &  $ 19.86$  &  $ 21.09$  &  $ 21.45$  &  0.2760  &  1  
2  3                    &  \nodata  &  \nodata  &  $ 174$  \\ 
189.319382  &  62.29261  &  189.319412  &  62.29264  &  $1230$  &  $ 20.16$  & 
$ 22.11$  &  $ 22.68$  &  $ 23.63$  &  $ 24.79$  &  $ 24.75$  &   1.1460  &     
3                    &  $ 11042$  &  $  2719$  &  $ 346$  \\ 
189.000534  &  62.17981  &  189.000717  &  62.17978  &  $1220$  &  $ 22.10$  & 
$ 24.32$  &  $ 24.63$  &  $ 25.08$  &  $ 25.58$  &  $ 24.18$  &   2.0020  &     
8          &  \nodata  &  \nodata  &  $ 131$  \\ 
189.013488  &  62.18639  &  189.013534  &  62.18636  &  $1210$  &  $ 18.53$  & 
$ 20.00$  &  $ 20.30$  &  $ 21.22$  &  $ 22.38$  &  $ 22.37$  &  0.6380  &     2
3        7            &  \nodata  &  $  23.9$  &  $ 124$  \\ 
189.290573  &  62.14475  &  189.290619  &  62.14481  &  $1120$  &  $ 19.34$  & 
$ 21.40$  &  $ 21.91$  &  \nodata  &  $ 24.19$  &  $ 23.94$  &  0.9020  &      
3                    &  \nodata  &  $  38.0$  &  \nodata  \\ 
189.392761  &  62.15878  &  189.392853  &  62.15878  &  $1110$  &  $ 17.40$  & 
$ 18.65$  &  $ 19.01$  &  \nodata  &  $ 20.82$  &  $ 21.27$  &  0.1890  &      
3                    &  \nodata  &  $   2.4$  &  $ 142$  \\ 
189.228271  &  62.07419  &  189.228210  &  62.07411  &  $1040$  &  $ 17.87$  & 
$ 18.92$  &  $ 19.26$  &  \nodata  &  $ 21.17$  &  $ 21.35$  &  0.2870  &      
3                    &  \nodata  &  $  10.8$  &  $   60$  \\ 
189.590714  &  62.28514  &  189.590683  &  62.28528  &  $1030$  &  $ 20.20$  & 
$ 21.34$  &  $ 21.90$  &  \nodata  &  $ 23.47$  &  $ 23.43$  &   1.5960 :  &  1 
 &  \nodata  &  \nodata  &  \nodata  \\ 
189.356552  &  62.32817  &  189.356339  &  62.32814  &  $1000$  &  $ 17.96$  & 
$ 18.13$  &  $ 18.32$  &  $ 18.73$  &  $ 19.66$  &  $ 20.68$  &  0.2780  &     2
3                    &  \nodata  &  $   2.1$  &  $ 122$  \\ 
189.426178  &  62.25508  &  189.426178  &  62.25508  &  $ 992$  &  $ 18.18$  & 
$ 18.44$  &  $ 18.57$  &  $ 18.96$  &  $ 19.70$  &  $ 20.42$  &  0.0700  &     2
3        7            &  $   1.4$  &  $  00.3$  &  \nodata  \\ 
189.212997  &  62.17525  &  189.212997  &  62.17522  &  $ 984$  &  $ 18.54$  & 
$ 19.93$  &  $ 20.36$  &  $ 21.17$  &  $ 22.71$  &  $ 23.25$  &  0.4100  &     2
3    5  6              &  \nodata  &  $   5.6$  &  $   95$  \\ 
189.194458  &  62.14256  &  189.194458  &  62.14258  &  $ 982$  &  $ 19.15$  & 
$ 20.65$  &  $ 21.18$  &  $ 22.13$  &  $ 22.85$  &  $ 22.94$  &  0.9710  &     2
3        7            &  \nodata  &  $  21.3$  &  $   80$  \\ 
189.081161  &  62.21464  &  189.081115  &  62.21461  &  $ 976$  &  $ 18.66$  & 
$ 19.72$  &  $ 20.06$  &  $ 20.93$  &  $ 22.36$  &  $ 23.15$  &  0.4734  &     2
3    5                &  \nodata  &  $   6.7$  &  $ 108$  \\ 
189.338242  &  62.21308  &  189.338684  &  62.21311  &  $ 966$  &  $ 18.03$  & 
$ 17.48$  &  $ 17.65$  &  $ 18.13$  &  $ 18.92$  &  $ 20.74$  &  0.1060  &     2
3    5                &  \nodata  &  \nodata  &  \nodata  \\ 
189.360474  &  62.34072  &  189.360413  &  62.34069  &  $ 932$  &  $ 21.24$  & 
$ 23.58$  &  $ 24.07$  &  $ 24.71$  &  $ 25.92$  &  $ 26.16$  &   2.3650  &  1  
 &  $ 4325.$  &  $  498.$  &  $ 102$  \\ 
188.991425  &  62.26025  &  188.991425  &  62.26022  &  $ 925$  &  $ 17.92$  & 
$ 19.25$  &  $ 19.74$  &  \nodata  &  $ 22.08$  &  $ 22.45$  &  0.3760  &      
3                    &  \nodata  &  \nodata  &  $ 209$  \\ 
189.278549  &  62.28392  &  189.278625  &  62.28392  &  $ 903$  &  $ 19.00$  & 
$ 19.86$  &  $ 19.88$  &  $ 19.84$  &  $ 20.08$  &  $ 19.36$  &   1.0190  &  1  
3  4                  &  $ 10153$  &  $  8434$  &  \nodata  \\ 
189.460556  &  62.19550  &  189.460327  &  62.19536  &  $ 889$  &  $ 17.17$  & 
$ 18.51$  &  $ 18.90$  &  \nodata  &  $ 20.90$  &  $ 21.46$  &  0.1890  &      
3                    &  \nodata  &  $   1.2$  &  $ 198$  \\ 
189.365570  &  62.17667  &  189.365387  &  62.17667  &  $ 871$  &  $ 17.40$  & 
$ 17.70$  &  $ 17.92$  &  $ 18.48$  &  $ 22.72$  &  $ 20.93$  &  0.2140  &      
3                    &  \nodata  &  $   1.3$  &  \nodata
\enddata
\end{deluxetable}

%
%
\begin{deluxetable}{ccrrrrrrrrrrrrrr}
\scriptsize
\tablewidth{0pt}
\tablecaption{ACS GOODS-N Region NUV$+$FUV Sample\label{tab3}}
\tablehead{R.A. & Decl. & NUV & FUV & $f_{24}$ & $K_s$ & F850LP & F775W & F606W & F435W & $U$ & $z_{spec}$ & source & $L_{2-8}$ & $L_{0.5-2}$ & $f_{20~{\rm cm}}$ \\
(J2000.0) & (J2000.0) & & & ($\mu$Jy) & & & & & & & & & \multicolumn{2}{c}{($10^{40}$~ergs~s$^{-1}$)} & ($\mu$Jy) \\  
(1) & (2) & (3) & (4) & (5) & (6) & (7) & (8) & (9) & (10) & (11) & (12) & (13) & (14) & (15) & (16)}
\startdata
189.278625  &  62.28391  &  $ 19.89$  &  $ 20.65$  &  $ 903$  &  $ 19.24$  & 
$ 19.86$  &  $ 19.88$  &  $ 19.84$  &  $ 20.08$  &  $ 19.36$  &   1.0190  &  1  
3  4                  &  $ 10032$  &  $  8334$  &  \nodata  \\ 
189.153595  &  62.19299  &  $ 20.06$  &  $ 20.37$  &  $ 732$  &  $ 17.75$  & 
$ 17.98$  &  $ 18.06$  &  $ 18.36$  &  $ 18.86$  &  $ 19.94$  &  0.0790  &     2
3    5                &  \nodata  &  $  0.0$  &  \nodata  \\ 
189.287094  &  62.29699  &  $ 20.30$  &  $ 26.87$  &  \nodata  &  $ 17.46$  & 
$ 17.02$  &  $ 17.31$  &  $ 18.09$  &  $ 18.68$  &  $ 17.70$  &  star   &      
3                    &  \nodata  &  \nodata  &  \nodata  \\ 
189.356339  &  62.32814  &  $ 20.49$  &  $ 21.25$  &  $1000$  &  $ 17.47$  & 
$ 18.13$  &  $ 18.32$  &  $ 18.73$  &  $ 19.66$  &  $ 20.65$  &  0.2780  &     2
3                    &  \nodata  &  $   2.4$  &  \nodata  \\ 
189.338516  &  62.21310  &  $ 20.70$  &  $ 21.18$  &  $ 966$  &  $ 16.75$  & 
$ 17.48$  &  $ 17.65$  &  $ 18.13$  &  $ 18.92$  &  $ 20.80$  &  0.1060  &     2
3    5                &  \nodata  &  $  0.2$  &  \nodata  \\ 
189.201248  &  62.24068  &  $ 20.78$  &  $ 21.15$  &  $ 460$  &  $ 18.58$  & 
$ 18.93$  &  $ 19.00$  &  $ 19.32$  &  $ 19.89$  &  $ 20.17$  &  0.1390  &     2
3  4  5                &  \nodata  &  $  0.4$  &  \nodata  \\ 
189.243088  &  62.16612  &  $ 20.82$  &  $ 21.23$  &  $ 518$  &  $ 16.95$  & 
$ 17.54$  &  $ 17.74$  &  $ 18.23$  &  $ 19.20$  &  $ 20.39$  &  0.1363  &  1   
3    5    7            &  \nodata  &  $  0.3$  &  \nodata  \\ 
189.469711  &  62.27448  &  $ 20.95$  &  $ 21.36$  &  $1720$  &  $ 17.71$  & 
$ 18.61$  &  $ 18.94$  &  $ 19.46$  &  $ 20.03$  &  $ 19.26$  &  0.3060  &     2
3                    &  $ 1381.$  &  $  813.$  &  \nodata  \\ 
189.103424  &  62.12196  &  $ 20.99$  &  $ 21.46$  &  $ 308$  &  $ 18.58$  & 
$ 18.80$  &  $ 18.89$  &  $ 19.25$  &  $ 19.85$  &  $ 20.30$  &  0.1140  &     2
3                    &  \nodata  &  \nodata  &  \nodata  \\ 
189.166962  &  62.14449  &  $ 21.19$  &  $ 21.57$  &  $ 327$  &  $ 18.00$  & 
$ 18.36$  &  $ 18.52$  &  $ 18.96$  &  $ 19.72$  &  $ 20.42$  &  0.0872  &     2
3                    &  \nodata  &  \nodata  &  \nodata  \\ 
189.173462  &  62.19215  &  $ 21.22$  &  $ 21.42$  &  \nodata  &  $ 19.74$  & 
$ 19.74$  &  $ 19.77$  &  $ 20.03$  &  $ 20.45$  &  $ 21.02$  &  0.0892  &  1  
2      5                &  \nodata  &  $  0.2$  &  \nodata  \\ 
189.256317  &  62.31183  &  $ 21.24$  &  $ 21.90$  &  $ 377$  &  $ 17.35$  & 
$ 18.15$  &  $ 18.40$  &  $ 18.94$  &  $ 20.13$  &  $ 21.49$  &  0.2330  &  1   
3                    &  \nodata  &  $   2.3$  &  \nodata  \\ 
189.189240  &  62.20376  &  $ 21.29$  &  $ 23.83$  &  \nodata  &  $ 22.66$  & 
$ 21.20$  &  $ 20.98$  &  $ 20.79$  &  $ 20.71$  &  $ 20.21$  &  star   &      
3    5                &  \nodata  &  \nodata  &  \nodata  \\ 
189.000701  &  62.23579  &  $ 21.35$  &  $ 21.80$  &  $ 330$  &  $ 18.23$  & 
$ 18.64$  &  $ 18.78$  &  $ 19.21$  &  $ 19.95$  &  $ 20.58$  &  0.1140  &  1  
2                      &  \nodata  &  \nodata  &  \nodata  \\ 
189.258499  &  62.18968  &  $ 21.37$  &  $ 21.70$  &  $ 191$  &  $ 18.54$  & 
$ 18.87$  &  $ 19.01$  &  $ 19.41$  &  $ 20.16$  &  $ 21.06$  &  0.1360  &     2
3    5  6  7            &  \nodata  &  $  0.2$  &  \nodata  \\ 
189.275757  &  62.28677  &  $ 21.46$  &  $ 22.14$  &  $ 596$  &  $ 17.48$  & 
$ 18.25$  &  $ 18.48$  &  $ 19.04$  &  $ 20.23$  &  $ 21.34$  &  0.2540  &     2
3                    &  $   5.9$  &  $  0.9$  &  \nodata  \\ 
189.389099  &  62.22782  &  $ 21.53$  &  $ 21.87$  &  $ 139$  &  $ 18.71$  & 
$ 18.95$  &  $ 19.07$  &  $ 19.47$  &  $ 20.21$  &  $ 20.92$  &  0.1050  &  1  
2          7            &  \nodata  &  \nodata  &  \nodata  \\ 
189.151459  &  62.11858  &  $ 21.60$  &  $ 22.35$  &  $ 296$  &  $ 18.82$  & 
$ 19.47$  &  $ 19.66$  &  $ 20.04$  &  $ 20.89$  &  $ 21.30$  &  0.2760  &     2
3                    &  \nodata  &  \nodata  &  \nodata  \\ 
189.212723  &  62.22241  &  $ 21.61$  &  $ 22.11$  &  \nodata  &  $ 19.30$  & 
$ 19.56$  &  $ 19.64$  &  $ 19.94$  &  $ 20.79$  &  $ 21.41$  &  0.2008  &     2
3    5                &  \nodata  &  \nodata  &  \nodata  \\ 
189.241898  &  62.29115  &  $ 21.63$  &  $ 26.71$  &  \nodata  &  $ 16.51$  & 
$ 16.55$  &  $ 16.99$  &  $ 17.80$  &  $ 18.39$  &  $ 17.71$  &  star   &  1    
 &  \nodata  &  \nodata  &  \nodata  \\ 
189.261353  &  62.26208  &  $ 21.67$  &  $ 22.94$  &  $ 473$  &  $ 18.51$  & 
$ 19.40$  &  $ 19.70$  &  $ 20.39$  &  $ 21.03$  &  $ 21.22$  &  0.5120  &     2
3    5    7            &  $ 2914.$  &  $  686.$  &  \nodata  \\ 
189.406876  &  62.29157  &  $ 21.68$  &  $-27.62$  &  \nodata  &  $ 16.59$  & 
$ 16.44$  &  $ 16.87$  &  $ 17.60$  &  $ 17.93$  &  $ 17.45$  &  star   &  1    
 &  \nodata  &  \nodata  &  \nodata  \\ 
189.052002  &  62.19455  &  $ 21.70$  &  $ 22.59$  &  $1240$  &  $ 18.04$  & 
$ 18.96$  &  $ 19.24$  &  $ 19.86$  &  $ 21.09$  &  $ 21.48$  &  0.2760  &  1  
2  3                    &  \nodata  &  \nodata  &  \nodata  \\ 
189.066177  &  62.21037  &  $ 21.71$  &  $ 22.53$  &  $ 396$  &  $ 18.74$  & 
$ 19.25$  &  $ 19.47$  &  $ 19.93$  &  $ 20.92$  &  $ 21.41$  &  0.2857  &  1   
5                &  \nodata  &  \nodata  &  \nodata  \\ 
189.426239  &  62.25509  &  $ 21.75$  &  $ 22.54$  &  $ 992$  &  $ 18.15$  & 
$ 18.44$  &  $ 18.57$  &  $ 18.96$  &  $ 19.70$  &  $ 20.41$  &  0.0700  &     2
3        7            &  $   1.5$  &  $  0.4$  &  \nodata  \\ 
189.399048  &  62.30290  &  $ 21.76$  &  $ 22.23$  &  \nodata  &  $ 20.81$  & 
$ 20.70$  &  $ 20.81$  &  $ 21.11$  &  $ 22.12$  &  $ 22.20$  &  0.2990  &     2
3                    &  \nodata  &  \nodata  &  \nodata  \\ 
189.299118  &  62.25406  &  $ 21.79$  &  $ 22.38$  &  $   88$  &  $ 19.79$  & 
$ 19.97$  &  $ 20.13$  &  $ 20.42$  &  $ 21.30$  &  $ 21.69$  &  0.2990  &     2
5                &  \nodata  &  \nodata  &  \nodata  \\ 
189.135498  &  62.28315  &  $ 21.80$  &  $ 23.32$  &  $ 735$  &  $ 17.82$  & 
$ 18.60$  &  $ 18.90$  &  $ 19.54$  &  $ 20.64$  &  $ 21.92$  &  0.4370  &     2
3                    &  \nodata  &  \nodata  &  \nodata  \\ 
188.953766  &  62.19576  &  $ 21.92$  &  $ 22.25$  &  \nodata  &  $ 21.98$  & 
$ 19.81$  &  $ 19.88$  &  $ 20.21$  &  $ 20.77$  &  $ 21.91$  &  0.1192  &     2
 &  \nodata  &  \nodata  &  \nodata  \\ 
188.948349  &  62.20189  &  $ 21.98$  &  $ 22.46$  &  \nodata  &  $ 19.72$  & 
$ 19.81$  &  $ 19.91$  &  $ 20.21$  &  $ 20.78$  &  $ 21.26$  &  0.0787  &     2
3                    &  \nodata  &  \nodata  &  \nodata  
\enddata
\end{deluxetable}

%
%
\begin{deluxetable}{ccrrrrrrrrrrr}
\scriptsize
\tablewidth{0pt}
\tablecaption{ACS GOODS-N Region Spectroscopic Redshifts\label{tab4}}
\tablehead{R.A. & Decl. & $K_s$ & F850LP & F775W & F606W & F435W & $U$ & $z_{spec}$ & source & $f_{24}$ & $L_{2-8~{\rm keV}}$ & $L_{0.5-2~{\rm keV}}$ \\ 
(J2000.0) & (J2000.0) & & & & & & & & & ($\mu$Jy) & \multicolumn{2}{c}{($10^{40}$~ergs~s$^{-1}$)} \\
(1) & (2) & (3) & (4) & (5) & (6) & (7) & (8) & (9) & (10) & (11) & (12) & (13)}
\startdata
188.922012  &  62.19910  &  $ 22.63$  &  $ 23.17$  &  $ 23.41$  &  $ 23.64$  & 
$ 23.71$  &  $ 23.32$  &   1.5367  &  1                          &  \nodata  & 
\nodata  &  \nodata  \\ 
188.923355  &  62.19751  &  $ 22.89$  &  $ 23.05$  &  $ 23.00$  &  $ 23.26$  & 
$ 23.22$  &  $ 22.72$  &   1.1073  &     2                       &  \nodata  & 
\nodata  &  \nodata  \\ 
188.925476  &  62.19721  &  $ 23.62$  &  $ 24.04$  &  $ 24.27$  &  $ 24.53$  & 
$ 24.45$  &  $ 24.15$  &   1.1410  &  1                          &  \nodata  & 
\nodata  &  \nodata  \\ 
188.927078  &  62.19710  &  $ 22.50$  &  $ 23.23$  &  $ 23.75$  &  $ 24.31$  & 
$ 24.71$  &  $ 24.42$  &   1.1452  &  1                          &  \nodata  & 
\nodata  &  \nodata  \\ 
188.930313  &  62.19232  &  $ 22.36$  &  $ 22.58$  &  $ 22.72$  &  $ 23.24$  & 
$ 23.90$  &  $ 24.11$  &  0.5583  &     2                       &  \nodata  & 
\nodata  &  \nodata  \\ 
188.931183  &  62.20426  &  $ 22.56$  &  $ 22.80$  &  $ 22.97$  &  $ 23.41$  & 
$ 24.49$  &  $ 24.42$  &  0.3269  &  1                          &  \nodata  & 
\nodata  &  \nodata  \\ 
188.933411  &  62.19930  &  $ 20.96$  &  $ 22.71$  &  $ 23.44$  &  $ 24.32$  & 
$ 24.84$  &  $ 25.02$  &   1.3151  &  1                          &  \nodata  & 
\nodata  &  \nodata  \\ 
188.934525  &  62.20675  &  $ 20.47$  &  $ 22.55$  &  $ 23.51$  &  $ 24.96$  & 
$ 25.89$  &  $ 27.11$  &   1.3140  &  1                          &  \nodata  & 
\nodata  &  \nodata  \\ 
188.938339  &  62.20606  &  $ 20.59$  &  $ 21.38$  &  $ 21.63$  &  $ 22.31$  & 
$ 23.39$  &  $ 23.67$  &  0.4734  &     2                       &  \nodata  & 
\nodata  &  \nodata  \\ 
188.939072  &  62.21012  &  $ 22.68$  &  $ 23.02$  &  $ 23.20$  &  $ 24.03$  & 
$ 24.57$  &  $ 24.68$  &  0.7133  &     2                       &  \nodata  & 
\nodata  &  \nodata  \\ 
188.939331  &  62.19835  &  $ 21.44$  &  $ 22.81$  &  $ 23.53$  &  $ 25.18$  & 
$ 26.54$  &  $ 24.33$  &  0.9599  &  1                          &  \nodata  & 
\nodata  &  \nodata  \\ 
188.939713  &  62.19888  &  $ 20.26$  &  $ 21.45$  &  $ 21.99$  &  $ 22.98$  & 
$ 23.72$  &  $ 23.67$  &  0.9590  &  1                          &  $ 225$  & 
\nodata  &  \nodata  \\ 
188.940292  &  62.19432  &  $ 20.73$  &  $ 21.10$  &  $ 21.25$  &  $ 21.69$  & 
$ 22.37$  &  $ 22.14$  &  0.4733  &  1   2                       &  \nodata  & 
\nodata  &  \nodata  \\ 
188.941605  &  62.20118  &  $ 18.76$  &  $ 20.15$  &  $ 20.72$  &  $ 22.30$  & 
$ 24.44$  &  $ 25.44$  &  0.7968  &  1                          &  \nodata  & 
\nodata  &  \nodata  \\ 
188.941803  &  62.19801  &  $ 21.37$  &  $ 21.71$  &  $ 21.94$  &  $ 22.26$  & 
$ 23.08$  &  $ 22.88$  &  0.2976  &  1                          &  \nodata  & 
\nodata  &  \nodata  \\ 
188.943405  &  62.20610  &  $ 22.86$  &  $ 23.57$  &  $ 24.10$  &  $ 24.52$  & 
$ 24.72$  &  $ 24.75$  &   1.3371  &  1                          &  \nodata  & 
\nodata  &  \nodata  \\ 
188.943649  &  62.20052  &  $ 21.66$  &  $ 21.72$  &  $ 21.86$  &  $ 22.43$  & 
$ 23.61$  &  $ 25.10$  &  star   &     2                       &  \nodata  & 
\nodata  &  \nodata  \\ 
188.948349  &  62.20189  &  $ 19.72$  &  $ 19.81$  &  $ 19.91$  &  $ 20.21$  & 
$ 20.78$  &  $ 21.26$  &  0.0787  &     2  3                     &  \nodata  & 
\nodata  &  \nodata  \\ 
188.949982  &  62.21469  &  $ 23.97$  &  $ 24.56$  &  $ 24.50$  &  $ 24.49$  & 
$ 24.55$  &  $ 23.60$  &   1.7980  &         4                   &  \nodata  & 
\nodata  &  \nodata  \\ 
188.950317  &  62.20343  &  $ 21.39$  &  $ 21.56$  &  $ 21.65$  &  $ 22.14$  & 
$ 22.84$  &  $ 22.84$  &  0.5084  &     2                       &  \nodata  & 
\nodata  &  \nodata  \\ 
188.951263  &  62.18684  &  $ 21.82$  &  $ 22.15$  &  $ 22.28$  &  $ 22.78$  & 
$ 23.31$  &  $ 23.47$  &  0.5065  &     2                       &  \nodata  & 
\nodata  &  \nodata  \\ 
188.951614  &  62.18248  &  $ 19.60$  &  $ 20.41$  &  $ 20.65$  &  $ 21.15$  & 
$ 22.26$  &  $ 22.34$  &  0.3000  &  1                          &  $   86$  & 
\nodata  &  \nodata  \\ 
188.952530  &  62.19215  &  $ 19.54$  &  $ 20.66$  &  $ 21.17$  &  $ 22.67$  & 
$ 24.44$  &  $ 24.61$  &  0.8429  &     2          7             &  \nodata  & 
\nodata  &  \nodata  \\ 
188.952576  &  62.19814  &  $ 22.36$  &  $ 22.55$  &  $ 22.86$  &  $ 23.52$  & 
$ 23.77$  &  $ 23.57$  &   1.0007  &  1   2                       &  \nodata  & 
\nodata  &  \nodata  \\ 
188.953720  &  62.21213  &  $ 22.58$  &  $ 23.28$  &  $ 23.44$  &  $ 23.96$  & 
$ 25.05$  &  $ 24.88$  &  0.4102  &     2                       &  \nodata  & 
\nodata  &  \nodata  \\ 
188.953766  &  62.19576  &  $ 21.98$  &  $ 19.81$  &  $ 19.88$  &  $ 20.21$  & 
$ 20.77$  &  $ 21.91$  &  0.1192  &     2                       &  \nodata  & 
\nodata  &  \nodata  \\ 
188.954742  &  62.18175  &  $ 22.65$  &  $ 23.30$  &  $ 23.45$  &  $ 24.09$  & 
$ 25.00$  &  $ 25.61$  &  0.5075  &     2                       &  \nodata  & 
\nodata  &  \nodata  \\ 
188.956726  &  62.21053  &  $ 19.56$  &  $ 20.51$  &  $ 20.87$  &  $ 21.86$  & 
$ 23.73$  &  $ 24.82$  &  0.4573  &  1             7             &  \nodata  & 
\nodata  &  \nodata  \\ 
188.958984  &  62.19461  &  $ 22.62$  &  $ 23.40$  &  $ 23.83$  &  $ 23.95$  & 
$ 23.99$  &  $ 23.63$  &   1.3274  &  1                          &  \nodata  & 
\nodata  &  \nodata  \\ 
188.960098  &  62.18470  &  $ 22.56$  &  $ 22.80$  &  $ 23.11$  &  $ 23.68$  & 
$ 23.87$  &  $ 23.53$  &   1.0872  &  1                          &  \nodata  & 
\nodata  &  \nodata
\enddata
\end{deluxetable}

%
%
\begin{deluxetable}{ccccc}
\scriptsize
\tablewidth{0pt}
\tablecaption{Spectroscopically Identified Fraction for the Optical and NIR Samples\label{tab5}}
\tablehead{Band & 98\% & 95\% & 90\% & 80\%}
\startdata
F435W  &  23.9  &  24.2  & 24.5 & 24.9 \\
F606W  &  23.6  &  23.9  & 24.2 & 24.5 \\
F775W  &  23.1  &  23.4  & 23.6 & 24.0 \\
F850LP  &  22.6  &  22.9  & 23.3 & 23.7 \\
$K_s$  &  20.7  &  21.0  & 21.5 & 22.2
\enddata
\end{deluxetable}

%
%
\begin{deluxetable}{ccccc}
\scriptsize
\tablewidth{0pt}
\tablecaption{Source of Spectroscopic Redshifts\label{tab6}}
\tablehead{Source & Number & Unique\tablenotemark{a} & Cumulative & Reference  \\ (1) & (2) & (3) & (4) & (5)}
\startdata
1  &  1506  &  842  & 1506 & This work; see also Cowie et al.\ (2004) \\
2  &  1389  &  557  & 2551 & TKRS; Wirth et al.\ (2004) \\
3  &  394  &  19   & 2581 & Our LRIS spectra \\
4  &  343  &  224  & 2853 & Reddy et al.\ (2006) \\
5  &  613  &   13 & 2879 & Cohen et al.\ (2000, 2001) \\
6  &  114  &   1  & 2880 & Lowenthal et al.\ (1997); Phillips et al.\ (1997) \\
7  &  243  &  2  &  2882 & Treu et al.\ (2005) \\
8  &   14  &  8  & 2890 &  Chapman et al.\ (2004, 2005); Swinbank et al.\ (2004) \\
9  &   50  &  3  &  2893 & Steidel et al.\ (2003) \\
10  &   44  &  5  & 2898 &  Dawson et al.\ (2001) \\
11  &  4  &  4  & 2902 & Pope et al.\ (2008) \\
12  &  0  &  0  & 2902 & Barger et al.\ (2003)\tablenotemark{b} \\
13  &  2  &  0  & 2902 & Daddi et al.\ (2008) \\
14  &  8  &  1  & 2903 & MOIRCS spectra
\enddata
\tablenotetext{a}{The number of sources
for which only that catalog has obtained redshift identifications.}
\tablenotetext{b}{We have not distinguished the observations made
in the Barger et al.\ (2003) paper (literature source 12) from our 
DEIMOS and LRIS observations of the ACS GOODS-N area (literature 
sources 1 and 3), which is why zeros appear in the number and unique 
columns for Barger et al.\ (2003). However, we retain Barger et al.\ (2003)
as a literature source in this table because it includes some ESI
identifications which appear in the 24~$\mu$m catalog outside the
ACS GOODS-N area.}
\end{deluxetable}

\end{document}